\documentclass[12pt]{article}
\pdfoutput=1
\usepackage{amsmath,amssymb,amsfonts}
\usepackage[dvipdfmx]{graphicx}
\usepackage{subfigure}
\usepackage[usenames, dvipsnames]{color}
\usepackage{hyperref}
\usepackage{cite}
\usepackage{verbatim} 
\usepackage[mathscr]{eucal} 

\makeatletter

\DeclareGraphicsExtensions{.pdf,.png,.jpg}

\newcommand{\C}{\mathbb{C}}
\renewcommand{\P}{\mathbb{P}}

\newcommand{\Z}{{\mathbb Z}}

\newcommand{\cO}{\mathcal{O}}
\newcommand{\cN}{\mathcal{N}}

\newcommand{\nn}{\nonumber}
\newcommand{\be}{\begin{equation}}
\newcommand{\ee}{\end{equation}}
\newcommand{\bea}{\begin{eqnarray}}
\newcommand{\eea}{\end{eqnarray}}

\renewcommand\paragraph{\@startsection{paragraph}{4}{\z@}%
            {-2.5ex\@plus -1ex \@minus -.25ex}%
            {1.25ex \@plus .25ex}%
            {\normalfont\normalsize\bfseries}}
            
\newcount\hour \newcount\minute
\hour=\time \divide \hour by 60
\minute=\time
\def\now{%
\ifnum \hour<13
  \ifnum \hour=0 \advance \hour by 12 \number\hour:\else \number\hour:\fi%
     \ifnum \minute<10 0\fi%
     \number\minute%
\ A.M.%
\else \advance \hour by -12 \number\hour:%
  \ifnum \minute<10 0\fi%
  \number\minute%
  \ P.M.%
\fi%
}

\makeatother

\begin{document}

\baselineskip=18pt  
\numberwithin{equation}{section}  
\allowdisplaybreaks  


%
%


\thispagestyle{empty}

\vspace*{-2cm} 
\begin{flushright}
IFT-UAM/CSIC-13-101
\\
RUNHETC-2013-22
\end{flushright}

\vspace*{0.5cm} 
\begin{center}
 {\LARGE Topological strings and 5d $T_N$ partition functions}\\

 \vspace*{1.5cm}
{Hirotaka Hayashi$^{1,4}$, Hee-Cheol Kim$^{2,4}$ and Takahiro Nishinaka$^3$}\\

 \vspace*{1.0cm}  
 $^1$  {\it Instituto de F\'isica Te\'orica UAM/CSIS,\\
 Cantoblanco, 28049 Madrid, Spain}\\
{\tt h.hayashi csic.es}\\
\vspace*{0.5cm}
$^2$ {\it Perimeter Institute for Theoretical Physics, \\
  Waterloo, Ontario, N2L 2Y5, Canada}\\
  {\tt hkim perimeterinstitute.ca}\\
\vspace*{0.5cm}
$^3$ {\it NHETC and Department of Physics and Astronomy, Rutgers University, \\
  Piscataway, NJ 08854, USA}\\
  {\tt nishinaka physics.rutgers.edu} \\
\vspace*{0.5cm}
$^4$ {\it School of Physics, Korea Institute for Advanced Study, \\
Seoul 130-722, Korea}\\

\vspace*{0.3cm}
\end{center}
\vspace*{.5cm}

\noindent
We 
evaluate the Nekrasov partition function of 5d gauge theories engineered by webs of 5-branes, using the refined topological vertex on the dual Calabi-Yau threefolds.
The theories include certain non-Lagrangian theories such as the $T_N$ theory.
 The refined topological vertex computation generically contains contributions from decoupled
M2-branes which are not charged under the 5d gauge symmetry engineered. 
We argue that, 
after eliminating them, the refined topological string partition function agrees with the 5d Nekrasov partition function.
We 
explicitly check 
this for the $T_3$ theory as well as $Sp(1)$ gauge theories with $N_f = 2, 3, 4$ flavors.  
In particular, our method leads to a new expression of the $Sp(1)$ Nekrasov partition functions without any contour integrals.
We 
also develop prescriptions to calculate the partition functions of 
theories obtained by Higgsing the $T_N$ theory.
We 
 compute the partition function of the $E_7$ theory 
via this prescription, and 
find the $E_7$ global symmetry enhancement. We 
finally discuss a potential application of the refined topological vertex to non-toric web diagrams.

\newpage

\tableofcontents



\section{Introduction}

$\cN\!=\!2$ supersymmetric gauge theories in four dimensions are particularly of interest since they exhibit interesting non-perturbative effects but still we can obtain their exact results.
The low energy prepotentials were exactly determined in \cite{Seiberg:1994rs, Seiberg:1994aj} for $SU(2)$ gauge theories with $N_f \leq 4$ flavors, and the computation was generalized to $SU(N)$ gauge theories with $N_f \leq 2N$ flavors  in \cite{Klemm:1994qs, Argyres:1994xh, Hanany:1995na, Argyres:1995wt}. The Seiberg--Witten formulation was further extended to class $\mathcal{S}$ theories in \cite{Gaiotto:2009we} which include non-Lagrangian theories like so-called $T_N$ theory. The exact prepotentials of $SU(N)$ gauge theories were also determined by a microscopic approach in \cite{Nekrasov:2002qd, Nekrasov:2003rj} where exact instanton partition functions were directly evaluated by a localization technique. Note that the localization method is not applicable to theories which do not have Lagrangian descriptions.

The computation of the $SU(N)$ instanton partition function usually involves the resolution of small instanton singularities in the instanton moduli space (see, for example, \cite{MR1711344} for a review). The resolution introduces a non-commutative parameter in the ADHM equations\footnote{This can be interpreted as introducing a Fayet--Illiopoulos parameter in the D-term equation on D0-branes in the D0--D4-brane system.}, and then we also have a $U(1)$ instanton solution from the deformed equations \cite{Nekrasov:1998ss}. Therefore, the resulting partition function is in fact the $U(N)$ instanton partition function rather than the $SU(N)$ instanton partition function. In fact, the $SU(N)$ instanton partition function has not been known except for $N\!=\!2$. Even in the known case of $SU(2) \cong Sp(1)$, the Nekrasov partition function involves complicated contour integrals, \cite{Marino:2004cn, Nekrasov:2004vw, Kim:2012gu} which makes the computation technically difficult in particular for $N_f = 6, 7$ \cite{Kim:2012gu}.

The difference between the $U(N)$
and $SU(N)$ instanton partition functions is in fact important, for example, for the AGT 
correspondence
 originally proposed in \cite{Alday:2009aq}. The correspondence can be only seen after eliminating an extra factor, which realizes a correct flavor symmetry of an $SU(2)$ gauge theory with $N_f\! = \!4$ flavors. Another important example is the generation of the $E_n$-type flavor symmetry in the five-dimensional superconformal index \cite{Kim:2012gu, Bashkirov:2012re}. It is crucial to use the five-dimensional $Sp(1)$ instanton partition function\footnote{It is also important to use the correct dual gauge group $O(k)$ instead of $SO(k)$ for the $k$-instanton moduli space.} in order to achieve the enhancement of the $E_n$-type global symmetry. Therefore, a systematic study of the difference between the $U(N)$
and $SU(N)$ instanton partition functions is clearly important for further applications.

String theory provides another way to get the Nekrasov partition functions, namely by using the topological vertex \cite{Iqbal:2002we,Aganagic:2003db} or its refinement \cite{Awata:2005fa,Iqbal:2007ii}. All genus amplitudes of topological string turn out to yield the five-dimensional Nekrasov partition functions due to the fact that five-dimensional gauge theories with eight supercharges may be geometrically engineered by M-theory compactifications on non-compact toric Calabi--Yau threefolds \cite{Klemm:1996bj, Morrison:1996xf, Douglas:1996xp, Katz:1996fh, Intriligator:1997pq, Katz:1997eq}. In this case also, the (refined) topological vertex reproduces the $U(N)$ instanton partition function \cite{Iqbal:2003ix, Iqbal:2003zz, Eguchi:2003sj, Hollowood:2003cv, Taki:2007dh} instead of the $SU(N)$ instanton partition function although the low energy effective field theory is expected to yield an $SU(N)$ gauge theory.

To understand the relation between $U(N)$ and $SU(N)$ partition functions, let us recall that the five-dimensional Nekrasov partition function is written as a trace over the space of BPS states. The 5d BPS states include W-bosons, massive hypermultiplets and instantons. In the geometric engineering, they are M2-branes wrapping compact two-cycles of the Calabi-Yau threefold.  However, in general, the BPS spectrum contains M2-branes which are {\it not} charged under the $SU(N)$ gauge symmetry engineered. They are only charged under decoupled $U(1)$ gauge symmetries, but still contribute to the topological string amplitudes.\footnote{The instanton in a five-dimensional gauge theory also carries a gauge charge due to a coupling between a current whose conserved charge is the instanton number and a dynamical gauge field.} Now we expect that, by eliminating contributions from such decoupled M2-branes, we obtain the $SU(N)$ partition function from the (refined) topological string amplitude. We explicitly check this for $SU(2)\cong Sp(1)$ gauge group with $N_f\leq 4$ flavors. Two-cycles which support decoupled M2-branes are easily 
 identified in the toric web diagram of the Calabi-Yau threefold. The contribution from decoupled M2-branes is always a {\it prefactor} of the topological string partition function, which we call the ``$U(1)$-factor.''
Therefore we argue that the $U(N)$ and $SU(N)$ partition functions are related just by multiplying/dividing the $U(1)$-factor. This particularly gives a closed formula for the five-dimensional $Sp(1)$ instanton partition function which does not involve any integration. Although it would be hard to infer such a contribution from the view point of field theory in particular in four-dimensions, the web diagram naturally implies the decoupled factor in the instanton partition function.

An important application of our method is the computation of exact partition functions of certain non-Lagrangian theories. It has been known that the web diagram can realize not only conventional gauge theories but also non-Lagrangian theories such as a five-dimensional version of $T_N$ theory \cite{Benini:2009gi}. Since $T_N$ theory is an isolated superconformal field theory and does not have a Lagrangian description,  we cannot obtain its exact partition function by the localization technique used in \cite{Nekrasov:2002qd, Nekrasov:2003rj}. However, there is no obstruction to use the refined topological vertex for the web diagram of $T_N$ theory. By eliminating a $U(1)$ factor for the $T_N$ diagram, we obtain the exact partition function of the five-dimensional $T_N$ theory. In particular, we will confirm that the partition function of $T_3$ theory becomes the $Sp(1)$ Nekrasov partition function with $N_f \!=\! 5$ flavors as expected. This example is particularly interesting since it exhibits an enhanced $E_6$ global symmetry at a special point in the moduli space \cite{Seiberg:1996bd}.

One of the limitation of the refined topological vertex computation is that we cannot apply the technique to a non-toric variety\footnote{A vertex formalism for unrefined topological string partition functions which can be applied to local non-toric del Pezzo surfaces has been discussed in \cite{Diaconescu:2005ik, Diaconescu:2005mv}.}. For example, an $SU(2)$ gauge theory with $N_f \!=\! 6$ flavors, which will realize an $E_7$ global symmetry at a special point in the moduli space, is geometrically engineered by a non-compact Calabi--Yau threefold whose base is a $dP_7$ surface, which is not toric \cite{Morrison:1996xf, Douglas:1996xp}. However, it has been also pointed out that the $E_7$ theory may be realized by an infrared description in a Higgs branch vacuum of $T_4$ theory \cite{Benini:2009gi} which is described by a web diagram which does not admit its dual toric fan. Motivated by this, we also develop prescriptions to evaluate partition functions of theories which are low energy descriptions of some Higgs branch vacua of the theories arising from non-toric web diagrams. The web diagram is powerful enough to visualize the Higgs branch, and it suggests a correct root of the Higgs branch. As for its consistency check, we compute the five-dimensional superconformal index of the $E_7$ theory and show the $E_7$ enhancement. Therefore, our approach can also evaluate partition functions of the theories arising from non-toric web diagrams. 
Putting this together, our technique can be applied to the computation of partition functions of theories from any web diagrams constructed in \cite{Benini:2009gi}.


The organization of the paper is as follow. In section \ref{sec:engineering}, we review how a certain class of gauge theories can be geometrically engineered by non-compact toric Calabi--Yau threefolds or their dual web diagrams. This includes non-Lagrangian theories such as $T_N$ theory. We then describe how to compute the partition function of the theories by using the refined topological vertex in section \ref{sec:topst}. We point out that  the refined topological vertex computation necessarily contains contributions from decoupled M2-branes which needs to be subtracted. In section \ref{sec:Sp1}, we explicitly compute partition functions of various examples such as $SU(2)$ gauge theories with $N_f = 2, 3, 4$ flavors as well as $T_{N=2, 3, 4}$ theories. We then verify that we get the correct $SU(2)$ partition functions after dividing by what we call $U(1)$ factors. In section \ref{sec:general-TN}, we generalize our method to $T_N$ theory and propose the five-dimensional partition function of $T_N$ theory. Some technical details are relegated to appendices. We also comment on the $SU(N)$ partition function with $N_f = 2N$ flavors in appendix \ref{app:SUN}. In section \ref{sec:Higgs}, we move on to the computation of the low energy partition functions in a Higgs vacuum of the $T_{N}$ theories. It is known that the Higgs branch vacuum expectation value leads to different class $\mathcal{S}$ theories in the far infrared. Although the corresponding web diagrams are no longer toric, we develop a method to compute such partition functions, and carry out non-trivial consistency checks.

\vspace*{5mm}

\noindent{\it Note added:}

We here note that the results in sections 4 and 5 have some overlap with \cite{Bao:2013pwa} which appeared in arXiv on the same day.

\section{Five-dimensional theories from M-theory}
\label{sec:engineering}

Five-dimensional supersymmetric gauge theories with eight supercharges can be obtained from M-theory compactifications on Calabi-Yau threefolds $\tilde{X}$ \cite{Klemm:1996bj, Morrison:1996xf, Douglas:1996xp, Katz:1996fh, Intriligator:1997pq, Katz:1997eq}. An $ADE$ singularity fibered over a curve $B$ inside $\tilde{X}$ yields an $ADE$ gauge group $G$ in the five-dimensional low energy effective field theories. The genus of the curve $B$ is related to the number of the adjoint hypermultiplets. We always consider $B = \P^1_b$ so that there is no adjoint hypermultiplet. 
The massless hypermultiplets in some representation under the $ADE$ gauge group are introduced when there are some singularity enhancement loci on $B$. The enhanced singularity type characterizes the representation of the hypermultiplets \cite{Bershadsky:1996nh, Katz:1996xe}.

After the resolution of the $ADE$ singularity over $B$, the non-Abelian gauge theories will be in their Coulomb branch. Therefore, the resolution parameters associated with resolved divisors $D_i, i=1, \cdots, \text{rank}(G)$ for the $ADE$ singularity are related to the vevs $a_i$ of the scalars in the vector multiplets. The resolution over $B$ may also resolve the enhanced singularities, which generates the Coulomb branch dependent mass terms for the hypermultiplets. One may also introduce a blow up divisor $H$ for resolving the enhanced singularity, maintaining the Calabi--Yau condition, which, on the other hand, corresponds to generating the classical mass term $m$ for the hypermultiplets. The resolution of the singularities is not generically unique and different resolutions correspond to different phases of the five-dimensional supersymmetric gauge theories \cite{Morrison:1996xf, Intriligator:1997pq}.

In five-dimensional gauge theories, there is in fact a peculiar global $U(1)$ symmetry. One can consider a current 
\be
j = \ast(F \wedge F), \label{instanton_current}
\ee
which is always conserved. The expression \eqref{instanton_current} implies that its charge is the instanton number. The mass parameter associated to the instanton flavor symmetry is related to the gauge coupling as $\frac{1}{g^2}$, which has dimension one in five-dimensional gauge theories. This global symmetry also plays an important role in five-dimensional gauge theories.

\subsection{$SU(2)$ gauge theories with $N_f$ flavors}
\label{sec:SU2}

Let us consider five-dimensional $SU(2)$ gauge theories with $N_f$ hypermultiplets in the fundamental representation. When $N_f < 8$, the metric of the Coulomb branch moduli space is always positive. 
Hence, we may consider a strong coupling limit where $\frac{1}{g^2} = 0$ of the theories. Then, the theories are supposed to be at a non-trivial fixed point with an enhanced global symmetry \cite{Seiberg:1996bd}. For a finite gauge coupling, we have a global symmetry $SO(2N_f) \times U(1)$ where the Cartans of $SO(2N_f)$ are associated with the mass parameters of the fundamental hypermultiplets. The other $U(1)$ is the instanton flavor symmetry \eqref{instanton_current} and it is associated with the gauge coupling $\frac{1}{g^2}$. At the fixed point, this global symmetry is enhanced to $E_{N_{f+1}}$-type\footnote{For $N_f + 1 \leq 5$, $E_{N_f + 1}$ is defined $E_5 = Spin(10), E_4 = SU(5), E_3 = SU(3) \times SU(2), E_2 = SU(2) \times U(1), E_1 = SU(2)$.}. Therefore, this type of five-dimensional gauge theories is of particular interest and we will focus on $N_f < 8$ hereafter.

Such a five-dimensional gauge theory can be realized when one considers an M-theory compactification on a Calabi--Yau threefold $\tilde{X}_{N_f}$ with an $A_1$ singularity over a $\P^1_b$. At $N_f$ points in $\P^1_b$, the $A_1$ singularity is enhanced to $A_2$ singularities, which amounts to introducing the $N_f$ fundamental hypermultiplets. As a resolution of such a Calabi--Yau threefold, we may consider a Calabi--Yau threefold which has a compact divisor $D = \P^1_f \times \P^1_b$\footnote{In general, we may consider a compact divisor where $\P^1_f$ is non-trivially fibered over $\P^1_b$. In the case of $SU(N)$ gauge theories, the non-trivial fibration is related to non-zero five-dimensional Chern-Simons couplings \cite{Intriligator:1997pq, Iqbal:2003zz}.} with $N_f$ points blown up. The infinite coupling limit  corresponds to the case where the entire $D$ collapses \cite{Morrison:1996xf, Douglas:1996xp}. In fact, the geometry implies why the global symmetry may be enhanced to the $E_{N_{f}+1}$-type.  The enhancement may come from the fact that the blow up of $\P^1_f \times \P^1_b$ at $N_f$ points $(1 \leq N_f \leq 7)$ is isomorphic to the blow up of $\P^2$ at $N_f+1$ points $(2 \leq N_f + 1 \leq 8)$. From the latter point of view, the Weyl group of $E_{N_{f+1}}$ acts on the curves in $H^{1,1}(D)$, and the $E_{N_{f+1}}$ group appears in a natural way.

\begin{figure}[tb]
\begin{center}
\includegraphics[width=80mm]{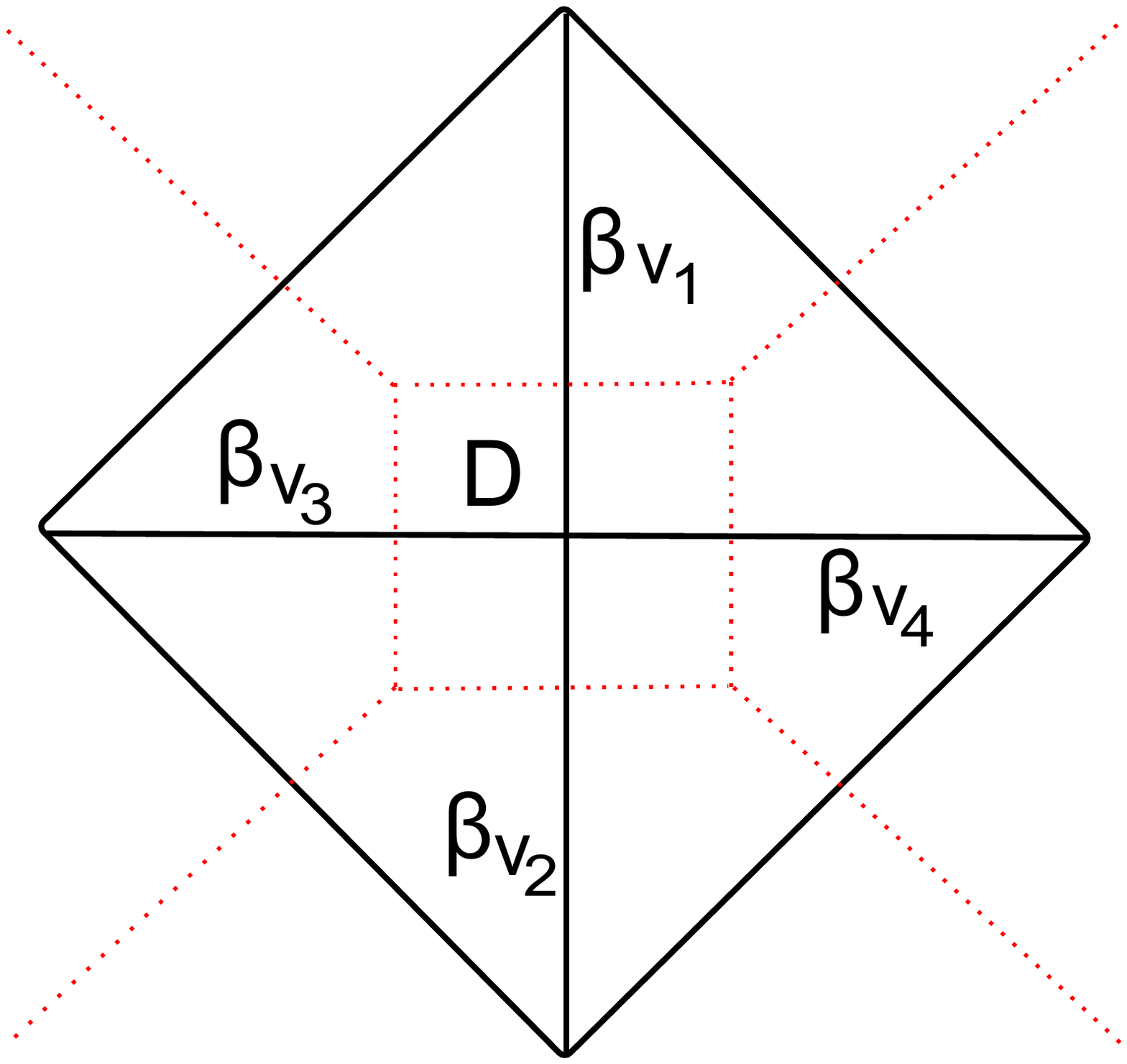}
\end{center}
\caption{The black solid lines represent the toric fan on the two-dimensional hyperplane for a local $\P^1_f \times \P^1_b$. The red dotted lines denote the dual diagram which can be viewed as a $(p, q)$ 5-brane web in type IIB string theory.}
\label{fig:toric1}
\end{figure}

\begin{figure}[tb]
\begin{center}
\includegraphics[width=160mm]{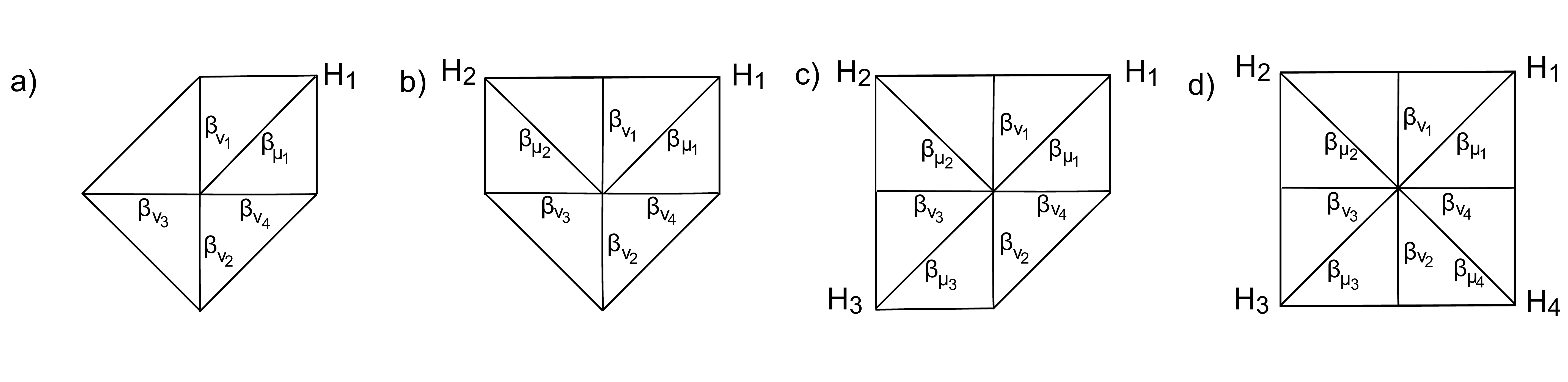}
\end{center}
\caption{The toric fans on the two-dimensional hyperplane for toric Calabi--Yau threefolds whose bases are the blow up of $\P^1_f \times \P^1_b$ at $(a): N_f = 1, (b): N_f = 2, (c): N_f = 3, (d): N_f = 4$ points. $H_i, i = 1, 2, 3, 4$ are the blow up divisors. We chose a particular triangulation which will be used in the later computation of the partition functions. M2-branes wrapping $\beta_{\nu_i}$ give fundamental hypermultiplets. }
\label{fig:toric2}
\end{figure}

Since we are interested in the field theory limit, we will always decouple gravity by considering a non-compact Calabi--Yau threefold $X$ \cite{Kachru:1995fv, Katz:1996fh}. Then, such a Calabi--Yau threefold $X$ can be in general constructed by a line bundle $\cO(K_D)$ fibered over a compact surface $D$, 
where $K_D$ denotes the canonical divisor of $D$. For $SU(2)$ gauge theories with $N_f$ flavors, $D$ is the blow up of $\P^1_f \times \P^1_b$ at $N_f$ points. In particular, when $N_f \leq 4$, the blow up of $\P^1_f \times \P^1_b$ at $N_f$ points can be described by toric varieties. 
Since the dimension of the total space is three, the dimension of the corresponding toric fan is also three. However, due to the Calabi--Yau condition, all vectors of the toric fan 
lie on a two-dimensional hyperplane. Therefore, one can describe the toric fan of a toric Calabi--Yau threefold as a two-dimensional toric fan which is related to a fan of the compact base $D$. We illustrate the pictures of the toric fans of a toric Calabi--Yau threefold with $D = \P^1_f \times \P^1_b$ and also the cases when $D$ is the blow up of $\P^1_f \times \P^1_b$ at $1 \leq N_f \leq 4$ points on the two-dimensional hyperplane in Figure \ref{fig:toric1}, \ref{fig:toric2} respectively. 

The black solid lines of Figure \ref{fig:toric1} denote a two-dimensional slice of a three-dimensional polyhedral toric fan 
for the toric Calabi-Yau manifold. 
A one-dimensional cone denotes a divisor which becomes a point in the two-dimensional toric fan. There is only one compact divisor $D = \P^1_f \times \P^1_b$ which corresponds to 
an interior point. 
The other points represent non-compact divisors. A two-dimensional face generated by two one-dimensional cones corresponds to a curve as an intersection between the two divisors. Then, the curves correspond to the internal lines on the two-dimensional hyperplane. We denote compact curves by $\beta_{\nu_i}, i=1, \cdots, 4$. We choose $\beta_{\nu_3}$ and $\beta_{\nu_4}$ as the fiber $\P^1_f$, and $\beta_{\nu_1}$ and $\beta_{\nu_2}$ as the base $\P^1_b$.

The parameters and the Coulomb branch moduli of the five-dimensional supersymmetric gauge theories are associated with the parameters of the geometry. Since we will use the toric Calabi--Yau threefolds $X_{N_f}$, we consider the K\"ahler parameters of the manifolds. We have $N_f + 2$ divisors, $H_0, D, H_i, i = 1, \cdots, N_f$ where $D = \P^1_f \times \P^1_b$ and the other divisors are non-compact. The K\"ahler form may be expanded by the divisors, and the expansion parameters of compact divisors become moduli and the expansion parameters of non-compact divisors become parameters in the five-dimensional effective field theory. $H_0$ is defined as a divisor which yields $H_0 \cdot D = \P^1_f$. In other words, its dual two-cycle is the base $\P^1_b$. 
Hence, the expansion parameter $t_b$ of the divisor $H_0$ is related to 
the classical gauge coupling 
$\frac{1}{g_{\text{clasical}}^2}$ of the five-dimensional gauge theory. Then, the K\"ahler form can be parameterized by 
\be
J = t_b H_0 + a D + m_i H _i\,,
\ee
where $a$ is the Coulomb branch parameter of the $SU(2)$ and $m_i$ is the classical mass parameter for the fundamental hypermultiplets. Therefore, the volume of a compact curve $\beta$ inside $X_{N_f}$ can be measured by
\be
\int_{\beta} J = t_b (H_0 \cdot \beta) + a (D \cdot \beta) + m_i (H_i \cdot \beta)\,. 
\ee 
Since M2-branes wrapping the curves $\beta$ represent BPS particles in the five-dimensional gauge theory, the volume of the curves is related to the charges of the particles. 

\subsection{$T_N$ theories}
\label{sec:TN}

So far we have seen that M-theory compactifications on the particular types of non-compact Calabi--Yau manifolds realize supersymmetric five-dimensional gauge theories. In fact, there is a different way to see how the five-dimensional gauge theories are generated when the non-compact Calabi--Yau manifolds are toric varieties. This can be seen when one moves to a dual diagram of the toric fan. The dual diagram of a toric variety for 
a local $\P^1_f \times \P^1_b$ which is projected onto a two-dimensional hyperplane is depicted by the red dotted lines in Figure \ref{fig:toric1}. On the hyperplane, the two-dimensional faces are replaced with dots, and the one-dimensional lines become the perpendicular lines. This diagram represents the degeneration of $T^2$ on the two-dimensional space. One of the two one-cycles shrinks along the red dotted lines, and the whole $T^2$ shrinks at the points on the two-dimensional hyperplane.

In fact, there is a corresponding physical picture for the dual toric diagram. Namely, the red dotted lines in Figure \ref{fig:toric1} give a web of $(p, q)$ 5-branes in type IIB string theory \cite{Aharony:1997bh, Leung:1997tw}. 
Let us first start from type IIB string theory on $S^1$. We also assume that there are a D5-brane and an NS5-brane which do not extend along the $S^1$ direction. When one performs a T-duality along the $S^1$, the D5-brane becomes a wrapped D6-brane and an NS5-brane becomes a Kaluza--Klein monopole in type IIA string theory. The Kaluza-Klein monopole is located where the T-dualized $S^1$ shrinks. When one promotes this setup to M-theory, both become Taub-NUT manifolds. The location of the Taub-NUT manifold coming from the wrapped D6-brane in type IIA string theory is then determined by the location where the M-theory $S^1$ shrinks. Therefore the locations of the two Taub-NUT manifolds are specified by different $S^1$'s shrink. Namely, the degeneration of two one-cycles in $T^2$ in M-theory represent the locations of the D5-brane and the NS5-brane respectively. One can also consider a situation where $(p, q)$ one-cycle shrinks and this yields a $(p, q)$ 5-brane which is a bound state of a D5-brane and NS5-brane. The degeneration of the $T^2$ may be realized by the degeneration of $T^2$ on the two-dimensional hyperplane in the dual toric diagram. Therefore, the dual toric diagram represents a web diagram of $(p, q)$ 5-branes.

When one views a toric diagram from a web of $(p, q)$ 5-branes, toric Calabi--Yau threefolds may yield interesting non-Lagrangian theories. For example, so-called $T_N$ theory has been constructed from a web of $(p, q)$ 5-brane in \cite{Benini:2009gi}. $T_N$ theory has been originally constructed in \cite{Gaiotto:2009we} by a compactification of M5-branes on a sphere with which three bunches of $N$ semi-infinite M5-branes intersect. Such a configuration gives rise to a four-dimensional superconformal field theory with at least $SU(N)^3$ flavor symmetry associated with the three bunches of the $N$ semi-infinite M5-branes. Furthermore, the theory has no marginal coupling as three points on a sphere have no moduli. Hence, $T_N$ theory represents an isolated SCFT.

\begin{figure}[tb]
\begin{center}
\includegraphics[width=80mm]{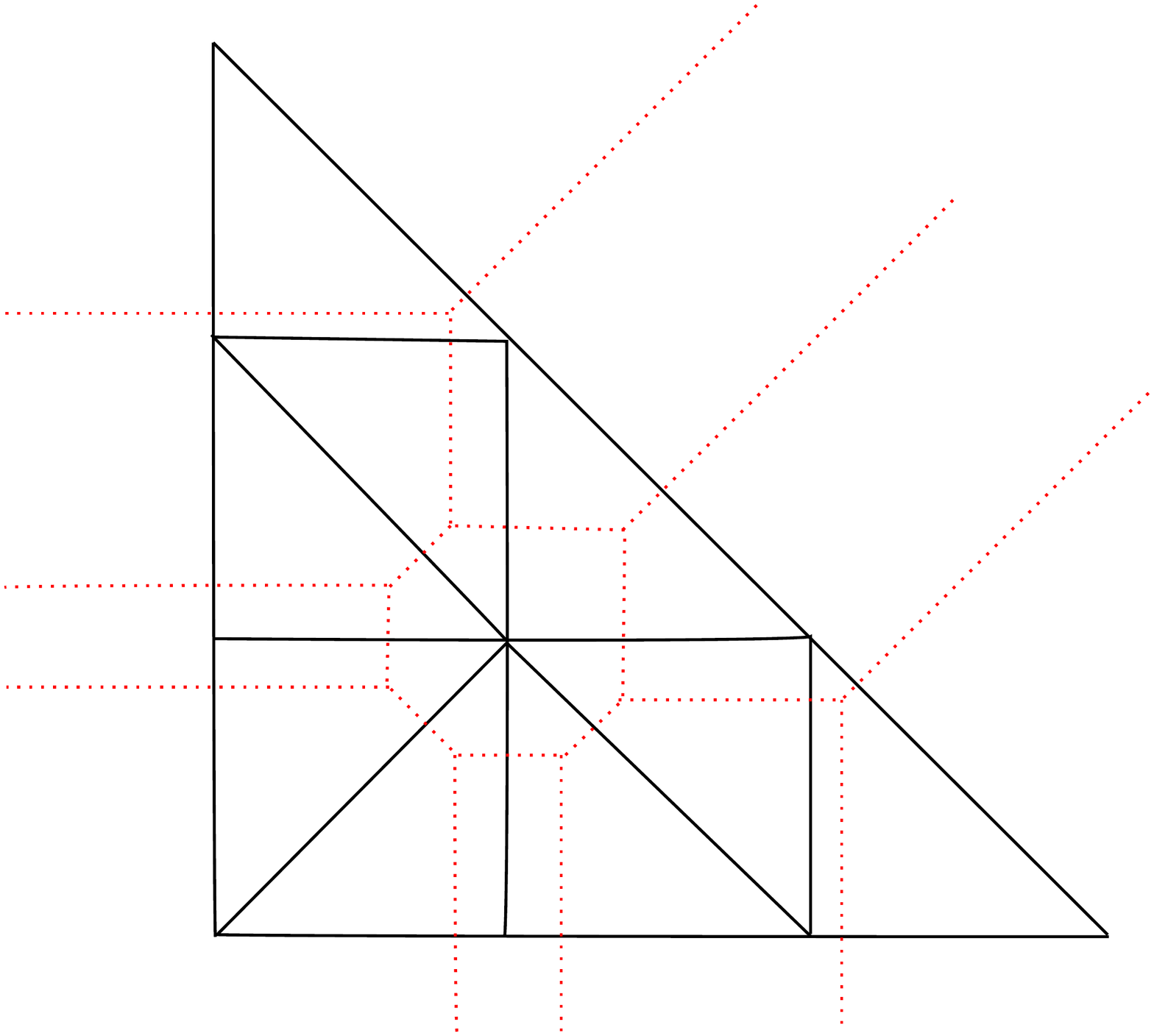}
\end{center}
\caption{The red dotted lines represent a web of $(p, q)$ 5-branes for $T_3$ theory. The black solid lines stand for the corresponding toric fan of a particular triangulation.}
\label{fig:toric3}
\end{figure}

Ref.~\cite{Benini:2009gi} has proposed that a five-dimensional version of $T_N$ theory can be realized by a web where $N$ D5-branes, $N$ NS5-branes and $N$ $(1, 1)$ 5-branes meet together. An $S^1$ compactification of the five-dimensional $T_N$ theory realizes the original $T_N$ theory in \cite{Gaiotto:2009we}. An example of a web for $T_3$ theory is depicted in Figure \ref{fig:toric3}.
From the toric point of view, the web diagram can be seen as the blow up of $\C^3/\Z_N \times \Z_N$. In fact, the number of the Coulomb branch moduli and the number of the Higgs branch moduli, which can be read off from the web diagram, completely agree with those of $T_N$ theory.

The dimension of the Coulomb branch moduli space can be computed by the number of local deformation of the web diagram which does not change the locations of the semi-infinite 5-branes, which turns out to be equal to the number of closed faces in the web diagram
\be
\text{dim}_{\C}(\mathcal{M}_{\text{Coulomb}}) = \frac{(N-1)(N-2)}{2}\,.
\label{Coulomb}
\ee

\begin{figure}[tb]
\begin{center}
\includegraphics[width=80mm]{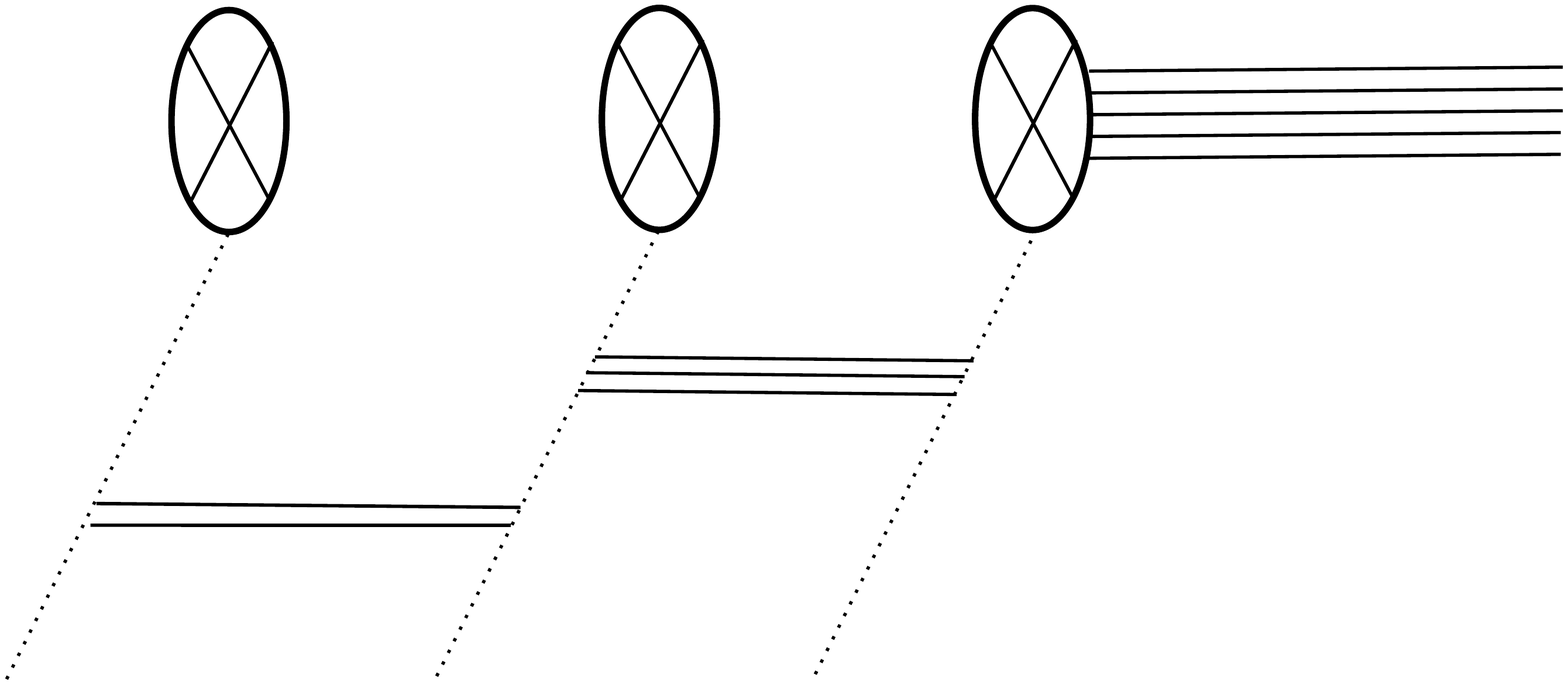}
\end{center}
\caption{The deformation of pieces of 5-branes between 7-branes. One black solid line represents a 5-brane and one $\otimes$ represents a 7-brane. The dotted line stands for the direction where 7-branes extend but the 5-brane does not extend.}
\label{fig:Higg}
\end{figure}

On the other hand, the Higgs branch of the $T_N$ theory may be understood when we terminate the semi-infinite 5-branes on 7-branes which are points in the two-dimensional hyperplane and share the Minkowski five-dimensional space. Then, a global symmetry can be realized as a symmetry on the 7-branes. In general a $(p, q)$ 5-brane can end on a orthogonal $(p, q)$ 7-brane without breaking supersymmetry. Hence we terminate one 7-brane at each end of the semi-infinite 5-branes. When $k$ 7-branes are attached to the parallel $k$ semi-infinite 5-branes at a point on the two-dimensional hyperplane, we have an $SU(k)$ global symmetry. The maximal Higgs branch can be seen when all parallel 5-branes are coincident. Then we have $N$ separate simple junctions, which yields an $N$ dimensional Higgs branch. When all the parallel 5-branes are overlapped, one can also strip off pieces of some of the 5-branes between the 7-branes into the direction where the 7-branes extend as in Figure \ref{fig:Higg}.
In this case, some of the 7-branes are separated from the web diagram on the two-dimensional hyperplane and the global symmetry gets reduced. Therefore, the deformation of the piece of the 5-branes also contributes to the Higgs branch. In general, we can consider a situation where several 5-branes are put on the same 7-brane. Let $k_i$ 5-branes be on a 7-brane where $\sum_{i=1}^J k_i = N$, and $k_1 \geq k_{2} \geq \cdots \geq k_J$. We denote such a configuration by $\{k_1, \cdots, k_J\}$. The number of the deformation of the pieces of the 5-branes in the configuration can be counted by \cite{Benini:2009gi}
\be
\text{dim}_{\mathbb{H}}(\mathcal{M}_{\text{Higgs}}^{ \{k_1, \cdots, k_J\}}) = \sum_{i=1}^J (i-1)k_i = -N + \sum_{i=1}^J ik_i\,.  
\label{onepuncture}
\ee
When each of the $N$ parallel 5-branes is put on a 7-brane, the contribution becomes
\be
\text{dim}_{\mathbb{H}}(\mathcal{M}_{\text{Higgs}}^{ \{1^N\}}) = -N + \sum_{i=1}^N i = \frac{1}{2}N(N-1)\,.
\ee
Therefore, putting these two contributions together and subtracting the overall center of mass motion, the dimension of the Higgs branch of $T_N$ theory is 
\be
\text{dim}_{\mathbb{H}}(\mathcal{M}_{\text{Higgs}}) = N - 1 + \frac{3}{2}N(N-1) = \frac{3N^2 - N - 2}{2}\,. 
\label{Higgs}
\ee

Note that, when some 5-branes are put on one 7-brane, we need to take care of the generalized s-rule in \cite{Benini:2009gi} in order to preserve the supersymmetry, which is an application of the original s-rule of \cite{Hanany:1996ie} to a web of $(p, q)$ 5-brane. In fact, there is a case where one 5-brane needs to jump over other 5-brane. Then, the diagram does not remain toric and becomes a so-called dot diagram by including a new white dot in the toric diagram. The white dot represents that the 5-branes on both sides of the white dot are on the same 7-brane. Furthermore, the s-rule in fact can propagate inside the web diagram due the property of a junction, and then a white dot can be an interior point.

\begin{figure}[tb]
\begin{center}
\includegraphics[width=80mm]{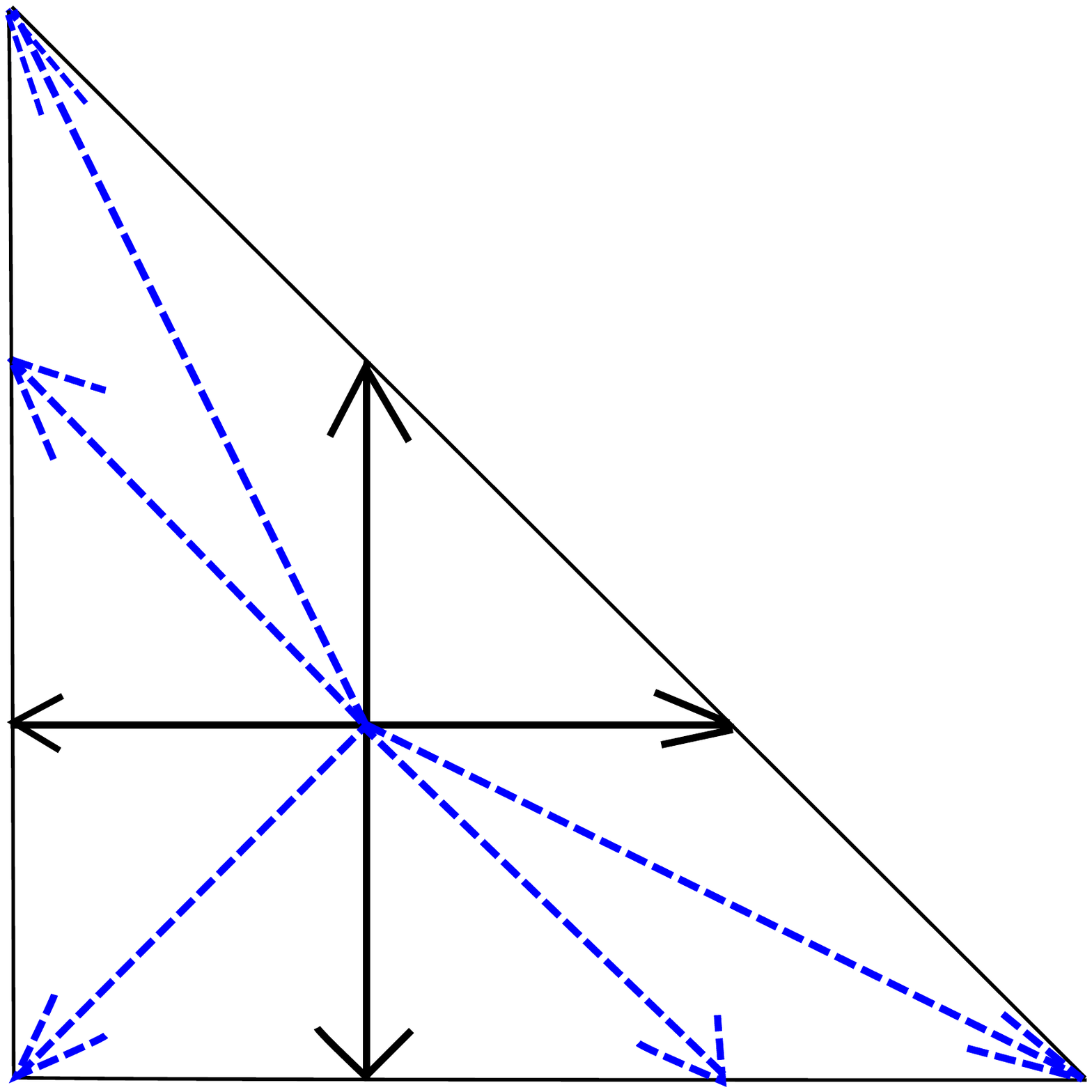}
\end{center}
\caption{The toric fan of $T_3$ under a particular triangulation. On the two-dimensional hyperplane, the original toric fan can be thought as a two-dimensional toric fan for the base manifold $D$. Then, the two-dimensional toric fan stands for the blow up of $\P^1 \times \P^1$ at five points. The five blow up divisors are depicted by five blue dashed arrows.}
\label{fig:5blowups}
\end{figure}

Let us look at $T_3$ theory more closely. $T_3$  theory has at least $SU(3)^3$ global symmetries and the dimensions of the moduli spaces are $\text{dim}_{\C}(\mathcal{M}_{\text{Coulomb}}) = 1$ and $\text{dim}_{\mathbb{H}}(\mathcal{M}_{\text{Higgs}}) = 11$ from \eqref{Coulomb} and \eqref{Higgs}. In a particular triangulation of the toric fan, the toric diagram on the two-dimensional hyperplane can be seen as $\P^1 \times \P^1$ with $5$ points blown up as in Figure \ref{fig:5blowups}.
The M-theory compactification on the blow up of $\P^1 \times \P^1$ at $5$ points gives rise to an $SU(2)$ gauge theory with five fundamental hypermultiplets as discussed in section \ref{sec:SU2}. Since this theory yields the $E_6$ global symmetry at the fixed point, $T_3$ theory is also supposed to exhibit an enhanced $E_6$ global symmetry. 
Indeed, the quaternionic dimension of the Higgs branch of $T_3$ theory agrees with the quaternionic dimension of the one-instanton moduli space of $E_6$. 

\begin{figure}[tb]
\begin{center}
\includegraphics[width=120mm]{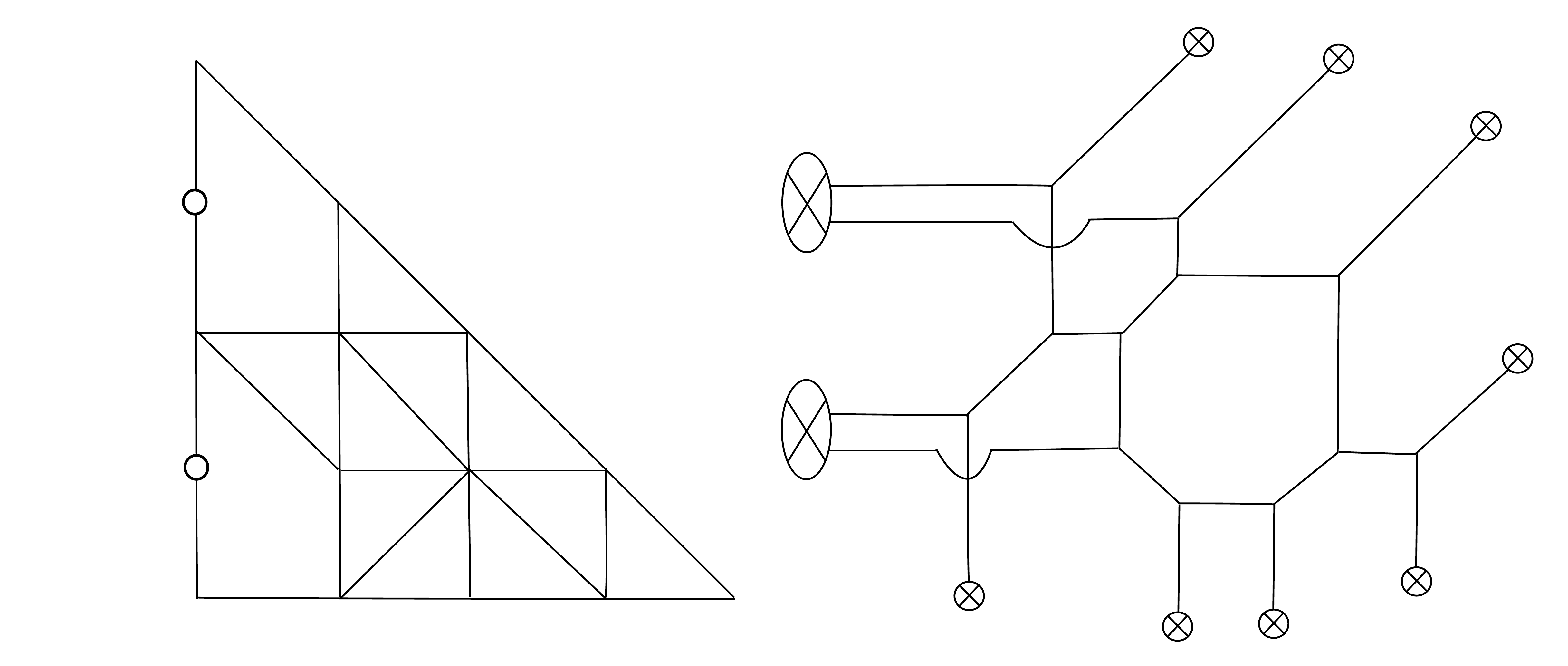}
\end{center}
\caption{The dot diagram (left) and the web diagram (right) of the $E_7$ theory. 
$\otimes$ in the right figure again represents a 7-brane.}
\label{fig:HiggsT4}
\end{figure}

It has been also suggested that a theory realized as a vacuum in a Higgs branch of $T_4$ theory generates an $E_7$ global symmetry \cite{Benini:2009gi}. When we put the two 5-branes and the other two 5-branes from one bunch of the four parallel 5-branes on two 7-branes as in Figure \ref{fig:HiggsT4}, then the global symmetry reduces to $SU(4)^2 \times SU(2)$. 
In this case, the dimension of the Coulomb branch moduli space is $\text{dim}_{\C}(\mathcal{M}_{\text{Coulomb}}) = 1$ as  there is only on interior point in the left figure of Figure \ref{fig:HiggsT4}. As for the dimension of the Higgs branch, the contribution from the 5-branes in the left part of the web diagram in the right figure of Figure \ref{fig:HiggsT4} is 
\be
\text{dim}_{\mathbb{H}}(\mathcal{M}_{\text{Higgs}}^{ \{2, 2\}}) = - 4 + \sum_{i=1}^2 2i= 2 \,,
\ee
from \eqref{onepuncture}. The contributions from the other two bunches of the parallel 5-branes are $6 + 6 = 12$. Hence, the total quaternionic dimension of the Higgs branch is  $\text{dim}_{\mathbb{H}}(\mathcal{M}_{\text{Higgs}}) = 4 - 1 + 2 + 6 + 6  = 17$. The quaternionic dimension of the Higgs branch then agrees with the quaternionic dimension of the one-instanton moduli space of $E_7$. Now the diagram is no longer toric and we need to introduce white dots in the dot diagram as in Figure \ref{fig:HiggsT4}.

\section{Topological strings and extra $U(1)$ factors}
\label{sec:topst}

As discussed in section \ref{sec:engineering}, a class of five dimensional gauge theories are geometrically engineered by toric Calabi-Yau threefolds $X$. We are particularly interested in the theories on $S^1\times \mathbb{R}^4$, which are regarded as the strong coupling limit of type IIA string theory on $X$. This relates the Nekrasov partition function of the 5d theory on $S^1\times \mathbb{R}^4$ with the topological string amplitude on $X$. 

\subsection{Topological string amplitude}

We first review how the topological string amplitude appears in this context. Although our $X$ is non-compact, we regard it as a non-compact limit of some compact Calabi-Yau threefold $\tilde{X}$. When dimensionally reducing on $\tilde{X}\times S^1$, we have a 4d, $\mathcal{N}=2$ supergravity. The BPS states in the 4d theory come from D$p$-branes wrapped on holomorphic compact $p$-cycles of $\tilde{X}$. Among others, D2-branes on compact two-cycles are regarded as ``electric'' BPS-states (such as W-bosons) and D4-branes on compact four-cycles are ``magnetic'' BPS states (such as monopoles).
The 4d effective action contains various F-terms. In particular, a class of F-terms is calculated in the topological string theory on the same Calabi-Yau threefold $\tilde{X}$. Let $\mathcal{F}$ be the topological string amplitude on $\tilde{X}$, and $\lambda$ be the topological string coupling.
When $\lambda$ is small, $\mathcal{F}$ is expanded as
\begin{eqnarray}
\mathcal{F} = \sum_{g=0}^\infty \mathcal{F}_g \lambda^{2g-2}\,.
\label{eq:amplitude}
\end{eqnarray}
Then the effective action of the supergravity contains F-terms of the form $\mathcal{F}_g R_+^2 (g_sF_+)^{2g-2}$ for $g>0$, where $R_+$ and $F_+$ are the self-dual part of the curvature and gravi-photon field strength, respectively.\footnote{The constant map contribution $\mathcal{F}_0$ gives rise to $(\partial_j\partial_i\mathcal{F}_0) F^i\wedge F^j$.} In other words, by identifying $\lambda=g_sF_+$, the topological string amplitude evaluates the gravi-photon corrected F-terms in the 4d $\mathcal{N}=2$ supergravity \cite{Antoniadis:1993ze, Bershadsky:1993cx}.

Since we are interested in 5d theories, we take the strong coupling limit $g_s\to \infty$. Since $g_s$ belongs to a 4d hypermultiplet, the F-term is independent of $g_s$. However, the same quantity $\mathcal{F}$ now has a different interpretation. Since in the strong coupling limit BPS D2-D0 states become very light, the low-energy theory is described by quantum fields associated with them. Then $\mathcal{F}$ can be evaluated as a sum of their one-loop amplitudes \cite{Gopakumar:1998ii,Gopakumar:1998jq}.
 Since the quantum fields are off-shell 4d short multiplets, they correspond to on-shell 5d BPS states, namely M2-branes wrapping holomorphic two-cycles in $\tilde{X}$. Let $\mathfrak{t}$ be the K\"ahler two-form of the Calabi-Yau threefold, and $\beta$ be the two-cycle wrapped by the M2-brane. The central charge of the M2-brane is given by $2\pi \mathfrak{t}\cdot \beta$. The M2-brane also carries spin $(j_L,j_R)$ with respect to $SO(4)\simeq SU(2)_L\times SU(2)_R$ acting on $\mathbb{R}^4$. Let $\mathcal{M}$ be the moduli space of deformations of the curve $\beta$ in $\tilde{X}$. Then $SU(2)_R$ is identified with the $SL(2)$ Lefschetz actions on the moduli space of deformations of $\beta$ in $\tilde{X}$ \cite{Gopakumar:1998jq}. On the other hand, $SU(2)_L$ is identified with the Lefschetz action on the moduli space of flat bundles over $\beta$, which is generically $T^{2g}$ if $\beta$ has genus $g$. Now, let $N_{j_L,j_R}^{(\beta)}$ be the BPS degeneracy of M2-branes wrapping $\beta$ which have spin $(j_L,j_R)$.
The one-loop amplitudes are then written in terms of $n_{j_L}^{(\beta)} = \sum_{j_R}(-1)^{2j_R}(2j_R+1) N_{j_L,j_R}^{(\beta)}$ as \cite{Gopakumar:1998ii, Gopakumar:1998jq}
\begin{eqnarray}
\mathcal{F} = -\sum_{\beta, j_L, k>0}\sum_{\ell = -j_L}^{j_L}(-1)^{2j_L}\frac{n_{j_L}^{(\beta)}}{k}\frac{q^{2\ell k}}{(q^{k/2}-q^{-k/2})^2}e^{-2\pi k \mathfrak{t}\cdot\beta}\,,
\label{eq:one-loop}
\end{eqnarray}
where $q=e^{i\lambda}$. Note that $n_{j_L}^{(\beta)}$ depends only on the left spin $j_L$ of M2-branes because $\mathcal{F}$ couples to the self-dual gravi-photon field strength $F_+$.

\subsection{Comparison to Nekrasov partition function} 
\label{subsec:comparison}

 Since $X$ engineers a gauge theory, the gravi-photon corrected F-term is also evaluated as the (logarithm of) Nekrasov partition function \cite{Nekrasov:2002qd}. This leads to the idea that, in the non-compact limit $\tilde{X}\to X$, the topological string partition function $\exp \mathcal{F}$ is identified with the Nekrasov partition function for $\epsilon_1=-\epsilon_2 = \lambda$. The special choice of $\epsilon_{1,2}$ comes from the fact that $\mathcal{F}_g$ couples to the self-dual part of the gravi-photon field strength. This relation between the topological string amplitude and Nekrasov partition function was checked in various examples \cite{Iqbal:2003ix, Iqbal:2003zz, Eguchi:2003sj, Hollowood:2003cv}. The expression \eqref{eq:one-loop} particularly implies that the logarithm of the Nekrasov partition function is written as a sum over M2-branes on holomorphic two-cycles $\beta$.

Note here that some of the two-cycles $\beta$ become non-compact in the limit $\tilde{X}\to X$. Then we should eliminate M2-branes on such $\beta$ from the 5d BPS spectrum because their masses are divergent. There is another class of M2-branes which are decoupled in the rigid limit $\tilde{X}\to X$. To see this, suppose that $X$ has $k$ compact four-cycles, which we denote by $D_i$ for $i=1,\cdots,k$. Then, 
the 5d gauge theory engineered by $X$ has a non-trivial $U(1)^k$ gauge symmetry.
The 5d gauge fields $A^{(i)}$ come from the M-theory three-form potential of the form
\begin{eqnarray}
A = \sum_{i=1}^{k} A^{(i)} \wedge \omega^{(i)}\,,
\end{eqnarray}
where $\omega^{(i)}$ is the harmonic two-form 
which is Poincar\'e dual to $D_i$. Now, an M2-brane on $\beta$ has the following coupling to the gauge fields:
\begin{eqnarray}
 \sum_{i=1}^{k}A^{(i)} \int_\beta \omega^{(i)}\,.
\end{eqnarray}
Namely the $i$-th electric charge of the M2-brane is identified with $\int_\beta \omega^i = D_i\cdot \beta$. If the intersection number $\beta\cdot D_i$ vanishes for all the compact four-cycles $D_i$, M2-branes wrapping $\beta$ has no electric charge of the 5d gauge theory engineered. This is the case if $\beta$ can be continuously moved to infinity. Such M2-branes are charged only under decoupled $U(1)$ gauge symmetries in the limit $\tilde{X}\to X$.

In the study of the 5d gauge theory we should eliminate such M2-branes from the 5d BPS spectrum because they are decoupled. This means that, when comparing the topological string partition function with the Nekrasov partition function, we have to omit in \eqref{eq:one-loop} all the contributions from such M2-branes; they do not contribute to the one-loop amplitude of the gauge theory. The contribution to be eliminated is a prefactor of $\exp\mathcal{F}$ of the form
\begin{eqnarray}
\prod_{\ell=-j_L}^{j_L}\prod_{k=1}^\infty(1-q^{2\ell+k}Q_\beta)^{(-1)^{2j_L}k\,n_{j_L}^{(\beta)}}\,.
\label{eq:decoupled}
\end{eqnarray}
We should eliminate this for any $j_L$ and $\beta$ such that $D_i\cdot \beta =0$ for all $D_i$. Here we used the shorthand notation $Q_\beta= e^{-2\pi it\cdot\beta}$.

\subsection{Five-dimensional $U(1)$-factor from geometry}

\begin{figure}
\begin{center}
\includegraphics[width=5cm]{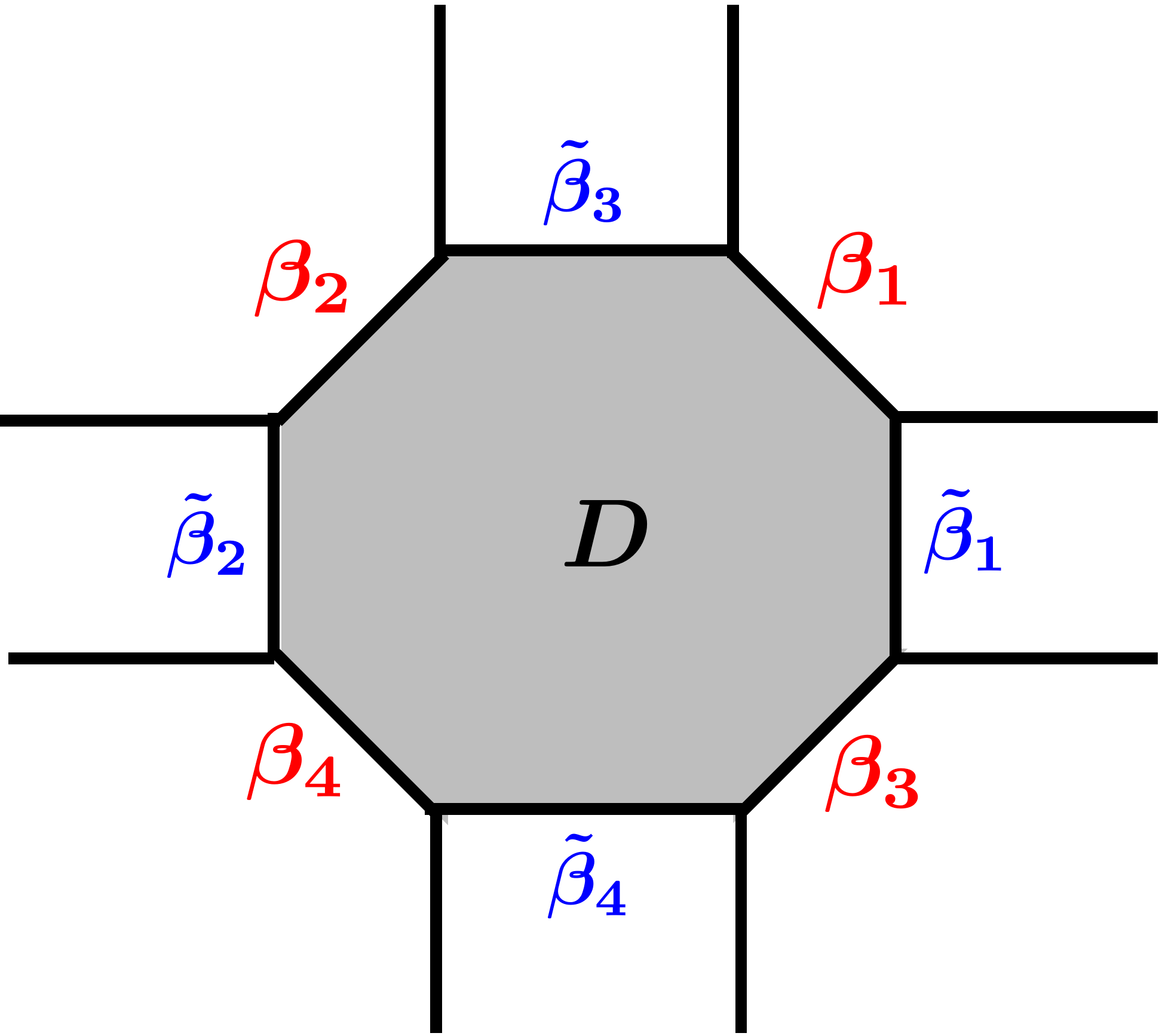}\qquad\qquad
\includegraphics[width=5cm]{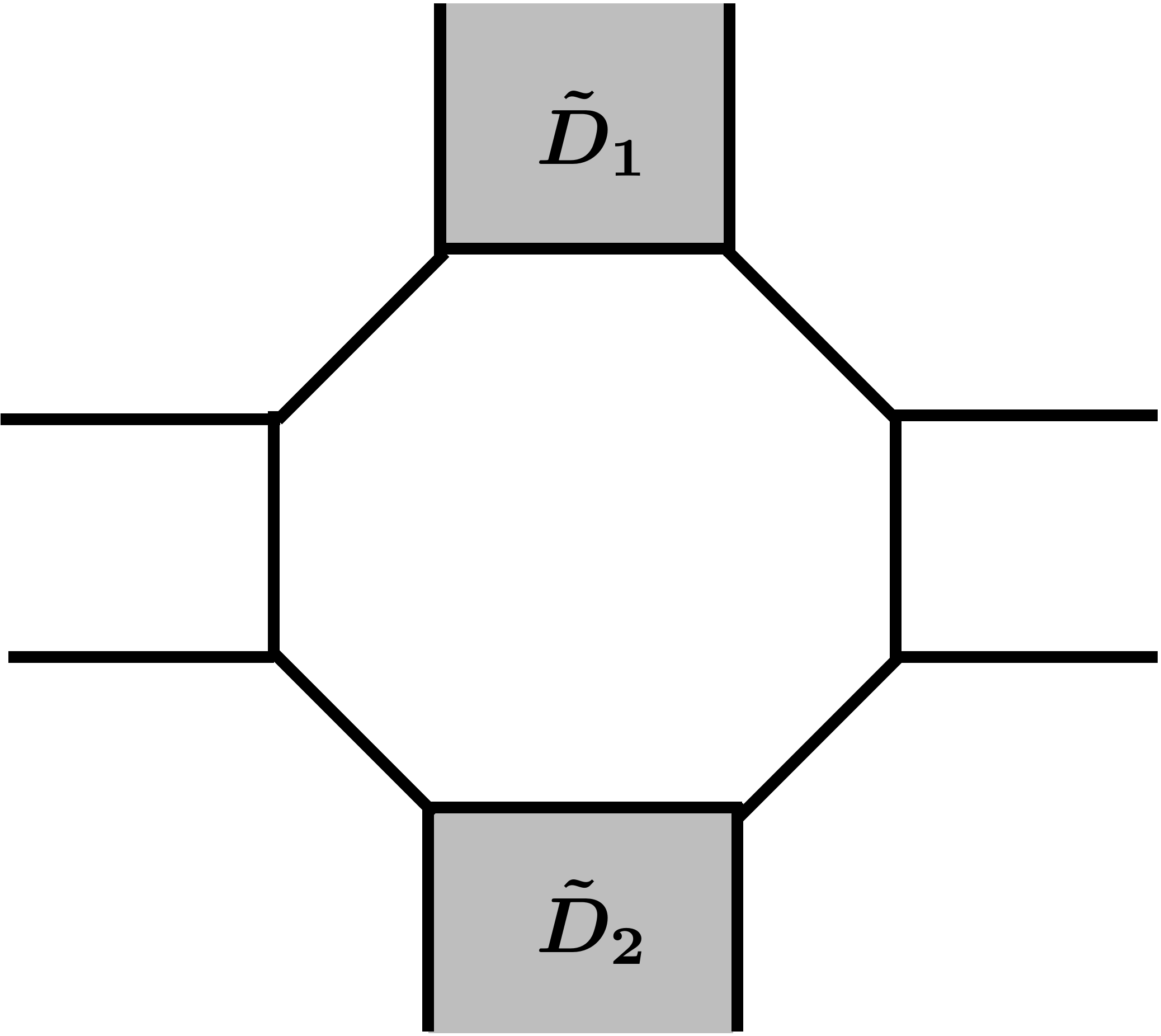}
\caption{Left: The toric web-diagram of the four-point blowup of local $\mathbb{P}^1_b\times \mathbb{P}^1_f$. In our notation, $\mathbb{P}^1_f = \beta_2 + \beta_4 + \tilde{\beta}_2 = \beta_1 + \beta_3 + \tilde{\beta}_1$ and $\mathbb{P}^1_b = \beta_1 +\beta_2 + \tilde{\beta}_3 = \beta_3 +\beta_4 +\tilde{\beta}_4$. We have a single compact four-cycle $D$. \; Right: Two non-compact four-cycles $\tilde{D}_1$ and $\tilde{D}_2$ are depicted.}
\label{fig:web-diagram-Nf=4}
\end{center}
\end{figure}

One subtlety here is that, when $X$ engineers an $SU(N)$ gauge theory, the topological string partition function $\exp \mathcal{F}$ is known to reproduce the Nekrasov partition function of a $U(N)$ gauge theory rather than $SU(N)$ \cite{Iqbal:2003ix, Iqbal:2003zz, Eguchi:2003sj, Hollowood:2003cv}. In this subsection, we explain the physical reason for this using the above argument of decoupled M2-branes.

For example, suppose that $X$ is the four-point blowup of local $\mathbb{P}^1_b\times \mathbb{P}^1_f$, whose toric web-diagram is shown in figure \ref{fig:web-diagram-Nf=4}. Let $\beta_i$ be the $i$-th blowup two-cycle. There are six independent K\"ahler parameters; $\mathfrak{t}_b, \mathfrak{t}_f$ for $\mathbb{P}_b$ and $\mathbb{P}_f$ respectively, and $\mathfrak{t}_i$ for $\beta_i$. This $X$ engineers an $SU(2)$ gauge theory with $N_f=4$ flavors. The Coulomb branch parameter $a$, masses of matter hypermultiplets $m_i$ and the gauge coupling $g^2$ are related to the K\"ahler parameters by 
$\mathfrak{t}_f = 2a,\,\mathfrak{t}_i = a-m_i$ and $\mathfrak{t}_b = 1/g^2 + 2a$.
The topological string partition function on $X$ is evaluated in terms of $Q_i = e^{-2\pi \mathfrak{t}_i}$, $Q_f=e^{-2\pi \mathfrak{t}_f}$ and $Q_b=e^{-2\pi \mathfrak{t}_b}$ as \cite{Aganagic:2002qg}
\begin{eqnarray}
\exp \mathcal{F} &=& \prod_{n=1}^{\infty}\frac{1}{(1-q^n\tilde{Q}_1)^n(1-q^n\tilde{Q}_2)^n}\times \prod_{n=1}^\infty\left(\frac{1}{1-q^n}\right)^{\frac{n\,\chi(X)}{2}}
\nonumber\\[1mm]
&&\quad \times \frac{\prod_{i=1}^{4}\prod_{n=1}^\infty(1-q^nQ_i)^{n}(1-q^nQ_fQ_i^{-1})^n}{\prod_{n=1}^\infty(1-q^nQ_f)^{2n}}\times Z_{\rm inst}^{\rm U(2)}(Q_b,Q_f,Q_i)\,,
\label{eq:Nf=4-unrefined}
\end{eqnarray}
where we used the short-hand notations $\tilde{Q}_1 = Q_fQ_1^{-1}Q_3^{-1},\, \tilde{Q}_2 = Q_fQ_2^{-1}Q_4^{-1}$.
Here $\chi(X)$ is the Euler characteristic of $X$, which cannot be determined unambiguously because $X$ is non-compact. The moral is that we set $\chi(X)$ to be twice the number of $U(1)$ vector multiplets. When we write $\tilde{Q}_i = \exp(-2\pi \mathfrak{t}\cdot\tilde{\beta}_i)$, the two-cycles $\tilde{\beta}_1$ and $\tilde{\beta}_2$ have vanishing intersections with $D$. This means that the first factor
\begin{eqnarray}
 \prod_{n=1}^\infty \frac{1}{(1-q^n\tilde{Q}_1)^n(1-q^n\tilde{Q}_2)^n}
\label{eq:decoupled-1}
\end{eqnarray}
is a contribution from decoupled M2-branes as discussed near \eqref{eq:decoupled}. In the study of 5d gauge theory we should omit this factor. The remaining part of $\exp \mathcal{F}$ turns out to coincide with the Nekrasov partition function of $U(2)$ gauge theory with four flavors \cite{Hollowood:2003cv}. In particular,  $Z_{\rm inst}^{U(2)}$ agrees with the instanton partition function.\footnote{The second and third factors of \eqref{eq:Nf=4-unrefined} coincide with the perturbative part of the Nekrasov partition. Whether the perturbative part describes $U(2)$ or $SU(2)$ gauge symmetry depends on $\chi(X)$ which cannot be determined unambiguously.} 

While $X$ engineers $SU(2)$ gauge theory, we have obtained $U(2)$ Nekrasov partition function. The reason for this is that we did not eliminate all the contributions from decoupled M2-branes. To see this, note first that there are two more independent two-cycles $\tilde{\beta}_3,\tilde{\beta}_4$ whose intersections with $D$ vanish. In terms of $\mathbb{P}^1_{b},\mathbb{P}^1_f$ and $\beta_i$, they are expressed as
\begin{eqnarray}
\tilde{\beta}_3 = \mathbb{P}^1_b - \beta_1 -\beta_2,\qquad \tilde{\beta}_4 = \mathbb{P}^1_b - \beta_3-\beta_4\,.
\end{eqnarray}
M2-branes wrapping these two-cycles are not charged under $SU(2)$ gauge symmetry engineered, and therefore to be decoupled. From the symmetry of the toric diagram,\footnote{When we consider the refined topological string, this symmetry does not exist.} we find that $\exp \mathcal{F}$ contains the factor
\begin{eqnarray}
\prod_{n=1}^\infty \frac{1}{(1-q^n\tilde{Q}_3)^n(1-q^n\tilde{Q}_4)^n}\,,
\label{eq:decoupled-2}
\end{eqnarray}
coming from M2-branes on $\tilde{\beta}_3$ or $\tilde{\beta}_4$. Since both $\tilde{Q}_3$ and $\tilde{Q}_4$ involve $Q_b$, this factor is included in $Z_{\rm inst}^{U(2)}(Q_b,Q_f,Q_i)$. We interpret that the inclusion of \eqref{eq:decoupled-2} is the reason why we have obtained the $U(2)$ Nekrasov partition function rather than $SU(2)$.

One way to verify this interpretation is to identify an additional $U(1)$ gauge field which couples to M2-branes on $\tilde{\beta}_3,\tilde{\beta}_4$. Let us consider two non-compact four-cycles $\tilde{D}_1$ and $\tilde{D}_2$ shown in figure \ref{fig:web-diagram-Nf=4}. There is a $U(1)$ gauge symmetry associated with
\begin{eqnarray}
D' = \tilde{D}_1- \tilde{D}_2\,,
\end{eqnarray}
which we denote by $U(1)_{D'}$. Since $\tilde{D}_1$ and $\tilde{D}_2$ are non-compact, $U(1)_{D'}$ is decoupled in the rigid limit $\tilde{X}\to X$.
The intersections $D'\cdot \tilde{\beta}_{1} = D'\cdot\tilde{\beta}_2 = 0$ and $D'\cdot \tilde{\beta}_3 = - D'\cdot\tilde{\beta}_4 = 2$ imply that M2-branes on $\tilde{\beta}_3,\tilde{\beta}_4$ are charged under $U(1)_{D'}$ but those on $\tilde{\beta}_1,\tilde{\beta}_2$ are not. Therefore, eliminating \eqref{eq:decoupled-1} from $\exp\mathcal{F}$ while keeping \eqref{eq:decoupled-2} is equivalent to keeping $U(1)_{D'}$ in addition to the $SU(2)$ gauge symmetry in five dimensions. What is this additional $U(1)_{D'}$ gauge symmetry?
Since $D'\cdot \mathbb{P}^1_f = 0$, the W-boson is neutral under $U(1)_{D'}$. On the other hand, $D'\cdot \beta_{1} = D'\cdot\beta_{2} = -D'\cdot \beta_{3} = -D'\cdot \beta_{4}=-1$ imply that $U(1)_{D'}$ couples to the four matter hypermultiplets with the same magnitude of electric charge.\footnote{To be more precise, an M2 on $\beta_i$ and an $\overline{\rm M2}$ on $\mathbb{P}^1_f - \beta_i$, which form a (anti-) fundamental representation of $SU(2)$, have the same electric charge of $U(1)_{D'}$.} This means that $U(1)_{D'}$ is identified with the center $U(1)$ of the $U(2)$ gauge symmetry, which verifies our interpretation of the difference between $U(2)$ and $SU(2)$.

From this argument we expect that, by eliminating both \eqref{eq:decoupled-1} and \eqref{eq:decoupled-2} from $\exp \mathcal{F}$, we obtain the Nekrasov partition function of $SU(2)$ gauge theory rather than $U(2)$. We will explicitly check this in section \ref{sec:SU}. The latter factor \eqref{eq:decoupled-2} describes the difference between the $U(2)$ and $SU(2)$ gauge theories. Recently, such a difference has attracted much attention in the study of the AGT relation \cite{Alday:2009aq}. In particular, it was proposed in \cite{Alday:2009aq} that the 4d Nekrasov partition functions of $U(2)$ and $SU(2)$ gauge theories are generally related by multiplying a so-called ``$U(1)$-factor.''\footnote{For further studies on the $U(1)$-factor in four dimensions, see \cite{Hollands:2010xa}.} Here, we have found the {\it five-dimensional} counterpart of the $U(1)$-factor, and given its interpretation in the geometric engineering. Note that our interpretation is applicable to any toric Calabi-Yau threefold $X$ which engineers a 5d gauge theory.

\subsection{Refined topological string}
\label{subsec:refined-vertex}

\begin{figure}
\begin{center}
\includegraphics[width=2.3cm]{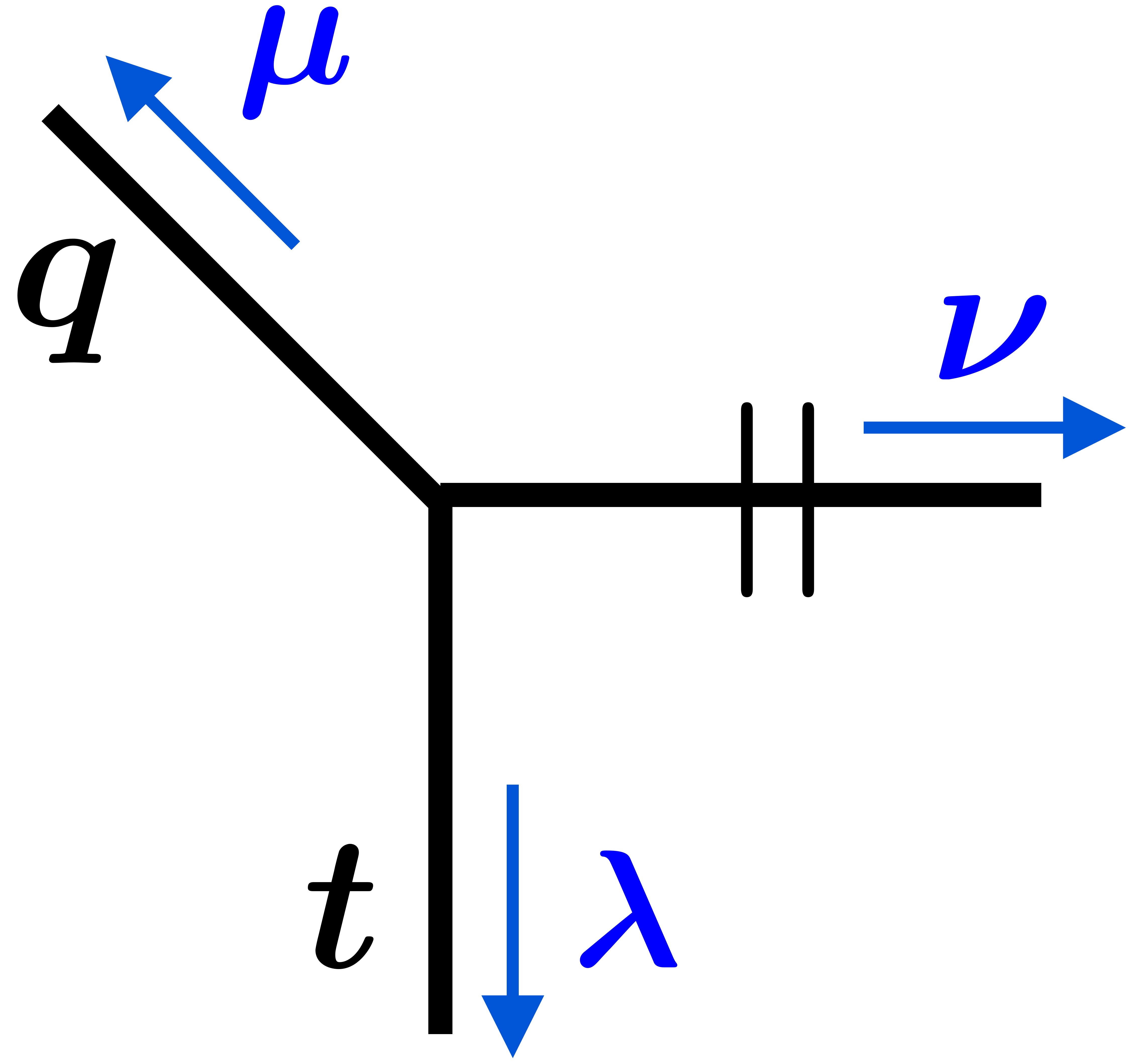}\qquad\quad
\includegraphics[width=3.7cm]{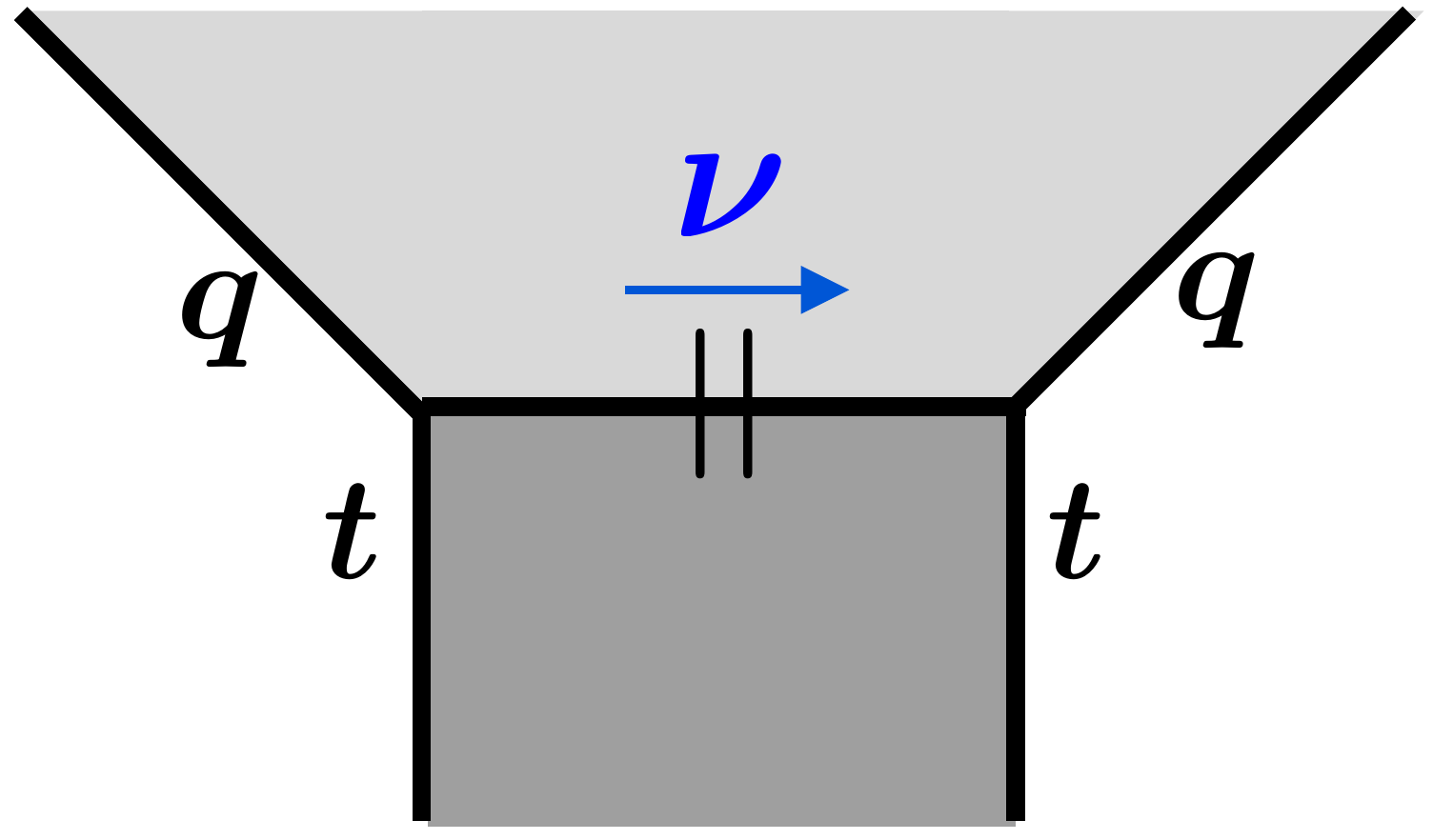}\qquad\quad
\includegraphics[width=3.7cm]{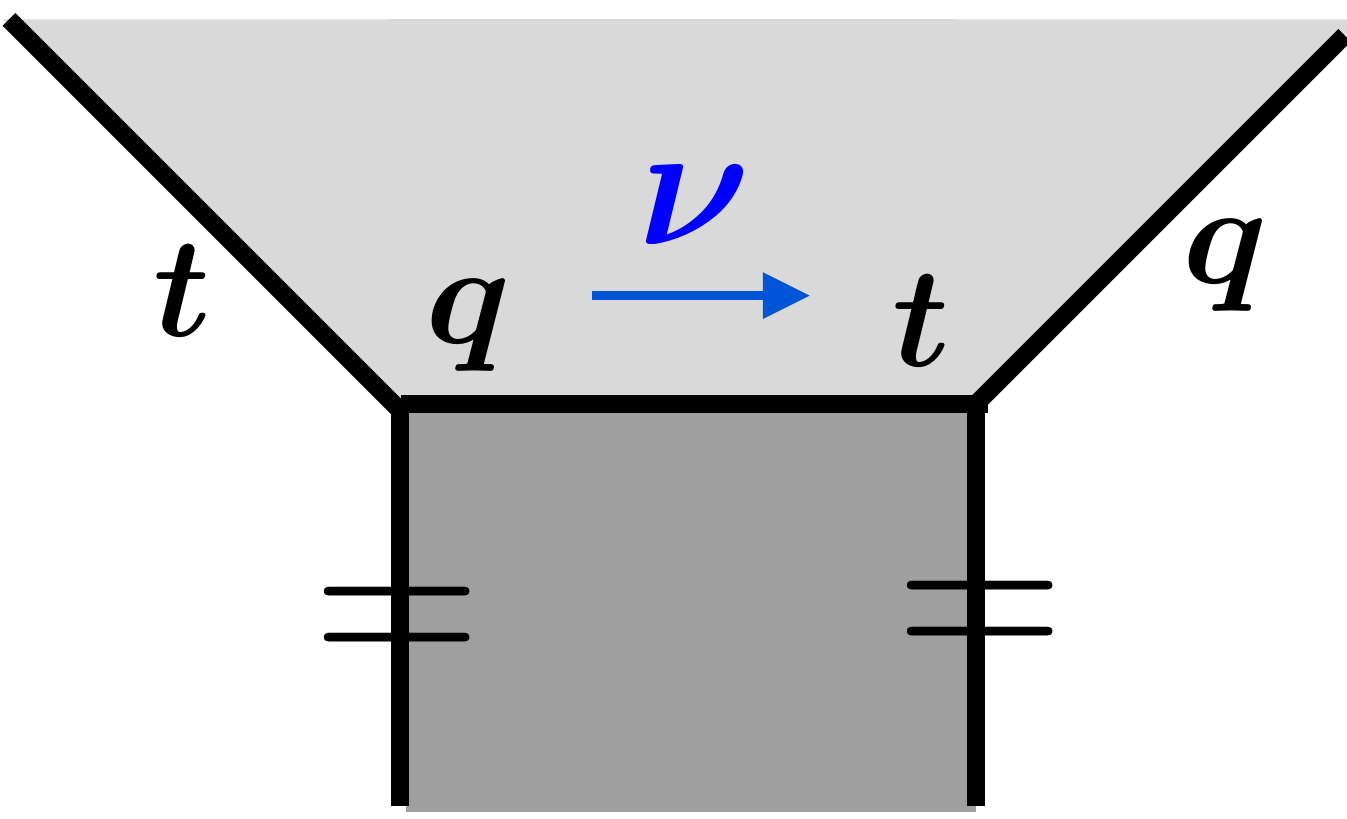}
\caption{Left: We assign $C_{\lambda\mu\nu}(t,q)$ to this vertex. The preferred direction is indicated by $||$. The vertex factor also depends on the positions of $q$- and $t$-directions. \; Middle: Near the two-cycle $\beta$, the toric Calabi-Yau $X$ is identified with the total space of $\mathcal{O}(m_\beta-1)\oplus\mathcal{O}(-m_\beta-1)\to\mathbb{P}^1$. This picture indicates the case of $m_\beta = 1$. When the toric web-diagram near $\beta$ is as in the picture, the framing factor is given by $f_\nu(t,q)^{m_\beta}$. Here the dark and light gray regions indicate the line bundles $\mathcal{O}(m_\beta-1)\to \mathbb{P}^1$ and $\mathcal{O}(-m_\beta -1)\to \mathbb{P}^1$, respectively. \; Right: When the web-diagram near $\beta$ is as in the picture, the framing factor is given by $\tilde{f}_\nu(t,q)^{m_\beta}$.}
\label{fig:vertex}
\end{center}
\end{figure}
So far we have discussed the un-refined topological string and its relation to the Nekrasov partition function with $\epsilon_1=-\epsilon_2$. In this paper, we are interested in the Nekrasov partition function for a general $\Omega$-background $\epsilon_1\neq -\epsilon_2$. It was argued in \cite{Hollowood:2003cv} that, when the gauge theory is engineered by some toric Calabi-Yau threefold $X$, the logarithm of the Nekrasov partition function is written as
\begin{eqnarray}
\mathcal{F}_{\rm ref} =  -\sum_{\beta,j_L,j_R}\sum_{k=1}^\infty\sum_{\ell=-j_L}^{j_L}\sum_{r=-j_R}^{j_R}(-1)^{2j_L+2j_R}\frac{N_{j_L,j_R}^{(\beta)}}{k}\frac{(tq)^{k\ell}(t/q)^{kr}}{(q^{k/2}-q^{-k/2})(t^{k/2}-t^{-k/2})}e^{-2\pi k \mathfrak{t}\cdot \beta} \,,
\label{eq:refined}
\end{eqnarray}
where $q=e^{-i\epsilon_2},\, t=e^{i\epsilon_1}$. Here $\mathcal{F}_{\rm ref}$ depends on the full spin spectrum  $N_{j_L,j_R}^{(\beta)}$ of M2-branes on $\beta$. Since \eqref{eq:refined} reduces to $\mathcal{F}$ if $t=q$, $\mathcal{F}_{\rm ref}$ is regarded as a generalization of the topological string amplitude. In this paper, we call $\exp \mathcal{F}_{\rm ref}$ the ``refined topological partition function,'' following the literature.

Although the world-sheet definition of the refined topological string is still mysterious (see \cite{Antoniadis:2010iq, Nakayama:2011be, Antoniadis:2013bja, Antoniadis:2013mna} for recent progress), it was proposed that $\exp \mathcal{F}_{\rm ref}$ is evaluated via the refinement \cite{Awata:2005fa, Iqbal:2007ii} of the topological vertex \cite{Aganagic:2003db}. In this paper, we use the so-called ``refined topological vertex'' proposed in \cite{Iqbal:2007ii}.
We first draw the toric web-diagram of the toric Calabi-Yau threefold $X$, which is decomposed into trivalent vertices and (internal and external) edges.
Each internal edge is associated with a Young diagram, and $\exp\mathcal{F}_{\rm ref}$ is written as a sum over all possible combinations of the Young diagrams up to a prefactor.\footnote{Here the prefactor is given by the refinement of the constant map contribution: $(M(t,q)M(q,t))^{-\chi(X)/4}$, where $M(t,q) = \prod_{i,j=1}^\infty(1-q^{i}t^{j-1})^{-1}$ is the refined MacMahon function and $\chi(X)$ is the Euler characteristic of $X$.}
What we sum up is the multiplication of factors from every edge and vertex. The vertex factor is given by
\begin{eqnarray}
C_{\lambda\mu\nu}(t,q) = t^{-\frac{||\mu^t||^2}{2}}q^{\frac{||\mu||^2 + ||\nu||^2}{2}}\widetilde{Z}_{\nu}(t,q) \sum_{\eta}\left(\frac{q}{t}\right)^{\frac{|\eta|+|\lambda|-|\mu|}{2}}s_{\lambda^t/\eta}(t^{-\rho}q^{-\nu})s_{\mu/\eta}(t^{-\nu^t}q^{-\rho})\,,
\label{eq:vertex}
\end{eqnarray}
where $\lambda,\mu$ and $\nu$ are the Young diagrams of the edges attached to the vertex. Here we assign $\emptyset$ to external edges. The function $\tilde{Z}_{\nu}(t,q)$ is written in terms of $\ell_{\nu}(i,j) = \nu_i - j$ and $a_{\nu}(i,j) = \nu_j^t - i$ as
\begin{align}
\tilde{Z}_{\nu}(t,q) = \prod_{s\in \nu}(1-q^{\ell_{\nu}(s)}t^{a_{\nu}(s)+1})^{-1}\,,
\end{align}
and $s_\nu(\textbf{x})$ is the Schur function.
 Since $C_{\lambda\mu\nu}(t,q)$ is not symmetric under permutations of the Young diagrams, we have to specify their ordering. For every vertex, we choose a ``preferred direction'' and call the other two directions $q$- and $t$-directions, respectively. We do this so that every internal edge connects two preferred directions or two un-preferred directions. We also impose that all the preferred directions are parallel in the toric web-diagram. Moreover, when connecting two un-preferred directions, we impose that one should be a $q$-direction and the other should be a $t$-direction. With this rule, we assign $C_{\lambda\mu\nu}(t,q)$ to a vertex in Figure \ref{fig:vertex}. Note that for a general toric web-diagram this assignment might be impossible. In that case we should also use another vertex factor \cite{Aganagic:2012hs, Iqbal:2012mt}, but we will not discuss such examples in this paper.

To describe the edge factor, we first define
\begin{align}
f_\nu(t,q) = (-1)^{|\nu|}t^{\frac{||\nu^t||^2}{2}}q^{-\frac{||\nu||^2}{2}}\,,\qquad \tilde{f}_\nu(t,q) = (-1)^{|\nu|}t^{\frac{||\nu^t||^2}{2}}q^{-\frac{||\nu||^2}{2}}(t/q)^{\frac{|\nu|}{2}}\,,
\end{align}
 where $|\nu|$ is the number of boxes in $\nu$. Let us consider an edge associated with a two-cycle $\beta$ and a Young diagram $\nu$.  The edge factor for the edge is given by
\begin{eqnarray}
e^{-2\pi \mathfrak{t}\cdot \beta|\nu|} f(\beta,\nu,t,q)\,,
\end{eqnarray}
where $f(\beta,\nu,t,q)$ is the so-called ``framing factor.'' Note that, near the two-cycle $\beta$, $X$ is locally identified with the total space of $\mathcal{O}(m_\beta-1)\oplus\mathcal{O}(-m_\beta-1)\to \mathbb{P}^1$ for some $m_\beta\in\mathbb{N}$. The framing factor depends on $m_\beta$, the positions of $q$- and $t$-edges, and whether the edge is a preferred or un-preferred direction. If the edge is a {\it preferred direction} as in the middle of figure \ref{fig:vertex}, the framing factor is given by
\begin{eqnarray}
f(\beta,\nu,t,q) = f_\nu(t,q)^{m_\beta}\,.
\end{eqnarray}
If the edge is an {\it un-preferred direction} as in the right of figure \ref{fig:vertex}, we set
\begin{eqnarray}
f(\beta,\nu,t,q) = \tilde{f}_\nu(t,q)^{m_\beta}\,.
\label{eq:unpreferred}
\end{eqnarray}
 Note that the framing factor \eqref{eq:unpreferred} for un-preferred directions is slightly different from that in \cite{Iqbal:2007ii,Iqbal:2012mt}, but we will see that \eqref{eq:unpreferred} leads to the correct 5d Nekrasov partition function with no ambiguity.

We finally note that the refinement of \eqref{eq:decoupled} is given by
\begin{eqnarray}
\prod_{\ell=-j_L}^{j_L}\prod_{r=-j_R}^{j_R}\prod_{i,j=1}^\infty \left(1- q^{\ell-r+i-1/2} t^{\ell+r+j-1/2} Q_\beta\right)^{(-1)^{2j_L+2j_R}\,N_{j_L,j_R}^{(\beta)}}\,.
\label{eq:U1-refined}
\end{eqnarray}
We interpret that, when comparing $\exp \mathcal{F}_{\rm ref}$ with the Nekrasov partition function, we have to divide out $\exp \mathcal{F}_{\rm ref}$ by this factor for all $(j_L,j_R)$ and $\beta$ such that $\beta\cdot D_i=0$ for every compact four-cycle $D_i$ of $X$. One subtlety here is that the refined topological vertex does not capture the whole $SU(2)_R$ spin multiplet of M2-branes on such $\beta$. To see this, let us again regard $X$ as a local limit of some compact Calabi-Yau threefold $\tilde{X}$. Before taking the limit $\tilde{X}\to X$, the moduli space $\mathcal{M}$ of the curve $\beta$ in the Calabi-Yau threefold is compact and K\"ahler. Then $SU(2)_R$ is identified with the $SL(2)$ Lefschetz action on $\mathcal{M}$ \cite{Gopakumar:1998jq}, which means that the $SU(2)_R$ spin multiplet is formed by elements of $H^*(\mathcal{M})$. In the local limit $\tilde{X}\to X$, however, $\mathcal{M}$ becomes non-compact and the full $SU(2)_R$ multiplet needs contributions from infinity of $\mathcal{M}$. On the other hand, the refined topological vertex only captures the local property of the Calabi-Yau threefold. Therefore we expect that, in calculations with refined topological vertex, the factor to be eliminated is
\begin{eqnarray}
\prod_{\ell=-j_L}^{j_L}\prod_{i,j=1}^\infty(1-q^{\ell-r+i-\frac{1}{2}}t^{\ell+r+j-\frac{1}{2}}Q_\beta)^{(-1)^{2j_L+ 2j_R}N_{j_L,j_R}^{(\beta)}}\,,
\end{eqnarray}
for some $r$ such that $-j_R\leq r\leq j_R$. The value of $r$ depends on $\beta$, and the choices of the preferred, $q$- and $t$-directions. In the rest of this paper, we encounter the cases of $r=\pm 1/2$ for $j_R=1/2$.\footnote{On the other hand, the $SU(2)_L$ multiplet is identified with the Lefschetz decompositions of the cohomology of the moduli space of flat bundles over $\beta$. Since this is insensitive to the non-compactness of $\mathcal{M}$, the topological vertex realizes the whole $SU(2)_L$ spin multiplet even if the M2-brane is wrapping $\beta$ with a non-compact moduli space.}

\section{$Sp(1)$ Nekrasov partition functions from topological string}
\label{sec:Sp1}

\subsection{Refined topological vertex computation}
\label{subsec:vertex}


In this section, we consider the refined topological string partition function on the blow-ups of local $\mathbb{P}^1\times \mathbb{P}^1$ and $\mathbb{C}^3/(\mathbb{Z}_N\times \mathbb{Z}_N)$ for $N=2,3,4$. As reviewed in section \ref{sec:engineering}, the former engineers 5d $SU(2)$ gauge theories with fundamental hypermutiplets while the latter is expected to engineer the 5d version of $T_N$-theory.
Since the Calabi-Yau threefolds are toric, the partition functions are evaluated via the refined topological vertex.

\subsubsection{
Local $\P^1 \times \P^1$ with $2 \leq N_f \leq 4$ blow ups}
\label{subsubsec:F0}

For the $SU(2)$ gauge theories with $N_f=0$ and $1$, two partition functions are already shown to be equivalent\footnote{The relation between the instanton partition functions of 
two $SU(2)$ gauge theories with $N_f\!=\!0$ at different Chern-Simons levels, $\kappa=0$
and $\kappa=2$, is discussed in \cite{Bergman:2013ala}.
Their result agrees with our prescription in section \ref{subsec:U1-factors}.} \cite{Iqbal:2012xm,Bergman:2013ala}.
Thus, we here focus on the partition functions of the $SU(2)$ gauge theories with $2\le N_f\le 4$ fundamental hypermultiplets.

We shall calculate the topological string partition function for $2\le N_f \le 4$ points blow up of $\mathbb{P}^1\times \mathbb{P}^1$ whose toric fans are depicted in Figure \ref{fig:toric2}.
One expects that the result agrees with the $SU(2)$ gauge theory partition function
with $N_f$ fundamental hypermultiplet since M-theory compactifications on the toric varieties lead to the $SU(2)$ gauge theory with $N_f$ flavors as discussed in section \ref{sec:SU2}.
However  we find that the topological vertex calculation yields the $U(2)$  gauge theory partition function instead of that of $SU(2)$. 

\begin{figure}
	\begin{center}
	\includegraphics[scale=0.3]{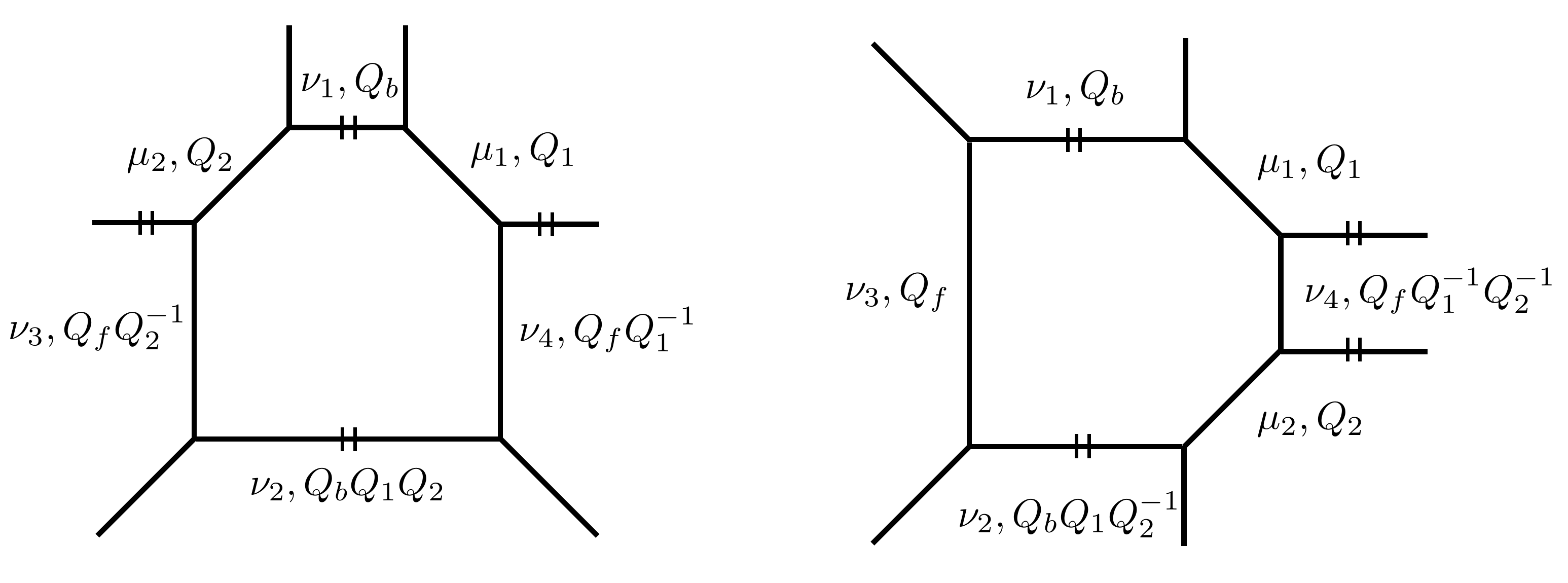}
	\caption{Toric diagrams of two points blow up of local $\mathbb{P}^1\times\mathbb{P}^1$. 
	The preferred direction is indicated by the double line.}
	\label{fig:Nf=2}
	\end{center}
\end{figure}

\paragraph{$N_f=2$}
For $N_f=2$, we 
consider two types of toric web diagrams as in  Figure~\ref{fig:Nf=2}.
The refined topological partition function for the first diagram of Figure~\ref{fig:Nf=2} can be written as
\bea
	\hspace{-0.5cm}Z^{(1)}_{N_f=2} \!\!&\!\!=\!\!&\!\! (M(t,q)M(q,t))^{1/2}Z^{(1)}(t,q,Q) \,,  \\
	\hspace{-0.5cm}Z^{(1)}(t,q,Q) \!\!&\!\!=\!\!&\!\! \sum_{\vec{\nu},\vec{\mu}}(-Q_b)^{|\nu_1|}(-Q_1)^{|\mu_1|}(-Q_2)^{|\mu_2|}
    (-Q_fQ^{-1}_1)^{|\nu_4|}(-Q_fQ^{-1}_2)^{|\nu_3|}(-Q_bQ_1Q_2)^{|\nu_2|} f_{\nu_1^t}(q,t)f_{\nu_2}(q,t)\nn \\
    &&\times 
    C_{\emptyset\mu_1\nu_1^t}(q,t)C_{\nu_4\mu_1^t\emptyset}(t,q)C_{\nu_4^t\emptyset\nu_2}(q,t)
    \times C_{\mu_2^t\emptyset\nu_1}(t,q)C_{\mu_2\nu_3^t\emptyset}(q,t)C_{\emptyset\nu_3\nu_2^t}(t,q) \,. \nn 
\label{eq:topst_Nf=2-1}
\eea
Here $Q_b, Q_f, Q_a$ are related to the K\"ahler parameters of the base $\mathbb{P}^1$, the fiber $\mathbb{P}^1$ and two exceptional curves from blowup, respectively.

By explicit calculation, one obtains
\bea
	\hspace{-0.8cm} Z^{(1)}_{N_f=2} 
			 \!\!&\!\!=\!\!&\!\! \left[\prod_{i,j=1}^\infty\!\!\frac{\prod_{a=1}^2(1-Q_aq^{i-\frac{1}{2}}t^{j-\frac{1}{2}}) (1-Q_fQ_a^{-1}q^{i-\frac{1}{2}}t^{j-\frac{1}{2}})}
			{(1\!-\!q^it^{j-1})^{\frac{1}{2}}(1\!-\!q^{i-1}t^{j})^{\frac{1}{2}}(1\!-\!Q_fq^it^{j-1})(1\!-\!Q_fq^{i-1}t^j)} \right]
			\sum_{\nu_1,\nu_2}(-Q_b)^{|\nu_1|}(-Q_bQ_1Q_2)^{|\nu_2|}q^{||\nu_1^t||^2+||\nu_2^t||^2} \nn \\
			\!\!&&\!\! \times \!\! \prod_{s\in\nu_1}\!\!\frac{\prod_{a=1}^2(1-Q_aq^{-i+\frac{1}{2}}t^{l_{\nu_1}(s)+\frac{1}{2}})}
			{(1\!-\!q^{a_{\nu_1}(s)+1}t^{l_{\nu_1}(s)})(1\!-\!q^{a_{\nu_1}(s)}t^{l_{\nu_1}(s)+1})
			(1\!-\!Q_fq^{a_{\nu_2}(s)+1}t^{l_{\nu_1}(s)})(1\!-\!Q_fq^{a_{\nu_2}(s)}t^{l_{\nu_1}(s)+1})} \nn \\
			\!\!&&\!\! \times \!\! \prod_{s\in\nu_2}\!\!\frac{\prod_{a=1}^2(1-Q_fQ_a^{-1}q^{i-\frac{1}{2}}t^{-l_{\nu_2}(s)-\frac{1}{2}})}
			{(1\!-\!q^{a_{\nu_2}(s)+1}t^{l_{\nu_2}(s)})(1\!-\!q^{a_{\nu_2}(s)}t^{l_{\nu_2}(s)+1})
			(1\!-\!Q_fq^{-a_{\nu_1}(s)-1}t^{-l_{\nu_2}(s)})(1\!-\!Q_fq^{-a_{\nu_1}(s)}t^{-l_{\nu_2}(s)-1})} \,. \nn \\
\eea
In order to compare this with the Nekrasov partition function, we reformulate this in terms of
the gauge theory parameters
\be
	q = e^{-\gamma_1+\gamma_2} \,, \quad t=e^{\gamma_1+\gamma_2} \,, \quad u = Q_bQ_1^{\frac{1}{2}}Q_2^{\frac{1}{2}}Q_f^{-\frac{1}{2}} \,, \quad Q_f = e^{-2i\lambda} \,, \quad Q_{a=1,2} = e^{-i\lambda+im_a}\,,
\label{parameters-Nf=2-1}
\ee
where $\lambda$ is the Coulomb branch parameter, $u$ is the instanton fugacity, $m_{a=1,2}$ are the mass parameters of the hypermultiplets and $i\epsilon_1 = \gamma_1+\gamma_2, i\epsilon_2=\gamma_1-\gamma_2$ denote the Omega deformation parameters for $\mathbb{R}^4$. 
In terms of these parameters, the partition function is written as
\bea
			Z^{(1)}_{N_f=2}\!\!&\!\!\equiv\!\!&\!\! Z_0^{(1)}Z^{(1)}_{\rm inst} \,, \nn \\
	\hspace{-0.5cm}Z_0^{(1)} \!\!&\!\!=\!\!&\!\! 
			\prod_{i,j=1}^\infty
			\frac{\prod_{a=1}^2(1-e^{-i\lambda+m_a}q^{i-\frac{1}{2}}t^{j-\frac{1}{2}})
			(1-e^{-i\lambda-m_a}q^{i-\frac{1}{2}}t^{j-\frac{1}{2}})}
			{(1-q^it^{j-1})^{\frac{1}{2}}(1-q^{i-1}t^{j})^{\frac{1}{2}}
			(1-e^{-2i\lambda}q^it^{j-1})(1-e^{-2i\lambda}q^{i-1}t^{j})} \,, \nn \\
			Z_{\rm inst}^{(1)} \!\!&\!\!=\!\!&\!\!
			\sum_{\vec{\nu}}u^{|\nu_1|+|\nu_2|}\prod_{\alpha=1}^2\prod_{s\in \nu_\alpha}
			\frac{e^{i(E_{\alpha\emptyset}+i\gamma_1)}\prod_{a=1}^2
			2i\sin\frac{E_{\alpha\emptyset}+i\gamma_1-m_a}{2}}
			{\prod_{\beta=1}^2(2i)^2\sin\frac{E_{\alpha\beta}}{2}\sin\frac{E_{\alpha\beta}+2i\gamma_1}{2}} \,,
\label{eq:part-U(2)-Nf=2-1}
\eea
where
\be	\label{definition-E}
	E_{\alpha\beta} = \lambda_\alpha - \lambda_\beta +i(\gamma_1+\gamma_2)\ell_{\nu_\alpha}(s)
		-i(\gamma_1-\gamma_2)(a_{\nu_\beta}(s)+1)\,,
\ee
and $\lambda_1 = -\lambda_2 = \lambda,\, \lambda_\emptyset = 0$.

Now the partition function (\ref{eq:part-U(2)-Nf=2-1}), which consists of two factors $Z_0^{(1)}$ and $Z_{\rm inst}^{(1)}$, can be identified with the Nekrasov partition function.
Firstly, the $Z_{\rm inst}^{(1)}$ factor perfectly agrees with the instanton contribution of the $U(2)$ gauge theory at Chern-Simons theory level $\kappa=+1$ with $N_f=2$ flavors obtained in \cite{Tachikawa:2004ur} after the identification (\ref{parameters-Nf=2-1}) of parameters in both gauge theory and string theory.
The product of sine factors in the numerator is exactly the contribution from two $U(2)$ fundamental hypermultiplets, while the product in the denominator is the vector multiplet contribution.
The phase factor in the numerator, i.e. $e^{\kappa i (E_{\alpha\emptyset}+i\gamma_1)}$, encodes the classical Chern-Simons level and it implies
that the Chern-Simons level of the gauge theory at hand is $\kappa=+1$.
On the other hand, the $Z_0^{(1)}$ factor corresponds to the perturbative contribution of the $SU(2)$
gauge theory partition function with 2 fundamental flavors, rather than $U(2)$ gauge group.
To identify this with that of the $U(2)$ gauge theory, an additional $U(1)$
vector multiplet contribution
\be
\prod_{i,j=1}^\infty(1-q^it^{j-1})^{-\frac{1}{2}}(1-q^{i-1}t^{j})^{-\frac{1}{2}} 
\ee
should be included in the string theory partition function. This is achieved by shifting $\chi(X)$ by two.\footnote{Recall that the Euler characteristic $\chi(X)$ of the non-compact Calabi-Yau $X$ is not unambiguously determined. Our moral is that we set $\chi(X)$ to be twice the number of vector multiplets.}
For this case therefore the partition function $Z_{N_f=2}^{(1)}$ agrees with the 5d Nekrasov
partition function of $U(2)$ gauge theory at CS-level $+1$ with two fundamental hypermultiplets.

\begin{figure}
\centering
\includegraphics[width=4cm]{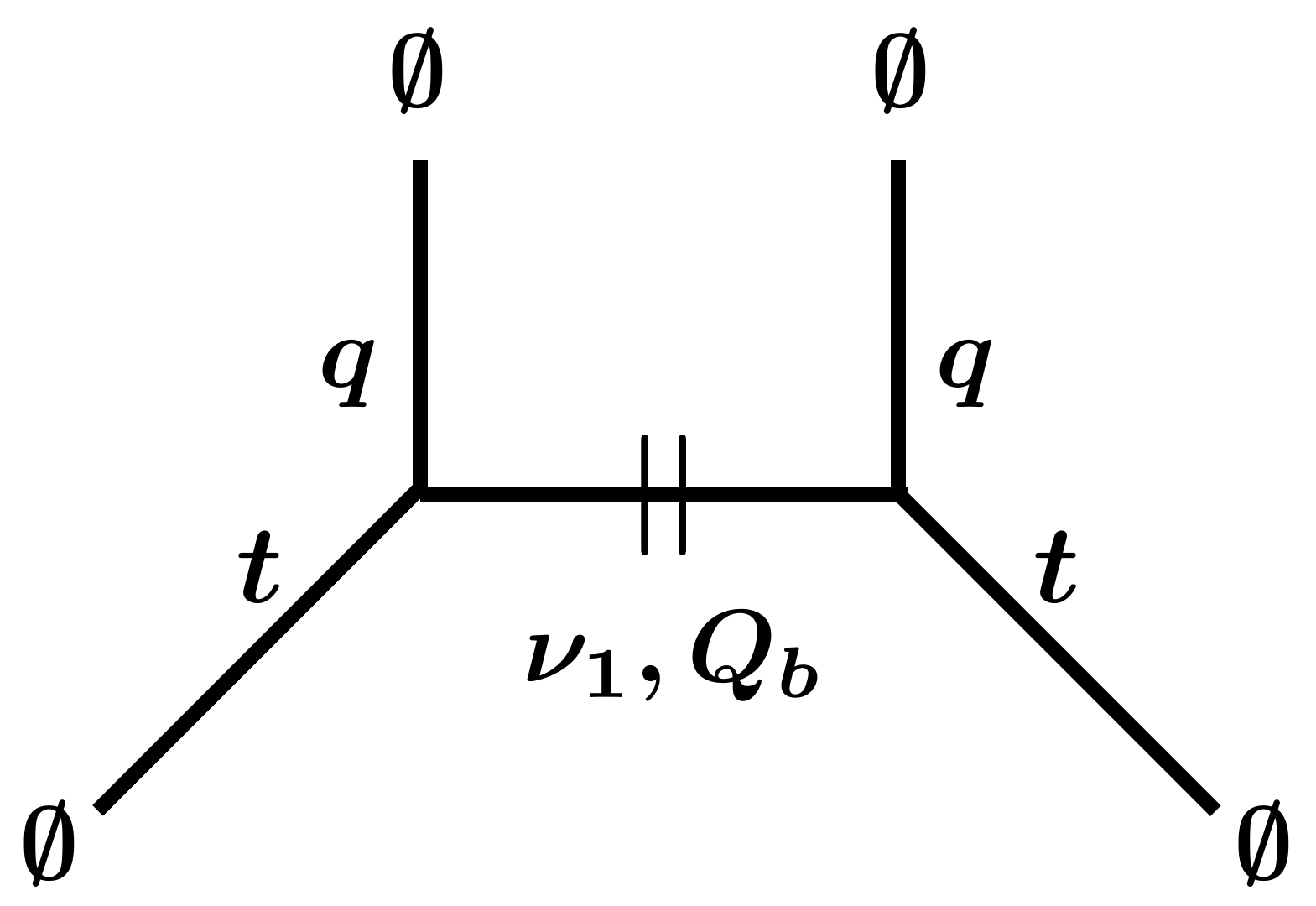}
\caption{The diagram giving rise to $Z_{U(1)}^{||}$.}
\label{fig:U1-Nf=2-1}
\end{figure}

Note here that $Z^{(1)}_{N_f=2}$ contains contributions from M2-branes with no $SU(2)$ electric charge. Such M2-branes are wrapping the two-cycle associated with $Q_b = ue^{-\frac{i}{2}(m_1+m_2)}$. The general argument in section \ref{sec:topst} implies that such M2-branes contribute
\begin{eqnarray}
\prod_{\ell=-j_L}^{j_L}\prod_{i,j=1}^\infty(1-Q_b \, q^{\ell-r+i-\frac{1}{2}}t^{\ell+r+j-\frac{1}{2}})^{(-1)^{2j_L + 2j_R}N_{j_L,j_R}^{(\beta)}} 
\label{eq:U1_Nf=2-1}
\end{eqnarray}
to $Z^{(1)}_{N_f=2}$ for some $r, j_L,j_R$ such that $-j_R\leq r\leq j_R$. We find $j_L=0$ because the two-cycle associated with $Q_b$ is genus zero. Moreover, the moduli space of deformations of the two-cycle is $\mathbb{C}$, which is regarded as a local limit of $\mathbb{P}^1$. This suggests that $j_R=1/2$. We can verify these by looking at the refined topological vertex calculation \eqref{eq:topst_Nf=2-1}. Since we are only interested in the factor \eqref{eq:U1_Nf=2-1}, we set all the fugacities to be zero except for $Q_b$. Then \eqref{eq:topst_Nf=2-1} reduces to
\begin{align}
Z_{U(1)}^{||} \equiv \sum_{\nu_1} \frac{Q_b^{|\nu_1|}q^{||\nu_1||^2}}{\prod_{s\in \nu_1}(1-q^{\ell_{\nu_1}(s)}t^{a_{\nu_1}(s) + 1}) ( 1- q^{\ell_{\nu_1}(s) + 1} t^{a_{\nu_1}(s)})}\,,
\end{align}
which is the amplitude associated with the diagram in figure \ref{fig:U1-Nf=2-1}.
Using the identity (6.5) of \cite{Iqbal:2008ra}, we can rewrite this as
\begin{align}
Z_{U(1)}^{||} = \prod_{k,\ell=1}^\infty (1-ue^{-\frac{i}{2}(m_1+m_2)}q^k t^{\ell-1})^{-1}\,.
\end{align}
This factor is of the form \eqref{eq:U1_Nf=2-1} with $N_{j_L,j_R}^{(\beta)} = \delta_{j_L,0}\, \delta_{j_R,\frac{1}{2}}$ and $r=-1/2$ for $j_R=1/2$. Note that $Z_{U(1)}^{||}$ contains the fugacity $u$ for instantons, which implies that it comes from $Z_\text{inst}^{(1)}$ in \eqref{eq:part-U(2)-Nf=2-1}.


Let us now turn to the second diagram of Figure \ref{fig:Nf=2}. The refined topological string partition function is now evaluated as
\bea
	Z_{N_f=2}^{(2)} \!\!&\!\!=\!\!&\!\! (M(t,q)M(q,t))^{1/2}Z^{(2)}(t,q,Q) \,,\nn \\
	\hspace{-0.5cm}Z^{(2)}(t,q,Q) \!\!&\!\!=\!\!&\!\! \sum_{\vec{\nu},\vec{\mu}}(-Q_b)^{|\nu_1|}(-Q_1)^{|\mu_1|}(-Q_2)^{|\mu_2|}
    (-Q_fQ^{-1}_1Q_2^{-1})^{|\nu_4|}(-Q_f)^{|\nu_3|}(-Q_bQ_1Q_2^{-1})^{|\nu_2|}\tilde{f}_{\nu_3}(t,q)\tilde{f}_{\nu_4^t}(t,q) \nn \\
    &&  \times C_{\emptyset\mu_1\nu_1^t}(q,t)C_{\nu_4\mu_1^t\emptyset}(t,q)C_{\mu_2\nu_4^t\emptyset}(t,q)C_{\mu_2^t\emptyset\nu_2}(q,t)
    \times C_{\nu_3^t\emptyset\nu_1}(t,q)C_{\emptyset\nu_3\nu_2^t}(t,q) \,.
\eea
A short calculation shows that
\bea
	Z^{(2)}_{N_f=2} \!\!&\!\!=\!\!&\!\!  Z_0^{(2)}Z_{\rm inst}^{(2)} \,,\nn \\
	Z_0^{(2)} \!\!&\!\!=\!\!&\!\! 
	\prod_{i,j=1}^\infty
	\frac{\prod_{a=1}^2(1-e^{-i\lambda+m_a}q^{i-\frac{1}{2}}t^{j-\frac{1}{2}})
	(1-e^{-i\lambda-m_a}q^{i-\frac{1}{2}}t^{j-\frac{1}{2}})}
	{(1-q^it^{j-1})^{\frac{1}{2}}(1-q^{i-1}t^{j})^{\frac{1}{2}}
	(1-e^{-2i\lambda}q^it^{j-1})(1-e^{-2i\lambda}q^{i-1}t^{j})
	(1-e^{-i(m_1+m_2)}q^{i-1}t^j)} \,, \nn \\
	Z_{\rm inst}^{(2)} \!\!&\!\!=\!\!&\!\!
	\sum_{\nu_1,\nu_2}u^{|\nu_1|+|\nu_2|}\prod_{\alpha=1}^2\prod_{s\in \nu_\alpha}
	\frac{
	\left(2i\sin\frac{E_{\alpha\emptyset}+i\gamma_1-m_1}{2}\right)
	\left(2i\sin\frac{-E_{\alpha\emptyset}-i\gamma_1-m_2}{2}\right)}
	{\prod_{\beta=1}^2(2i)^2\sin\frac{E_{\alpha\beta}}{2}\sin\frac{E_{\alpha\beta}+2i\gamma_1}{2}} \,.
\eea
Here we introduce another identification of the K\"ahler parameters different from the previous case.
\be\label{U(1)-Nf=2}
	u = -Q_bQ_1^{\frac{1}{2}}Q_2^{-\frac{1}{2}}Q_f^{-\frac{1}{2}} \,,\quad Q_f =e^{-2i\lambda}
	\,, \quad Q_{a=1,2} = e^{-i\lambda+im_a} \,.
\ee 
This result is rather different from the previous result for the first diagram of Figure \ref{fig:Nf=2}.
Firstly, there is an extra contribution to the perturbative part $Z^{(2)}_0$:
\be
	Z_{U(1)}^{=}\equiv \prod_{i,j=1}^\infty(1-e^{-i(m_1+m_2)}q^{i-1}t^j)^{-1}.
\label{eq:U1_Nf=2-2}
\ee
Also the instanton part $Z^{(2)}_{\rm inst}$ is different.
The numerator of the instanton part for this case implies that the $U(2)$
gauge theory now has zero Chern-Simons level and couples to two hypermultiplets,
one in the fundamental and the other one in the anti-fundamental representation, which is very different from the previous interpretation. Therefore $Z^{(2)}/Z_{U(1)}^{=}$ gives the Nekrasov partition function of $U(2)$ gauge theory with a fundamental and an anti-fundamental matter fields, and the vanishing Chern-Simons level.
Thus the topological vertex computations of the two toric diagrams
yield the two different 
Nekrasov partition functions.
Note here that \eqref{eq:U1_Nf=2-2} is precisely the contribution from M2-branes on the two-cycle associated with $Q_fQ_1^{-1}Q_2^{-1}$. Such M2-branes are not charged under the $SU(2)$ gauge symmetry, and to be decoupled. This gives a physical interpretation of the ratio $Z_{N_f=2}^{(2)}/Z_{U(1)}^{=}$. Note also that there is no other contributions from decoupled M2-branes in the second diagram of Figure \ref{fig:Nf=2}.

Despite the above difference between $Z_{N_f=2}^{(1)}$ and $Z_{N_f=2}^{(2)}$, we find that
they are in fact the same apart from decoupled factors which are independent of the Coulomb branch parameter. Two partition functions are related by
\begin{align}
Z^{(1)}_{N_f=2}\big/Z_{U(1)}^{||} = Z^{(2)}_{N_f=2}\big/Z_{U(1)}^{=}\,.
\label{eq:two_Nf=2}
\end{align}
The physical reason for this relation will be clear in the next subsection.

\begin{figure}[h!]
	\begin{center}
	\includegraphics[scale=0.3]{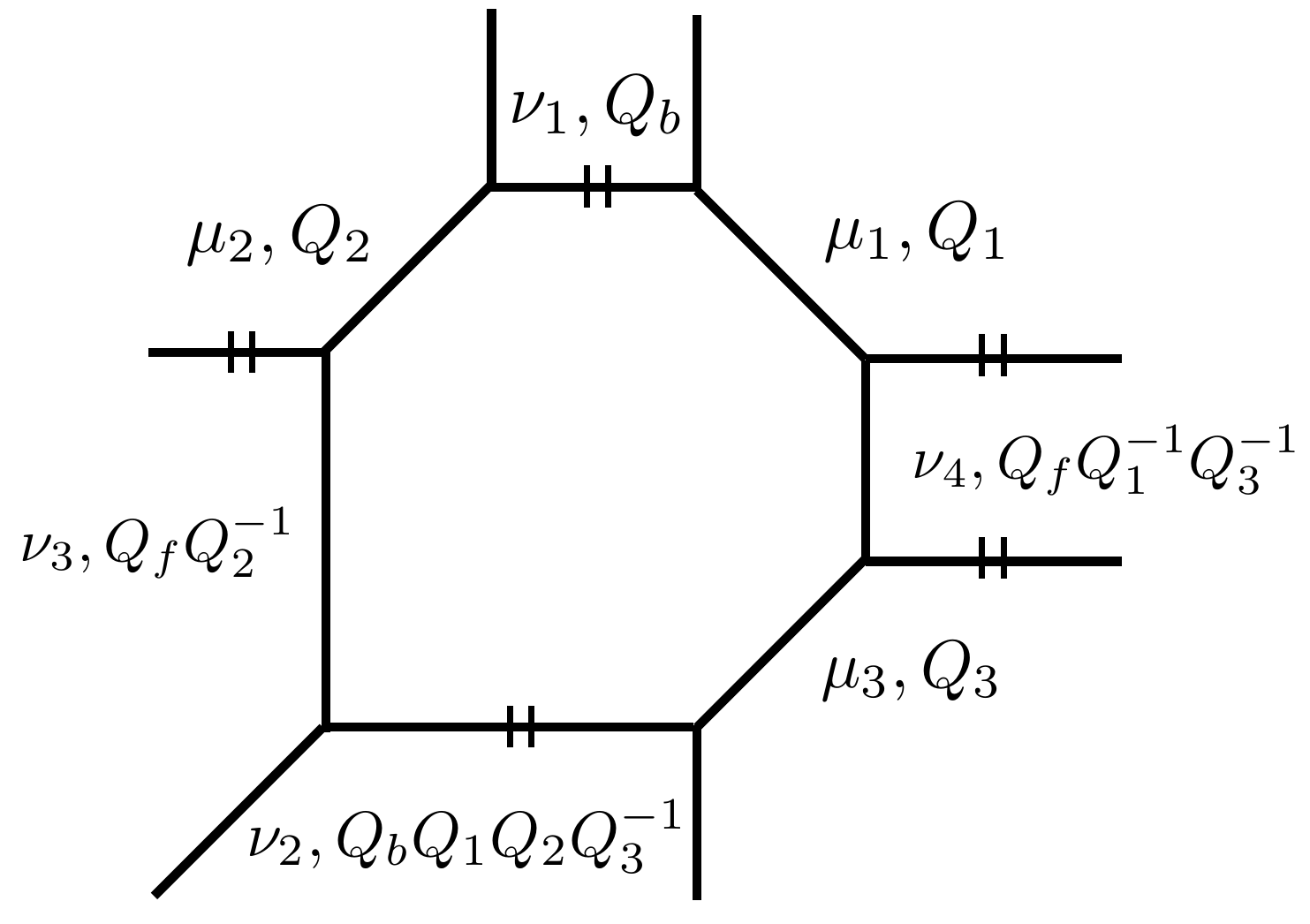}
	\caption{Toric diagram of three points blow up of local $\mathbb{P}^1\times\mathbb{P}^1$.}
	\label{fig:Nf=3}
	\end{center}
\end{figure}

\paragraph{$N_f=3$}
The toric diagram for the blow up of $\mathbb{P}^1\times \mathbb{P}^1$ at $N_f=3$ points is shown in Figure~\ref{fig:Nf=3}.
The corresponding topological string partition function is given by
\bea
	\hspace{-1.1cm}Z_{N_f=3} \!\!&\!\!=\!\!&\!\!(M(t,q)M(q,t))^{1/2}Z(t,q,Q) \,,  \\
	\hspace{-1.1cm}Z(t,q,Q) \!\!&\!\!=\!\!&\!\! \sum_{\vec\nu,\vec\mu}\!
	(-Q_b)^{|\nu_1|}(-Q_1)^{|\mu_1|}(-Q_2)^{|\mu_2|}
    (-Q_3)^{|\mu_3|}
    (-Q_fQ^{-1}_1Q_3^{-1})^{\!|\nu_4|}(-Q_fQ^{-1}_2)^{\!|\nu_3|}(-Q_bQ_1Q_2Q_3^{-1})^{\!|\nu_2|} \nn \\
    &&\times f_{\nu_1^t}(q,t)\tilde{f}_{\nu_4^t}(t,q) 
    C_{\emptyset\mu_1\nu_1^t}(q,t)C_{\nu_4\mu_1^t\emptyset}(t,q)C_{\mu_3\nu_4^t\emptyset}(t,q)C_{\mu_3^t\emptyset\nu_2}(q,t)
     C_{\mu_2^t\emptyset\nu_1}(t,q)C_{\mu_2\nu_3^t\emptyset}(q,t) C_{\emptyset\nu_3\nu_2^t}(t,q) \,.\nn
\eea
The final expression we obtain is
\bea\label{eq:part-U(2)-Nf=3}
	Z_{N_f=3} \!\!&\!\!=\!\!&\!\! Z_0 Z_{\rm inst} \,, \nn \\
	Z_0 \!\!&\!\!=\!\!&\!\! \prod_{i,j=1}^\infty
			\frac{\prod_{a=1}^3(1-e^{-i\lambda+m_a}q^{i-\frac{1}{2}}t^{j-\frac{1}{2}})
			(1-e^{-i\lambda-m_a}q^{i-\frac{1}{2}}t^{j-\frac{1}{2}})}
			{(1-q^it^{j-1})^{\frac{1}{2}}(1-q^{i-1}t^{j})^{\frac{1}{2}}
			(1-e^{-2i\lambda}q^it^{j-1})(1-e^{-2i\lambda}q^{i-1}t^{j})
			(1-e^{-i(m_1+m_3)}q^{i-1}t^j)} \,, \nn \\
	Z_{\rm inst} \!\!&\!\!=\!\!&\!\!\sum_{\nu_1,\nu_2}u^{|\nu_1|+|\nu_2|}\prod_{\alpha=1}^2\prod_{s\in \nu_\alpha}
    \frac{e^{\frac{i}{2}(E_{\alpha\emptyset}+i\gamma_1)}
    \!\left(\prod_{a=1}^22i\sin\frac{E_{\alpha\emptyset}+i\gamma_1-m_a}{2}\right)
    \!\!\left(2i\sin\frac{-E_{\alpha\emptyset}-i\gamma_1-m_3}{2}\right)}
    {\prod_{\beta=1}^2(2i)^2\sin\frac{E_{\alpha\beta}}{2}\sin\frac{E_{\alpha\beta}+2i\gamma_1}{2}}\,.
\eea
We again relate the K\"ahler parameters of the string theory with the gauge theory parameters such that
\be
	u = Q_bQ_1^{\frac{1}{2}}Q_2^{\frac{1}{2}}Q_3^{-\frac{1}{2}}Q_f^{-\frac{1}{4}}
	\,, \quad Q_f=e^{-2i\lambda} \,, \quad Q_{a=1,2,3} = e^{-i\lambda+im_a}
\ee
This partition function can be also identified with the Nekrasov partition function.
The factor $Z_{\rm inst}$ is precisely the 5d instanton partition function of the $U(2)$ gauge theory at Chern-Simons level $\kappa=+\frac{1}{2}$,
which is deduced from the phase factor $e^{\frac{i}{2}(E_{\alpha\emptyset}+i\gamma_1)}$
in the numerator, with two fundamantal and one
anti-fundamental hypermultiplets with masses $m_{a=1,2,3}$.
The factor $Z_0$ divided by the following factor
\be\label{U(1)-Nf=3}
	Z_{U(1)}^{=} \equiv \prod_{i,j=1}^\infty (1-e^{-i(m_1+m_3)}q^{i-1}t^j)^{-1} \,.
\ee
gives the perturbative contribution to the $U(2)$ gauge theory partition function
with three fundamental flavors. Therefore $Z_{N_f=3}/Z_{U(1)}^{=}$ coincides with the full Nekrasov partition function of $U(2)$ gauge theory with three flavors and Chern-Simons level $1/2$.

Note here that $Z_{U(1)}^{=}$ is the contribution from M2-branes on the two-cycle associated with $Q_fQ_1^{-1}Q_3^{-1}$. Since such M2-branes are not charged under the $SU(2)$ gauge symmetry, it is reasonable to take the ratio $Z_{N_f=3}/Z_{U(1)}^{=}$. However, there is still a contribution from decoupled M2-branes wrapping the two-cycle associated with $Q_b=ue^{-\frac{i}{2}(m_1+m_2-m_3)}$. In the same way as in the $N_f=2$ case, such contribution is evaluated by setting all the fugacities to be zero except for $Q_b$. The result is
\begin{eqnarray}
Z_{U(1)}^{||} \equiv \prod_{k,\ell=1}^\infty (1-ue^{-\frac{i}{2}(m_1+m_2-m_3)}q^kt^{\ell-1})^{-1}.
\end{eqnarray}
This contribution is still included in $Z_{N_f=3}/Z_{U(1)}^{=}$.

\begin{figure}[tb]
	\begin{center}
	\includegraphics[scale=0.35]{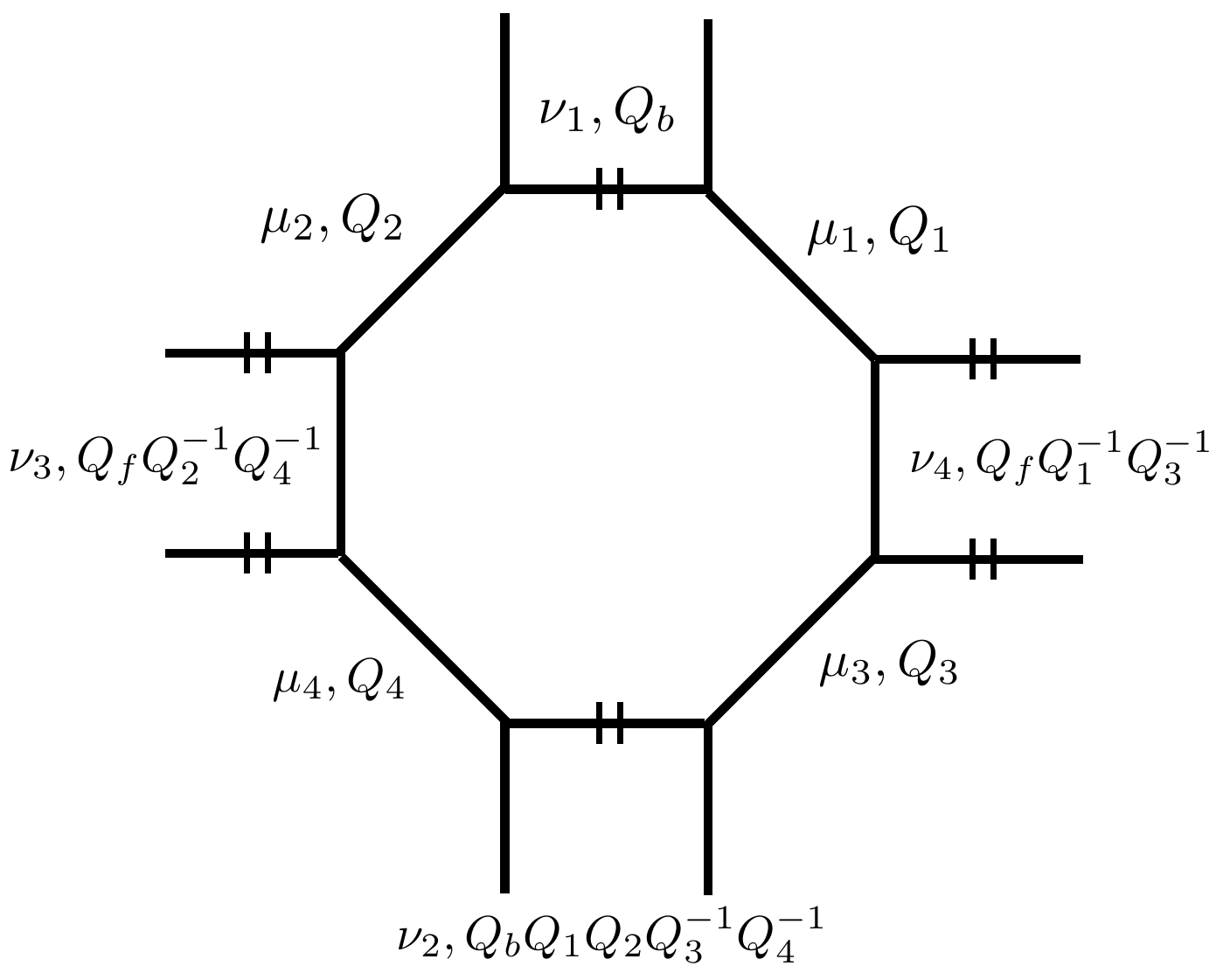}
	\caption{Toric diagram of four points blow up of local $\mathbb{P}^1\times\mathbb{P}^1$.}
\label{fig:Nf=4}
	\end{center}
\end{figure}

\paragraph{$N_f=4$}
\label{sec:Nf4}
Let us study the final example of this subsection. The toric diagram for the blow up of $\mathbb{P}^1\times \mathbb{P}^1$ at $N_f=4$ points is shown in Figure~\ref{fig:Nf=4}.
The corresponding topological string partition function is given by
\bea
	Z_{N_f=4} \!\!&\!\!=\!\!&\!\!(M(t,q)M(q,t))^{1/2}Z(t,q,Q) \,, \\
	\hspace{-0.8cm}Z(t,q,Q) \!\!&\!\!=\!\!&\!\!  \sum_{\vec\nu,\vec\mu}
	(-Q_b)^{|\nu_1|}(-Q_1)^{|\mu_1|}(-Q_2)^{|\mu_2|}(-Q_3)^{|\mu_3|}(-Q_4)^{|\mu_4|}
    (-Q_fQ^{-1}_1Q_3^{-1})^{|\nu_4|}(-Q_fQ^{-1}_2Q_4^{-1})^{|\nu_3|} \nn \\
    &&\times (-Q_bQ_1Q_2Q_3^{-1}Q_4^{-1})^{|\nu_2|}
    f_{\nu_1^t}(q,t)f_{\nu_2^t}(t,q)\tilde{f}_{\nu_3^t}(q,t)\tilde{f}_{\nu_4^t}(t,q) \nn \\
    &&\times C_{\emptyset\mu_1\nu_1^t}(q,t)C_{\nu_4\mu_1^t\emptyset}(t,q)
    C_{\mu_3\nu_4^t\emptyset}(t,q)C_{\mu_3^t\emptyset\nu_2}(q,t)
    C_{\mu_2^t\emptyset\nu_1}(t,q)C_{\mu_2\nu_3^t\emptyset}(q,t)
    C_{\emptyset\mu_4\nu_2^t}(t,q) C_{\nu_3\mu_4^t\emptyset}(q,t)\,. \nn
\eea
An explicit computation shows that the the partition function reduces to
\bea\label{eq:part-U(2)-Nf=4}
	Z_{N_f=4} \!\!&\!\!=\!\!&\!\! Z_0 Z_{\rm inst} \,,  \\
	Z_0 \!\!&\!\!=\!\!&\!\! \prod_{i,j=1}^\infty
			\frac{\prod_{a=1}^4(1-e^{-i\lambda+im_a}q^{i-\frac{1}{2}}t^{j-\frac{1}{2}})
			(1-e^{-i\lambda-im_a}q^{i-\frac{1}{2}}t^{j-\frac{1}{2}})}
			{(1-q^it^{j-1})^{\frac{1}{2}}(1-q^{i-1}t^{j})^{\frac{1}{2}}
			(1-e^{-2i\lambda}q^it^{j-1})(1-e^{-2i\lambda}q^{i-1}t^{j})} \,, \nn \\
		&& \times \prod_{i,j=1}^\infty(1-e^{-i(m_1+m_3)}q^{i-1}t^j)^{-1}
				(1-e^{-i(m_2+m_4)}q^{i}t^{j-1})^{-1} \,,\nn \\
	Z_{\rm inst} \!\!&\!\!=\!\!&\!\!
	 \sum_{\nu_1,\nu_2}u^{|\nu_1|+|\nu_2|}\prod_{\alpha=1}^2\prod_{s\in \nu_\alpha}
    \frac{
    \!\left(\prod_{a=1}^22i\sin\frac{E_{\alpha\emptyset}+i\gamma_1-m_a}{2}\right)
    \!\!\left(\prod_{a=3}^42i\sin\frac{-E_{\alpha\emptyset}-i\gamma_1-m_a}{2}\right)}
    {\prod_{\beta=1}^2(2i)^2\sin\frac{E_{\alpha\beta}}{2}\sin\frac{E_{\alpha\beta}+2i\gamma_1}{2}} \,, \nn
\eea
in terms of the gauge theory parameters 
\be
	u = Q_bQ_1^{\frac{1}{2}}Q_2^{\frac{1}{2}}Q_3^{-\frac{1}{2}}Q_4^{-\frac{1}{2}}
	 \,, \quad Q_f = e^{-2i\lambda} \,, \quad Q_{a=1,2,3,4} = e^{-i\lambda+im_a} \,.
\ee
When we define
\be\label{U(1)-Nf=4}
Z_{U(1)}^{=} \equiv \prod_{i,j=1}^\infty(1-e^{-i(m_1+m_3)}q^{i-1}t^j)^{-1}
				(1-e^{-i(m_2+m_4)}q^{i}t^{j-1})^{-1},
\ee
$Z_{N_f=4}/Z_{U(1)}^{=}$ turns out to coincide with the 
Nekrasov partition function of the $U(2)$
gauge theory with two fundamental and two anti-fundamental hypermultiplets without a classical
Chern-Simons term.

Note here that $Z_{U(1)}^{=}$ is the contribution from M2-branes on the two-cycles associated with $Q_fQ_1^{-1}Q_3^{-1}$ and $Q_fQ_2^{-1}Q_4^{-1}$. Since such M2-branes have no electric charge of $SU(2)$ gauge symmetry, it is reasonable to consider the ratio $Z_{N_f=4}/Z_{U(1)}^{=}$. However, there is still a contribution from decoupled M2-branes on the two-cycles associated with $Q_b=ue^{-\frac{i}{2}(m_1+m_2-m_3-m_4)}$ and $Q_bQ_1Q_2Q_3^{-1}Q_4^{-1}=ue^{+\frac{i}{2}(m_1+m_2-m_3-m_4)}$. Such a contribution is evaluated by setting all the other fugacities to be zero, which leads to
\begin{eqnarray}
Z_{U(1)}^{||} \equiv \prod_{k,\ell=1}^\infty \left(1-ue^{-\frac{i}{2}(m_1+m_2-m_3-m_4)}q^{k}t^{\ell-1}\right)^{-1}\left(1-ue^{+\frac{i}{2}(m_1+m_2-m_3-m_4)}q^{k-1}t^{\ell}\right)^{-1}.
\end{eqnarray}
This factor is still included in $Z_{N_f=4}/Z_{U(1)}^{=}$.

\subsubsection{$(p, q)$ webs of $T_{N=2, 3, 4}$ theories}
\label{subsubsec:TN}

For 5d $Sp(1)$ gauge theories with $N_f\geq 5$ flavors, the relevant Calabi-Yau threefold is not toric. However, as we will see later, they are closely related to the 5d $T_N$-theory which is engineered by a toric Calabi-Yau threefold. Therefore we now evaluate the refined topological string partition function associated with the 5d $T_N$ theory, especially for $N=2,3,4$.
The results then show their relations to the Nekrasov partition functions of certain linear quiver theories.

\paragraph{Warm up: $T_2$ theory}

\begin{figure}[tb]
	\begin{center}
	\includegraphics[scale=0.3]{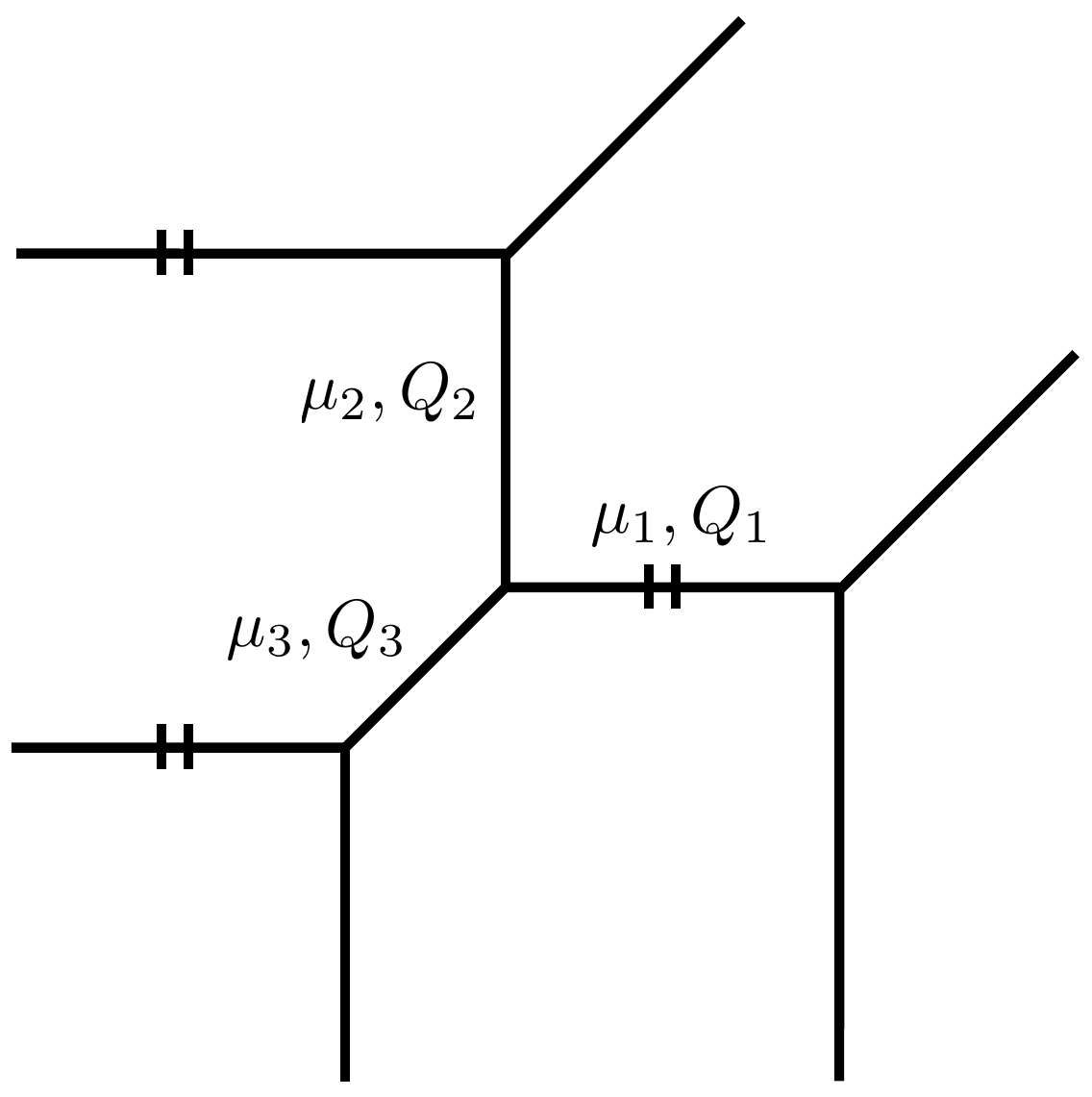}
	\caption{Toric diagram of $T_2$ theory.}
\label{fig:T2}
	\end{center}
\end{figure}

The toric Calabi-Yau threefold engineering $T_2$ theory is the resolved $\mathbb{C}^3/(\mathbb{Z}_2\times \mathbb{Z}_2)$, whose toric diagram is shown
in Figure~\ref{fig:T2}. 
This multi-junction system has $SU(2)^3$ flavor symmetry realized by two 
external lines in each of the left, bottom and upper-right directions.
The corresponding 5d theory is given by eight free chiral superfields transforming under the flavor symmetry.
The Higgs branch of the theory is complex eight dimensional.
The K\"ahler parameters $Q_1,Q_2,Q_3$ of the geometry are associated to the mass parameter of those chirals.
The Coulomb branch of this theory is absent since we have no compact four-cycle. 

Given the web diagram, the topological string partition function is simply given by \cite{Karp:2005vq,Sulkowski:2006jp, Kozcaz:2010af}
\bea
	\tilde{Z}_{T_2} \!\!&\!\!=\!\!&\!\! \sum_{\mu_1,\mu_2,\mu_3} (-Q_1)^{|\mu_1|}(-Q_2)^{|\mu_2|}(-Q_3)^{|\mu_3|}
	C_{\emptyset\emptyset\mu_1^t}(q,t)C_{\emptyset\mu_2\emptyset}(q,t)
	C_{\mu_3\emptyset\emptyset}(q,t)C_{\mu_3^t\mu_2^t\mu_1}(t,q)  \nonumber\\
	\!\!&\!\!=\!\!&\!\!\prod_{i,j=1}^\infty\frac{(1-e^{-im_2}q^{i-\frac{1}{2}}t^{j-\frac{1}{2}})(1-e^{im_3}q^{i-\frac{1}{2}}t^{j-\frac{1}{2}})}{1-e^{-im_2+im_3}q^{i}t^{j-1}} 
	\sum_{\mu_1}u^{|\mu_1|}
	\prod_{s\in \mu_1}\frac{\sin\frac{E_{1\emptyset}-m_2+i\gamma_1}{2}\sin\frac{E_{1\emptyset}-m_3+i\gamma_1}{2}}
	{\sin\frac{E_{11}}{2}\sin\frac{E_{11}+2i\gamma_1}{2}} \,,\nn
\\
\label{eq:T2}
\eea
where 
\be
	\lambda_1=\lambda_\emptyset = 0 \,, \quad Q_1Q_2^{\frac{1}{2}}Q_3^{\frac{1}{2}} = -u
	\,, \quad  Q_2=e^{-im_2} \,, \quad Q_3 = e^{im_3} \,.
\ee
One can easily check that the partition function $Z_{T_2}$ is the Nekrasov partition function
of the $U(1)$ gauge theory with 2 fundamental flavors whose mass parameters are $m_1,m_2$,
up to the extra free factor in the denominator of the first infinite product.
We can also use the identity \cite{Kozcaz:2010af}
\bea\label{eq:T2-identity}
	&&\sum_\mu (-Q_1)^{|\mu|}q^{\frac{||\mu_t||^2}{2}}t^{\frac{||\mu||^2}{2}}
	\prod_{s\in\mu}\frac{(1-Q_2q^{a_\emptyset(s)+\frac{1}{2}}t^{l_\mu(s)+\frac{1}{2}})
	(1-Q_3q^{-a_\emptyset(s)-\frac{1}{2}}t^{-l_\mu(s)-\frac{1}{2}})}
	{(1-q^{a_\mu(s)+1}t^{l_\mu(s)})(1-q^{a_\mu(s)}t^{l_\mu(s)+1})} \nn \\
	&=& \prod_{i,j=1}^\infty\frac{(1-Q_1q^{i-\frac{1}{2}}t^{j-\frac{1}{2}})(1-Q_1Q_2Q_3q^{i-\frac{1}{2}}t^{j-\frac{1}{2}})}
	{(1-Q_1Q_2q^{i-1}t^{j})(1-Q_1Q_3q^{i}t^{j-1})}
\eea
to simplify the partition function as
\be
	\tilde{Z}_{T_2} = \prod_{i,j=1}^\infty\frac{(1-Q_1q^{i-\frac{1}{2}}t^{j-\frac{1}{2}})
	(1-Q_2q^{i-\frac{1}{2}}t^{j-\frac{1}{2}})
	(1-Q_3q^{i-\frac{1}{2}}t^{j-\frac{1}{2}})
	(1-Q_1Q_2Q_3q^{i-\frac{1}{2}}t^{j-\frac{1}{2}})}
	{(1-Q_1Q_2q^{i-1}t^{j})(1-Q_2Q_3q^{i}t^{j-1})(1-Q_1Q_3q^{i}t^{j-1})} \,.
\ee
The numerator of this formulation shows the contribution of the 4 free hypermultiplets of the
$T_2$ theory as expected.

Let us define
\begin{align}
Z_{U(1)}^{=} \equiv \prod_{i,j=1}^\infty (1-Q_2Q_3q^it^{j-1})^{-1}&, \qquad Z_{U(1)}^{||} \nn \equiv \prod_{i,j=1}^\infty (1-Q_3Q_1q^it^{j-1})^{-1},
\\
Z_{U(1)}^{/\!/} \equiv \prod_{i,j=1}^\infty& (1-Q_1Q_2q^{i-1}t^{j})^{-1}.
\end{align}
Then $\tilde{Z}_{T_2}/(Z_{U(1)}^{=}Z_{U(1)}^{||}Z_{U(1)}^{/\!/})$ coincides with the partition function of the $T_2$-theory.
According to the argument in section \ref{sec:topst}, the factors $Z_{U(1)}^{=},Z_{U(1)}^{||}$ and $Z_{U(1)}^{/\!/}$ are identified with the contribution from decoupled M2-branes on two-cycles associated with $Q_2Q_3$, $Q_3Q_1$ and $Q_1Q_2$, respectively. All the two-cycles can be continuously moved to infinity, and therefore it is reasonable to eliminate them. Now let us consider $\tilde{Z}_{T_2}/Z_{U(1)}^{=}$. From \eqref{eq:T2}, it follows that $\tilde{Z}_{T_2}/Z_{U(1)}^{=}$ coincides with the Nekrasov partition function of the $U(1)$ gauge theory with two fundamental matters. This strongly suggests that the $T_N$-theory partition function is related to the partition function of some simpler gauge theory with Lagrangian description.

\begin{figure}[tb]
	\begin{center}
	\includegraphics[scale=0.3]{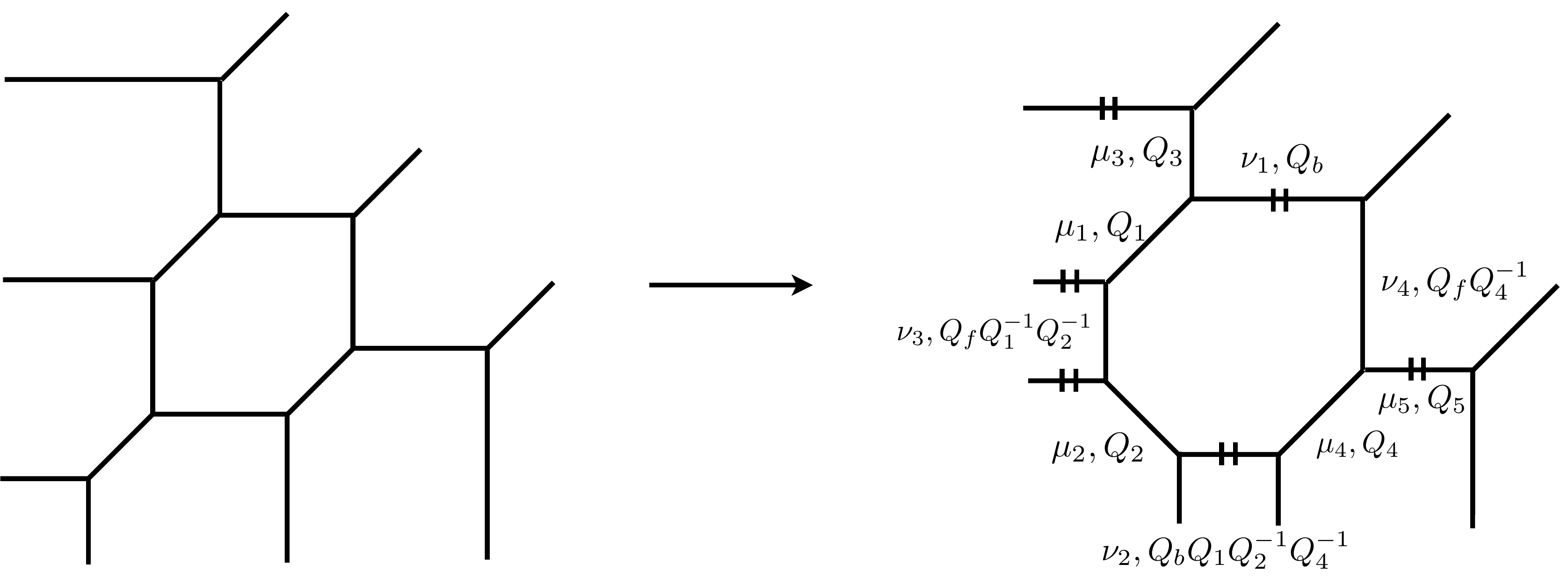}
	\caption{Toric diagram of $T_3$ theory.}
\label{fig:T3}
	\end{center}
\end{figure}

\paragraph{$T_3$ theory}
We now consider the $T_3$ theory engineered by $\mathbb{C}^3/(\mathbb{Z}_3\times \mathbb{Z}_3)$. The relevant toric web diagram is shown in Figure~\ref{fig:T3}.
The theory has a global symmetry $SU(3)^3$, which is visible from the diagram, and  
in fact the symmetry is enhanced to $E_6$ \cite{Gaiotto:2009we}.
The Coulomb branch of this theory is one dimensional corresponding to the size of the hexagon
in the center of the diagram.
The theory also has a Higgs branch of complex dimension 22, which is the same dimension as
the dimension of the one instanton moduli space of $E_6$ gauge theory.
These properties of the $T_3$ theory allow us to relate it with $SU(2)$ gauge theory with $N_f\!=\!5$ fundamental hypermultiplets in 5d.\footnote{As mentioned already, for a 5d $Sp(1)$ gauge theory with $N_f\geq 5$, the relevant Calabi-Yau threefold is {\it not} toric with a generic complex structure. However, in the case of $N_f=5$, there is a special choice of complex structure with which the Calabi-Yau threefold becomes toric.}
We will compare the partition functions of
two theories in the next section, which provides a strong evidence for the relation.

The diagram leads to the $T_3$ partition function
\bea
	\tilde{Z}_{T_3} \!\!&\!\!=\!\!&\!\! (M(t,q)M(q,t))^{1/2}Z(t,q,Q) \,, \nn \\
	\hspace{-0.5cm}Z(t,q,Q) \!\!&\!\!=\!\!&\!\! \sum_{\vec\nu,\vec\mu} (-Q_1)^{|\mu_1|}(-Q_2)^{|\mu_2|}(-Q_3)^{|\mu_3|}
  (-Q_4)^{|\mu_4|}(-Q_5)^{|\mu_5|}(-Q_b)^{|\nu_1|}(-Q_bQ_1Q_2^{-1}Q_4^{-1})^{|\nu_2|} \nn \\
	&& \times	(-Q_fQ_1^{-1}Q_2^{-1})^{|\nu_3|}(-Q_fQ_4^{-1})^{|\nu_4|} 
	 f_{\nu_2^t}(t,q)\tilde{f_{\nu_3^t}}(q,t)
		C_{\emptyset\nu_4\nu_1^t}(q,t)C_{\mu_4\nu_4^t\mu_5^t}(t,q)
		C_{\emptyset\emptyset\mu_5}(q,t)C_{\mu_4^t\emptyset\nu_2}(q,t) \nn \\
	&& \times C_{\mu_1^t\mu_3^t\nu_1}(t,q)C_{\emptyset\mu_3\emptyset}(q,t)C_{\mu_1\nu_3^t\emptyset}
		(q,t)C_{\nu_3\mu_2^t\emptyset}(q,t)C_{\emptyset\mu_2\nu_2^t}(t,q) \,.
\eea
A short calculation gives the simple result
\bea
\label{T3-partition}
	\hspace{-1cm}\tilde{Z}_{T_3} \!\!&\!\!=\!\!&\!\! Z_0\cdot Z_{\rm inst} \,, \nn \\
	\hspace{-1cm}Z_0 \!\!&\!\!=\!\!&\!\!\! \prod_{i,j=1}^\infty
			\!\!\frac{(1\!-\!e^{i\lambda-im_3}q^{i-\frac{1}{2}}t^{j-\frac{1}{2}})\!
			(1\!-\!e^{-i\lambda-im_3}q^{i-\frac{1}{2}}t^{j-\frac{1}{2}})\!\!
			\prod_{a=1,2,4}(1\!-\!e^{-i\lambda+im_a}q^{i-\frac{1}{2}}t^{j-\frac{1}{2}})\!
			(1\!-\!e^{-i\lambda-im_a}q^{i-\frac{1}{2}}t^{j-\frac{1}{2}})}
			{(1-q^it^{j-1})^{\frac{1}{2}}(1-q^{i-1}t^{j})^{\frac{1}{2}}
			(1-e^{-2i\lambda}q^it^{j-1})(1-e^{-2i\lambda}q^{i-1}t^{j})}  \nn \\
		\hspace{-1cm}&& \times \prod_{i,j=1}^\infty(1-e^{-im_1+im_2}q^{i}t^{j-1})^{-1}
				(1-e^{im_1-im_3}q^{i}t^{j-1})^{-1}(1-e^{im_2-im_3}q^{i}t^{j-1})^{-1}  \,, \nn \\
	\hspace{-1cm}Z_{\rm inst} \!\!&\!\!=\!\!&\!\!\sum_{\nu_1,\nu_2,\mu_5}u_2^{|\nu_1|+|\nu_2|}u_1^{|\mu_5|}
		\left[\prod_{\alpha=1}^2\prod_{s\in\nu_\alpha}
		\frac{\left(\prod_{a=1}^32i\sin\frac{E_{\alpha\emptyset}-m_a+i\gamma_1}{2}\right)
		(2i\sin\frac{E_{\alpha 5}-m_4+i\gamma_1}{2})}
		{\prod_{\beta=1}^2(2i)^2\sin\frac{E_{\alpha\beta}}{2}\sin\frac{E_{\alpha\beta}+2i\gamma_1}{2}} \right. \nn \\
		&& \times \left.\prod_{s\in \mu_5}
		\frac{\prod_{\alpha=1}^22i\sin\frac{E_{5\alpha}+m_4+i\gamma_1}{2}}
		{(2i)^2\sin\frac{E_{55}}{2}\sin\frac{E_{55}+2i\gamma_1}{2}}\right] \,,
\eea
where we used the parametrization
\bea
	&& Q_bQ_1^{\frac{1}{2}}Q_2^{-\frac{1}{2}}Q_3^{\frac{1}{2}}Q_4^{-\frac{1}{2}} = -u_2, \ \
	 Q_f = e^{-2i\lambda} ,\ \ Q_5 = -e^{i\lambda}u_1 ,\nn \\
	 &&  Q_1 = e^{-i\lambda+im_1} , \ \ Q_2 = e^{-i\lambda-im_2} , \ \ 
	Q_3 = e^{i\lambda-im_3} ,\ \ Q_4 = e^{-i\lambda-im_4} \,,
\eea
and $\lambda_\emptyset = \lambda_5 = 0$.
Let us define
\be
 Z_{U(1)}^{=} \equiv \prod_{i,j=1}^\infty(1-e^{-im_1+im_2}q^{i}t^{j-1})^{-1}
				(1-e^{im_1-im_3}q^{i}t^{j-1})^{-1}(1-e^{im_2-im_3}q^{i}t^{j-1})^{-1} \,.
\ee
Then it is easy to see that 
$\tilde{Z}_{T_3}/Z_{U(1)}^{=}$ is precisely the $U(2)\times U(1)$
gauge theory partition function with three hypermultiplets in the fundamental representation of $U(2)$
and one bi-fundamental hypermultiplet. 
The instanton part $Z_{\rm inst}$ takes the form of a instanton summation for two gauge groups
with instanton fugacities $u_1$ and $u_2$.

Note here that $Z_{U(1)}^{=}$ is the contribution from M2-branes on the two-cycles associated with $Q_fQ_1^{-1}Q_2^{-1},\, Q_1Q_3$ and $Q_fQ_2^{-1}Q_3$. Since such M2-branes are decoupled in the 5d gauge theory, it is reasonable to take the ratio $\tilde{Z}_{T_3}/Z_{U(1)}^{=}$. The three two-cycles are associated with the pairs of {\it horizontal} parallel external lines in figure \ref{fig:T3}. There are also two-cycles associated with pairs of {\it bottom} or {\it upper-right} parallel external lines. Since they have vanishing intersections with the compact four-cycle, they also support decoupled M2-branes. We denote by $Z_{U(1)}^{||}$ and $Z_{U(1)}^{/\!/}$ the contributions from such M2-branes associated with the bottom and upper-right external lines, respectively. Their explicit expressions are obtained in section \ref{sec:general-TN}.

\paragraph{$T_4$ theory}

\begin{figure}[tb]
	\begin{center}
	\includegraphics[width=100mm]{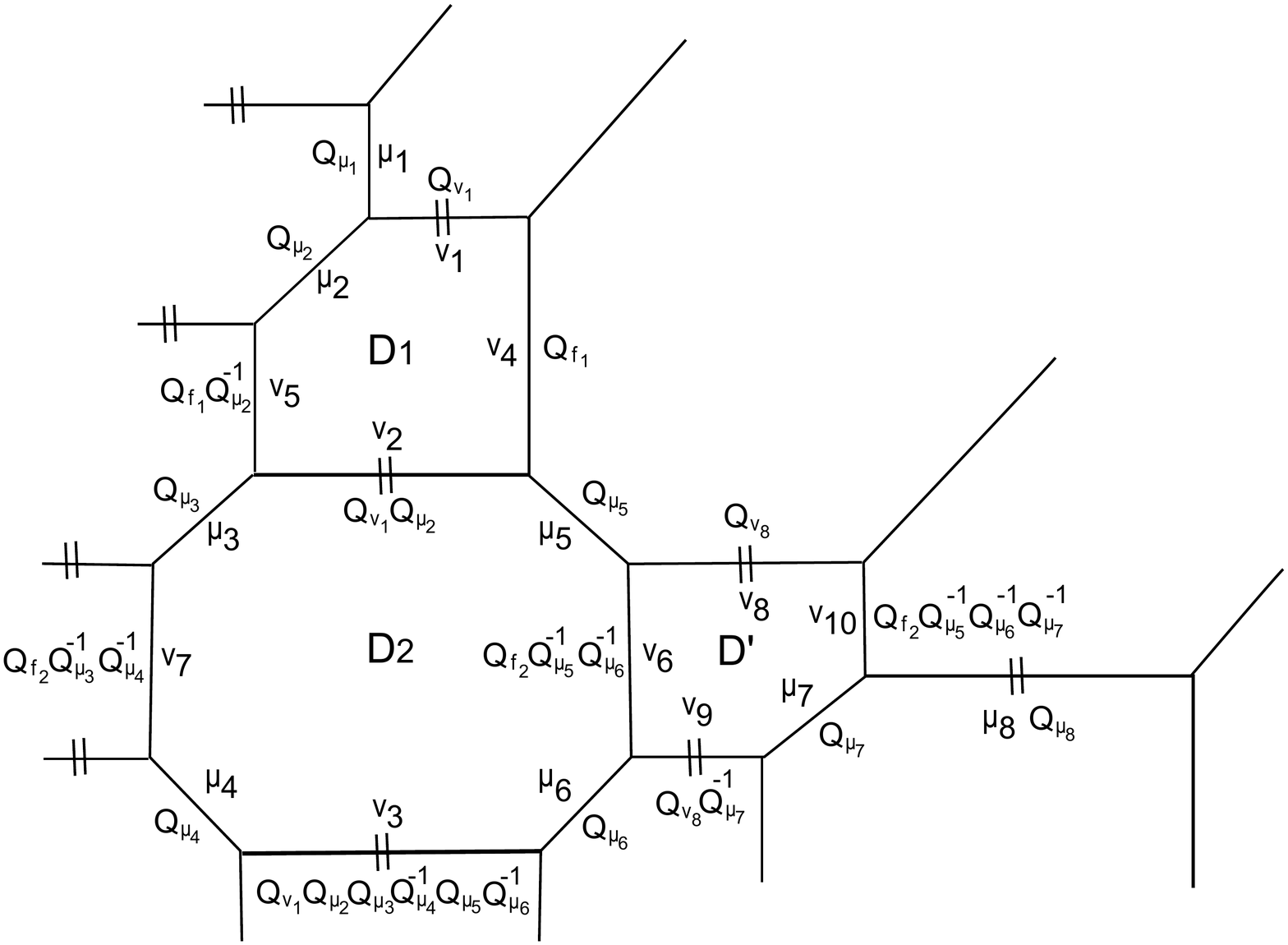}
	\caption{A toric diagram of $T_4$ theory of a particular triangulation. $D_1, D_2$ and $D'$ stand for the compact divisors corresponding to the three closed loops in the web diagram.}
\label{fig:T4}
	\end{center}
\end{figure}

The computation of the partition function of the $T_4$ theory can be done in the same manner. 
The relevant Calabi-Yau threefold is $\mathbb{C}^3/(\mathbb{Z}_4\times \mathbb{Z}_4)$, whose toric web-diagram is depicted in Figure \ref{fig:T4}. 

The topological 
string partition function is evaluated as
\be
\tilde{Z}_{T_4} = (M(t, q)M(q, t))^{\frac{3}{2}}Z(t, q, Q)\,,
\ee
where
\bea
Z(t, q, Q) &=& \sum_{\nu, \mu}(-Q_1)^{|\mu_1|} (-Q_2)^{|\mu_2|} (-Q_3)^{|\mu_3|} (-Q_4)^{|\mu_4|} (-Q_5)^{|\mu_5|} (-Q_6)^{|\mu_6|} (-Q_7)^{|\mu_7|} (-Q_8)^{|\mu_8|}  \nn \\
&&(-Q_{b_1})^{|\nu_1|}(-Q_{b_1}Q_2)^{|\nu_2|}(-Q_{b_1}Q_2 Q_3 Q_4^{-1} Q_5 Q_6^{-1})^{|\nu_3|}(-Q_{f_1})^{|\nu_4|} (-Q_{f_1}Q_2^{-1})^{|\nu_5|} \nn \\
&&(-Q_{f_2}Q_5^{-1}Q_6^{-1})^{|\nu_6|}(-Q_{f_2}Q_{\mu_3}^{-1} Q_{\mu_4}^{-1})^{|\nu_7|}(-Q_{b_2})^{|\nu_8|}(-Q_{b_2}Q_7^{-1})^{|\nu_9|}(-Q_{f_2}Q_5^{-1}Q_6^{-1}Q_7^{-1})^{|\nu_{10}|} \nonumber\\
&& f_{\nu_2^t}(q,t) f_{\nu_3^t}(t, q) \tilde{f}_{\nu_4^t}(q, t) \tilde{f}_{\nu_6^t}(t, q) \tilde{f}_{\nu_7^t}(q, t) f_{\nu_8^t}(t, q) \nonumber\\
&&C_{\emptyset\mu_1^t\emptyset}(q, t) C_{\mu_2\mu_1\nu_1^t}(t, q) C_{\mu_2^t\nu_5\emptyset}(q, t) C_{\mu_3^t\nu_5^t\nu_2}(t, q) C_{\mu_3\nu_7^t\emptyset}(q, t) C_{\nu_7\mu_4^t\emptyset}(q, t) C_{\emptyset\mu_4\nu_3^t}(t, q)\nonumber\\
&&C_{\emptyset\nu_4^t\nu_1}(q, t) C_{\nu_4\mu_5\nu_2^t}(q, t) C_{\nu_6\mu_5^t\nu_8^t}(t, q) C_{\mu_6\nu_6^t\nu_9}(t, q) C_{\mu_6^t\emptyset\nu_3}(q, t)\nonumber\\
&&C_{\emptyset\nu_{10}^t\nu_8}(q, t) C_{\mu_7^t\nu_{10}\mu_8^t}(t, q) C_{\mu_7\emptyset\nu_9^t}(q, t) C_{\emptyset\emptyset\mu_8}(q, t)\,. \label{T4}
\eea
There are three compact divisors $D_1, D_2$ and $D'$ corresponding to the three loops in Figure \ref{fig:T4}. The expansion parameters of the K\"ahler form $J$ by the three compact divisors are the Coulomb branch parameters and we take
\be
J \supset a_1D_1 + a_2D_2 + a'D'\,. \label{coulomb-T4}
\ee
The intersection between the two-cycles in Figure \ref{fig:T4} and \eqref{coulomb-T4} gives the Coulomb branch parameter dependence of the particles which arise from M2-branes wrapping on the two-cycles. With this information, we choose the parameters in the low energy effective theory in the following way. 
\bea
&&Q_1 = e^{ia_1-im_1}, Q_2 = e^{-ia_1 + im_2}, Q_3 = e^{i(a_1-a_2) + im_3}, Q_4= e^{-ia_2 - im_4}\,,\label{paraT4-1}\\
&& Q_5 = e^{i(a_1-a_2+a') + i\tilde{m}_1}, Q_6 = e^{i(-a_2+a') - i\tilde{m}_1}, Q_7=e^{-ia' - i\tilde{m}_2}, Q_8 = -u_3e^{ia'},   \label{paraT4-2}\\
&&Q_{f_1} = e^{i(-2a_1+a_2)}, Q_{f_2} = e^{i(a_1-2a_2)}\,, \label{paraT4-3}\\
&&Q_{\nu_1}Q_1^{\frac{1}{2}}Q_2^{\frac{1}{2}}Q_3^{\frac{1}{2}}Q_4^{-\frac{1}{2}}Q_5^{\frac{1}{2}}Q_6^{-\frac{1}{2}} = -u_1, \quad Q_{\nu_8}Q_5Q_6^{\frac{1}{2}}Q_7^{-\frac{1}{2}}Q_{f_1}^{\frac{1}{2}} = u_2 \,.\label{paraT4-4}
\eea

Under the parameterization \eqref{paraT4-1}--\eqref{paraT4-4}, the straightforward calculation shows that Eq.~\eqref{T4} yields
\be
\tilde{Z}_{T_4} = Z_{U(1)}^{=} \cdot Z_0 \cdot Z_{\text{inst}} \,. \label{NekrasovT4}
\ee
where 
\bea
Z_0=&\prod_{i,j=1}^\infty& \Big\{ \frac{Z_{0}^{\text{numerator}}}{(1-q^it^{j-1})^{3/2}(1-q^{i-1}t^{j})^{3/2}[\prod_{\alpha < \beta}^3(1-e^{-i(\lambda_{\alpha}-\lambda_{\beta})}q^it^{j-1})(1-e^{-i(\lambda_{\alpha}-\lambda_{\beta})}q^{i-1}t^{j})]} \nonumber \\
&&\times \frac{1}{(1-e^{-2i\lambda'}q^it^{j-1})(1-e^{-2i\lambda'}q^{i-1}t^{j})}\Big\} \,. 
\eea
Here, $Z_{0}^{\text{numerator}}$ is 
\bea
Z_{0}^{\text{numerator}}\!\!&\!\!=\!\!&\!\! \left[\prod_{\alpha=1}^3(1-e^{i\lambda_{\alpha}-im_1}q^{i-\frac{1}{2}}t^{j-\frac{1}{2})}\right](1-e^{-i\lambda_1 + im_2}q^{i-\frac{1}{2}}t^{j-\frac{1}{2}})(1-e^{i\lambda_2 - im_2}q^{i-\frac{1}{2}}t^{j-\frac{1}{2}})\nn\\
&&(1-e^{i\lambda_3 - im_2}q^{i-\frac{1}{2}}t^{j-\frac{1}{2}})\Big[\prod_{a=3}^4(1-e^{-i\lambda_1 + im_a}q^{i-\frac{1}{2}}t^{j-\frac{1}{2}})(1-e^{-i\lambda_2 + im_a}q^{i-\frac{1}{2}}t^{j-\frac{1}{2}})\nn\\
&&(1-e^{i\lambda_3 - im_a}q^{i-\frac{1}{2}}t^{j-\frac{1}{2}})\Big]\Big[\prod_{\alpha'=1}^2(1-e^{-i(\lambda_1 - \lambda^{\prime}_{\alpha'}) + i\tilde{m}_1}q^{i-\frac{1}{2}}t^{j-\frac{1}{2}})(1-e^{-i(\lambda_1 - \lambda^{\prime}_{\alpha'}) + i\tilde{m}_2}q^{i-\frac{1}{2}}t^{j-\frac{1}{2}})\nn\\
&&(1-e^{i(\lambda_3 - \lambda^{\prime}_{\alpha'}) - i\tilde{m}_1}q^{i-\frac{1}{2}}t^{j-\frac{1}{2}})\Big](1-e^{-i\lambda' + i\tilde{m}_2}q^{i-\frac{1}{2}}t^{j-\frac{1}{2}})(1-e^{-i\lambda' - i\tilde{m}_2}q^{i-\frac{1}{2}}t^{j-\frac{1}{2}}) \,.
\label{T4perturbative}
\eea
This corresponds to the perturbative contribution. The numerator of \eqref{T4perturbative} can be recast into a simpler form 
\be
\left[\prod_{\alpha=1}^3\prod_{a=1}^4\!(1\! -\! e^{-i\lambda_{\alpha}+ im_a}q^{i-\frac{1}{2}}t^{j-\frac{1}{2}})\!\right]\!\!\left[\prod_{\alpha=1}^3\prod_{\alpha'=1}^2\!(1 \!-\! e^{-i(\lambda_{\alpha} - i\lambda'_{\alpha'}) + i\tilde{m}_1}q^{i-\frac{1}{2}}t^{j-\frac{1}{2}})\!\right]\!\!\left[\prod_{\alpha'=1}^2\!(1\!-\!e^{-i\lambda'_{\alpha'} +i\tilde{m}_2}q^{i-\frac{1}{2}}t^{j-\frac{1}{2}})\!\right],
\ee 
if we ignore divergent terms.
The 
factor $Z_{U(1)}^{=}$ 
is given by
\bea
Z_{U(1)}^{=} &=& \prod_{i,j=1}^\infty(1-e^{-i(m_1-m_2)}q^it^{j-1})^{-1}(1-e^{-i(m_1-m_3)}q^it^{j-1})^{-1}(1-e^{-i(m_1-m_4)}q^it^{j-1})^{-1}\nonumber\\
&&(1-e^{-i(m_2-m_3)}q^it^{j-1})^{-1}(1-e^{-i(m_2-m_4)}q^it^{j-1})^{-1}(1-e^{-i(m_3-m_4)}q^it^{j-1})^{-1},
\eea
and finally $Z_{\text{inst}}$ is the instanton partition function 
\bea
Z_{\text{inst}} &=& \sum_{\nu_1, \nu_2, \nu_3, \tilde{\nu}_1, \tilde{\nu}_2, \mu_8}u_1^{|\nu_1|+|\nu_2| + |\nu_3|}u_2^{|\tilde{\nu}_1| + |\tilde{\nu}_2|} u_3^{|\mu_8|}\nonumber\\
&&\Big[\prod_{\alpha=1}^3\prod_{s\in\nu_\alpha}\frac{\left(\prod_{a=1}^42i\sin\frac{E_{\alpha\emptyset}-m_a+i\gamma_1}{2}\right)\left(\prod_{\alpha'=1}^22i\sin\frac{E_{\alpha \alpha'}- \tilde{m}_1+i\gamma_1}{2}\right)}{\prod_{\beta=1}^3(2i)^2\sin\frac{E_{\alpha\beta}}{2}\sin\frac{E_{\alpha\beta}+2i\gamma_1}{2}}\nonumber \\
&&\prod_{\alpha'=1}^2\prod_{s\in\tilde{\nu}_{\alpha'}}\frac{\left(\prod_{\alpha=1}^32i\sin\frac{E_{\alpha' \alpha} + \tilde{m}_1+i\gamma_1}{2}\right)\left(2i\sin\frac{E_{\alpha' \mu_8}- \tilde{m}_2+i\gamma_1}{2}\right)}{\prod_{\beta'=1}^2(2i)^2\sin\frac{E_{\alpha'\beta'}}{2}\sin\frac{E_{\alpha'\beta'}+2i\gamma_1}{2}}\nonumber\\
&&\prod_{s\in \mu_8}\frac{\prod_{\alpha'=1}^22i\sin\frac{E_{\mu_8\alpha'} + \tilde{m}_2+i\gamma_1}{2}}{(2i)^2\sin\frac{E_{\mu_8\mu_8}}{2}\sin\frac{E_{\mu_8\mu_8}+2i\gamma_1}{2}}\Big].
\eea
The Coulomb branch parameters $\lambda_\alpha$ and $\lambda'_{\alpha'}$ are
\bea
&&\lambda_1 = a_1, \quad \lambda_2 = -a_1 + a_2, \quad \lambda_3 = -a_2,\\
&&\lambda'_1 = \lambda' = a', \quad \lambda'_2 = -\lambda' = -a', \quad \lambda_{\mu_8} = 0,
\eea
and we also defined $\tilde{\nu}_1 := \nu_8$ and $\tilde{\nu}_2 := \nu_9$.

Here the ratio $\tilde{Z}_{T_4}/Z_{U(1)}^{=}$ is nothing but the Nekrasov partition function of $U(3) \times U(2) \times U(1)$ gauge theory with four hypermultiplets in the fundamental representation of $U(3)$, a bi-fundamental hypermultiplet of $U(3) \times U(2)$ and a bi-fundamental hypermultiplet of $U(2) \times U(1)$ except for the perturbative $U(1)$ Cartan contribution.\footnote{The lack of some perturbative $U(1)$ parts is not a contradiction because our Euler characteristic $\chi(X)$ originally has an ambiguity. We can shift $\chi(X)$ by six to recover them.} Note that $Z_{U(1)}^{=}$ is the contribution from decoupled M2-branes on the two-cycles associated with pairs of {\it horizontal} parallel external lines in figure \ref{fig:T4}. We similarly define $Z_{U(1)}^{||}$ and $Z_{U(1)}^{/\!/}$ for the {\it bottom} and {\it upper-right} parallel external lines, respectively. The explicit expression for them are obtained in section \ref{sec:general-TN}.

\subsection{$U(1)$ factors} 
\label{subsec:U1-factors}

In the previous subsection, we evaluated the refined topological string partition function $Z_\textrm{top}$ on various toric Calabi-Yau threefolds $X$. We have particularly found that if $X$ engineers $Sp(1)$ gauge theories then $Z_\textrm{top}/Z_{U(1)}^{=}$ reproduces the Nekrasov partition functions of $U(2)$ gauge theories rather than $Sp(1)$. Moreover, if $X$ engineers the $T_{N}$-theory for $N=2,3,4$,  $Z_\textrm{top}/Z_{U(1)}^{=}$ gives the Nekrasov partition function of $U(N-1)\times U(N-2)\times \cdots U(1)$ linear quiver gauge theory, rather than the $T_N$-theory. The reason for this is that we have not eliminated all the decoupled M2-branes from $Z_\textrm{top}$.

As discussed in subsection \ref{subsec:comparison}, $Z_\textrm{top}$ contains various contributions from M2-branes which are not charged under the 5d gauge symmetry engineered. The factor $Z_{U(1)}^{=}$ is a part of such contributions, but there are still contributions from decoupled M2-branes which are {\it not} included in $Z_{U(1)}^{=}$. Such M2-branes give a further prefactor $Z_{U(1)}^\textrm{others}$ of $Z_\textrm{top}$. To be specific, in \ref{subsubsec:F0} we have
\begin{eqnarray}
Z_{U(1)}^\textrm{others} = Z_{U(1)}^{||}
\end{eqnarray}
and in \ref{subsubsec:TN}
\begin{eqnarray}
Z_{U(1)}^\textrm{others} = Z_{U(1)}^{||}Z_{U(1)}^{/\!/}.
\end{eqnarray}
Then the total factor to be eliminated is 
\begin{align}
Z_{U(1)} = Z_{U(1)}^{=} Z_{U(1)}^\textrm{others}.
\end{align}
We call the total $Z_{U(1)}$ the ``U(1)-factor.''

{\it Now we claim that $Z_\textrm{top}/Z_{U(1)}$ correctly reproduces the Nekrasov partition function of $Sp(1)$ gauge theories and $T_N$-theories.} In the next subsection, we will explicitly show this for the $Sp(1)$ gauge theory with $N_f=2,3,4$ flavors as well as the $T_3$-theory. Note that this claim naturally explains the relation \eqref{eq:two_Nf=2} in the case of two-point blowup of the local $\mathbb{P}^1\times \mathbb{P}^1$. Namely the both sides of \eqref{eq:two_Nf=2} give the same Nekrasov partition function of the $SU(2)$ gauge theory with two fundamental matters; for $SU(2)$ there is no distinction between ${\bf 2}$ and $\overline{\bf 2}$.

\begin{figure}
\begin{center}
\includegraphics[width=3.5cm]{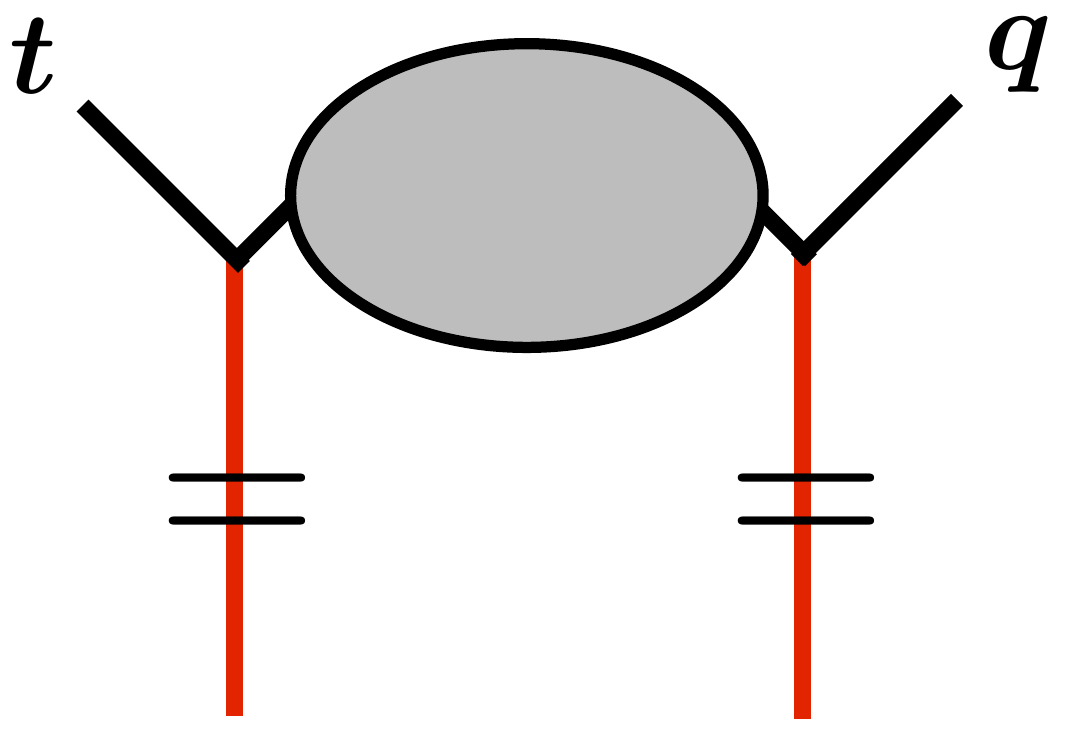}\qquad\qquad\qquad
\includegraphics[width=3.5cm]{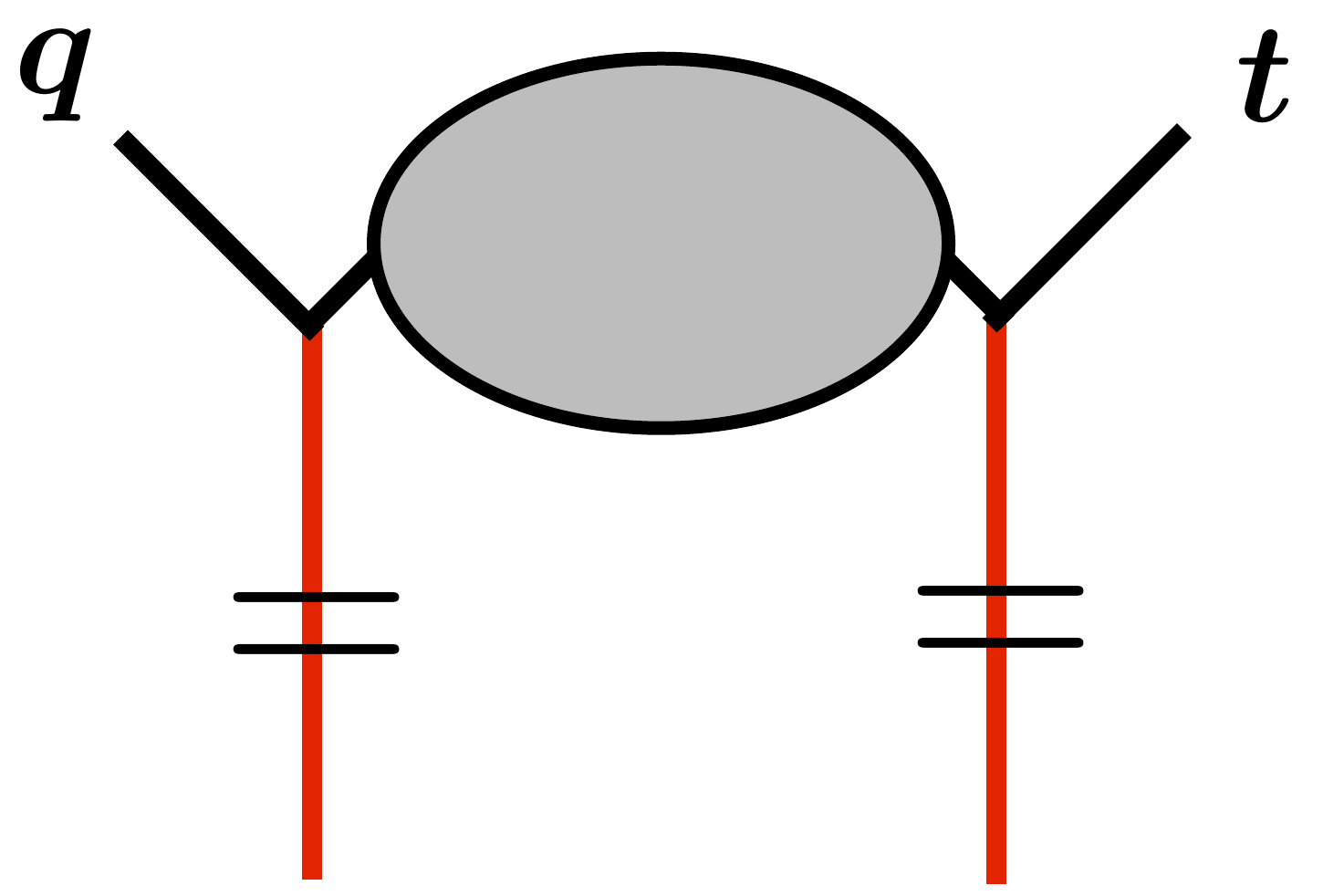}
\caption{When the two external lines are preferred directions, $r_\beta  =+1/2$ or $r_\beta = -1/2$, depending on the relative positions of the $q$- and $t$-directions. In each picture, the red lines are two parallel external lines associated with the two-cycle $\beta$, and the gray region is arbitrary. The left and right situations give $r_\beta = +1/2$ and $r_\beta = -1/2$, respectively.}
\label{fig:U(1)}
\end{center}
\end{figure}

The detail of the $U(1)$-factor depends on the toric Calabi-Yau three-fold $X$, but there is a general expression:
\begin{align}
 Z_{U(1)} = \prod_{\beta\,|\, \beta\cdot D_i=0}\prod_{m,n=1}^\infty(1-q^{m-(1/2+r_\beta)}t^{n-(1/2-r_\beta)}Q_\beta)^{-1}.
\label{eq:U1-revise-3}
\end{align}
Here $\beta$ runs over compact two-cycles whose intersection number with every compact four-cycle vanishes. Therefore $Q_\beta = \exp(-2\pi \mathfrak{t}\cdot \beta)$ is independent of the Coulomb branch parameters of the 5d gauge theory, while it could contain fugacities for gauge theory instantons. In the bracket of \eqref{eq:U1-revise-3}, we set $r_\beta = + 1/2$ or $r_\beta = -1/2$. The sign of $r_\beta$ is determined as follows. Suppose that $\beta$ is a two-cycle whose intersection vanishes with every compact four-cycle. In the toric web-diagram, such $\beta$ is associated with a pair of parallel external lines.\footnote{In other words, the ``north'' and ``south'' poles of $\beta$ are attached to the two parallel external lines.} Now, if the parallel external lines are $t$-directions, we set $r_\beta = +1/2$. If they are $q$-directions then we set $r_\beta = -1/2$. If they are preferred directions, it depends on the relative positions of the neighboring $q$- and $t$-directions. Namely $r_\beta =+1/2$ or $r_\beta = -1/2$ if we have the left or right situation in figure \ref{fig:U(1)}, respectively.
Note that, in all our examples in subsection \ref{subsec:vertex}, $Z_{\rm U(1)}^{=}$ is a part of \eqref{eq:U1-revise-3} coming from $\beta$ for {\it horizontal} parallel external lines. On the other hand, $Z_{U(1)}^\textrm{others}$ comes from $\beta$ for {\it non-horizontal} parallel external lines.

We can explain the expression \eqref{eq:U1-revise-3} from the viewpoint of the moduli space of M2-branes. 
Recall that the $U(1)$-factor is the contribution of decoupled M2-branes which are neutral under the 5d gauge symmetry engineered. As discussed in section \ref{sec:topst}, such a contribution is written as
\begin{align}
\prod_{\beta \,|\, \beta\cdot D_i = 0}\prod_{(j_L,j_R)}\prod_{\ell=-j_L}^{j_L}\prod_{r=-j_R}^{j_R}\prod_{m,n=1}^\infty(1- q^{\ell-r+m-1/2}t^{\ell + r +n-1/2}Q_\beta)^{(-1)^{2j_L+2j_R}N_{j_L,j_R}^{(\beta)}},
\label{eq:U1-revise}
\end{align}
where the integer $N_{j_L,j_R}^{(\beta)}$ is the BPS degeneracy of M2-branes wrapping on $\beta$ with spins $(j_L,j_R)$. 
 The BPS degeneracy $N_{j_L,j_R}^{(\beta)}$ has a nice interpretation in the moduli space of M2-branes \cite{Gopakumar:1998jq}. To see this, let $\mathcal{M}$ be the moduli space of the deformations of $\beta$ in the Calabi-Yau three-fold $X$. Let us also denote by $\widehat{\mathcal{M}}$ the moduli space of deformations of $\beta$ with a flat $U(1)$ bundle on it. There is a fiber structure
\begin{align}
\widehat{\mathcal{M}} \to \mathcal{M}
\label{eq:fiber}
\end{align}
where the fiber is the moduli space of flat bundles over $\beta$.
In physical language, $\mathcal{M}$ is the moduli space of a single D2-brane on $\beta$ while $\widehat{\mathcal{M}}$ is that of a single M2-brane on $\beta$.\footnote{When regarding an M2-brane as a bound state of a D2-brane and D0-branes, the $U(1)$ flux is identified with the D0-charge.} Recall that the left and right spins $(j_L,j_R)$ are charges associated with $SU(2)_L\times SU(2)_R$. In terms of the moduli space of the M2-brane, the actions of $SU(2)_R$ and $SU(2)_L$ are identified with $SL(2)$ Lefschetz actions in the base and fiber directions of \eqref{eq:fiber}, respectively \cite{Gopakumar:1998jq}. Then the cohomology $H^*(\widehat{\mathcal{M}})$ has the following decomposition \cite{Gopakumar:1998jq, Katz:1999xq}:
\begin{align}
H^{*}(\widehat{\mathcal{M}}) = \bigoplus_{j_L,j_R} N_{j_L,j_R}^{(\beta)} [j_L,j_R],
\end{align}
where $[j_L,j_R]$ is a subspace of $H^*(\mathcal{M})$ corresponding to the spin multiplet $(j_L,j_R)$.
Although $H^*(\widehat{\mathcal{M}})$ is highly non-trivial for general $\beta$, it is quite simple for our $\beta$ whose intersection number with every compact four-cycle vanishes.
Recall that our $\beta$ is associated with two parallel external lines in the web-diagram. To be more precise, $\beta$ is a genus zero curve whose area is given by the distance between the two external lines. Since the moduli space of flat bundles over a genus zero curve is a single point, the fiber of \eqref{eq:fiber} is a single point for our $\beta$. Therefore only $j_L=0$ is allowed. On the other hand, we can continuously move our $\beta$ along the parallel external lines in the web-diagram. This means that the moduli space $\mathcal{M}$ of our $\beta$ is isomorphic to $\mathbb{C}$; the boundary at infinity in $\mathbb{C}$ corresponds to moving $\beta$ to infinity in $X$. Since $\mathcal{M}$ is non-compact, $H^{*}(\mathcal{M})$ does not have a Lefschetz decomposition. To overcome this difficulty, let us compactify $\mathcal{M}$ by adding a point at infinity. The compactified moduli space is isomorphic to $\mathbb{P}^1$, whose cohomologies have a Lefschetz decomposition corresponding to a single $j_R=1/2$ multiplet. Hence, if the moduli space $\mathcal{M}$ is compactified, we have
\begin{align}
H^{*}(\widehat{\mathcal{M}}) = \Big[0,\frac{1}{2}\Big].
\label{eq:spin-content}
\end{align}
This implies that the BPS degeneracy of M2-branes wrapping on our $\beta$ is given by $N_{j_L,j_R}^{(\beta)} = \delta_{j_L,0}\,\delta_{j_R,\frac{1}{2}}$. Then the $U(1)$ factor \eqref{eq:U1-revise} turns out to be
\begin{align}
\prod_{\beta\,|\, \beta\cdot D_i=0}\prod_{r=-\frac{1}{2}}^{\frac{1}{2}}\prod_{m,n=1}^\infty(1-q^{m-(1/2+r)}t^{n-(1/2-r)}Q_\beta)^{-1}.
\label{eq:U1-revise-2}
\end{align}
Note here that, to obtain this result, we have compactified $\mathcal{M}$ by adding a point at infinity. The point at infinity in $\mathcal{M}$ corresponds to an M2-brane at infinity in the Calabi-Yau three-fold $X$. However, when $Z_{\text{top}}$ is computed via the refined topological vertex, all such contributions from infinity are omitted. Then $r$ in \eqref{eq:U1-revise-2} does not run over the whole {\it right} spin components. We should rather fix $r$ to be $+1/2$ or $-1/2$.\footnote{Here the {\it right} spin components for $r=\pm 1/2$ correspond to $H^0(\mathbb{P}^1)$ and $H^2(\mathbb{P}^1)$. When we decompactify $\mathbb{P}^1$ to recover $\mathcal{M}\simeq \mathbb{C}$, the second cohomology becomes trivial. This means that, if we omit the contribution from the point at infinity in $\mathbb{\mathcal{M}}$, we lose one {\it right} spin component.} Hence, {\it in the computation via the refined topological vertex,} the $U(1)$-factor is given by \eqref{eq:U1-revise-3}. The sign of $r_\beta$ depends on which {\it right} spin component of the M2-brane on $\beta$ is captured by the refined topological vertex. This explains the expression \eqref{eq:U1-revise-3}.



\subsection{$Sp(1)$ Nekrasov partition functions}
\label{sec:SU}
In this section, we will compare the topological string partition functions evaluated in the
previous subsection with the field theory partition functions of $Sp(1)\cong SU(2)$ gauge theories and show how two partition functions can be identified.
The Nekrasov partition functions for $Sp(N)$ gauge theories are evaluated using localization
in four-dimensions in \cite{Marino:2004cn,Nekrasov:2004vw} and in five-dimensions in \cite{Kim:2012gu}.

In \cite{Kim:2012gu}\footnote{The same index computation is done in \cite{Vafa:2012fi, Iqbal:2012xm} using the refined topological vertex method}, the Nekrasov partition function is used to compute the superconformal
index of 5d superconformal field theories including the $SU(2)$ theories we are interested in this section.
The 5d superconformal index (or the partition function on $S^1\times S^4$) is defined by
\be
	I = {\rm tr}\Big[(-1)^Fe^{-2(j_1+R)\gamma_1}e^{-2j_2}e^{-i\sum_iH_im_i}u^k\Big]\,,
\ee
where $j_1, j_2$ are the two Cartan generators of $SU(2)_1\times SU(2)_2 \subset SO(5)$ isometry on $S^4$, $R$ is the Cartan generator of the $SU(2)$ R-symmetry, $H_i$ are the 
flavor charges and $k$ is the instanton number.
The corresponding fugacities are $e^{-\gamma_1},e^{-\gamma_2},e^{-im_i}$ and $u$, respectively. 
In the localization computation of the index, the Nekrasov partition function corresponds to
the contribution localized at the north (or south) pole of $S^4$ and the full superconformal
index is given by product of the contributions from the north and south poles.
\be
	I =\int[d\lambda] Z_{\rm Nekra}(\lambda,\gamma_1,\gamma_2,m_i,u)\cdot 
	Z_{\rm Nekra}(-\lambda,\gamma_1,\gamma_2,-m_i,u^{-1})\,.
\ee
The index has the integration over the holonomy $\lambda$ which corresponds
to the Coulomb branch parameter in the Nekrasov partition function.
The measure $[d\lambda]$ includes the Haar measure and the Weyl factor of the gauge group G.

The perturbative contribution from the vector multiplet at the north pole is given by 
\be\label{eq:1-loop-vector}
	Z_{\rm pert}^{\rm vm}=\prod_{i,j=1}^\infty\left[(1-q^it^{j-1})^r(1-q^{i-1}t^{j})^r
	\prod_{\alpha \in {\rm root}}(1-e^{i\alpha\cdot\lambda}q^{i}t^{j-1})
	(1-e^{i\alpha\cdot\lambda}q^{i-1}t^{j})\right]^{-\frac{1}{2}}\,,
\ee
where $r$ is the rank of the gauge group, and 
the hypermultiplet contribution is 
\be\label{eq:1-loop-hyper}
	Z_{\rm pert}^{\rm hm}=\prod_{i,j=1}^\infty
	\prod_{\alpha \in R}(1-e^{i\alpha\cdot\lambda-im}q^{i-\frac{1}{2}}t^{j-\frac{1}{2}})\,,
\ee
where $m$ denotes the chemical potential for the flavor symmetry and $R$ denotes the weight vector of the gauge group in the representation $R$.
To get the instanton contribution we can use the contour integral formulae (3.58), (3.61)
and (3.62) in \cite{Kim:2012gu}.

Note that, when we compare the perturbative partition functions, we often ignore divergent prefactors $\mathcal{N}_1,\mathcal{N}_2$ given by
\bea\label{eq:divergent-factor}
	\mathcal{N}_1 &=& \prod_{i,j=1}^\infty\frac{(1-Qq^{i-\frac{1}{2}}t^{i-\frac{1}{2}})}
	{(1-Q^{-1}q^{i-\frac{1}{2}}t^{j-\frac{1}{2}})}=\prod_{i,j=1}^\infty(-Q q^{-i+\frac{1}{2}}t^{-j+\frac{1}{2}}) \,, \nn \\
	\mathcal{N}_2 &=& \prod_{i,j=1}^\infty\frac{(1-Qq^{i}t^{i-1})}{(1-Q^{-1}q^{i-1}t^{i})}= \prod_{i,j=1}^\infty(-Q q^{-i+1}t^{-j}) \,.
\eea
Two Nekrasov partition functions can be different up to the multiplication of the divergent
factors. It seems that there is an ambiguity on how to factorize the perturbative contribution
of the superconformal index.
However we expect that such prefactors cancels when we multiply two factorized partition
functions and no ambiguity remains in the superconformal index.

\subsubsection{$N_f = 2,3,4$}
We can now compare the results in subsection \ref{subsec:vertex} with the $SU(2)$ gauge theory
partition functions.
It turns out that, after stripping off the ``$U(1)$ factors'', the topological string partition functions $Z^{top}$ for $N_f=2,3,4$ are exactly
the same as the $SU(2)$ Nekrasov partition function with $N_f=2,3,4$ fundamental hypermultiplets. 
\be\label{eq:TSA-Nekra-relation}
	Z^{top}_{N_f} \equiv \tilde{Z}^{top}_{N_f}/Z_{U(1)} = Z^{SU(2)}_{\rm Nekra} \,.
\ee
Here, $\tilde{Z}^{top}$ is the origianl partition function from the topological vertex computation
and $Z_{U(1)}$ is the $U(1)$ factor which we can read off from the web diagram following the rule in the previous subsection.

Let us first check this equivalence for the perturbative contribution. The perturbative contributions $Z_0^{N_f}$
of the topological string partition functions are given 
 in (\ref{eq:part-U(2)-Nf=2-1}), (\ref{eq:part-U(2)-Nf=3}), (\ref{eq:part-U(2)-Nf=4}) for the $N_f=2,3,4$ cases, respectively.
As explained earlier, the perturbative conrtribution of the $SU(2)$ gauge theory partition function with $N_f$ fundamental hypermultiplets is given by
\be
	Z_{\rm pert}^{N_f}\!\! =\!\! \prod_{i,j=1}^\infty\!\frac{\prod_{a=1}^{N_f}(1-e^{i\lambda - im_a}q^{i-\frac{1}{2}}t^{j-\frac{1}{2}})}
	{\big[(1\!-\!q^it^{j-1})(1\!-\!q^{i-1}t^j)(1\!-\!e^{2i\lambda}q^{i}t^{j-1})
	(1\!-\!e^{2i\lambda}q^{i-1}t^{j})(1\!-\!e^{-2i\lambda}q^{i}t^{j-1})
	(1\!-\!e^{-2i\lambda}q^{i-1}t^{j})\big]^{\frac{1}{2}}}
\ee  
with mass parameters $m_{a=1,2,\cdots,N_f}$.
If we ignore the divergent factor mentioned in eqn (\ref{eq:divergent-factor}) and
remove the $U(1)$ factor $Z_{U(1)}^{=}$ for the {\it horizontal} parallel external lines,
one can easily check that the $Z_0^{N_f}$'s are the same as the gauge theory results :
\be
	Z_{\rm pert}^{N_f=2,3,4} = Z_0^{N_f=2,3,4}/ Z_{U(1)}^{=} \,, \quad (N_f=2,3,4) \,,
\ee
where the horizontal $U(1)$ factors are given by
\bea
	 N_f=2\ &:& \quad Z_{U(1)}^{=} = 1 \,, \\
	 N_f=3\ &:& \quad Z_{U(1)}^{=} = \prod_{i,j=1}^\infty (1-e^{-i(m_1+m_3)}q^{i-1}t^j)^{-1} \,,\nn \\
	 N_f=4\ &:& \quad Z_{U(1)}^{=} = \prod_{i,j=1}^\infty(1-e^{-i(m_1+m_3)}q^{i-1}t^j)^{-1}
				(1-e^{-i(m_2+m_4)}q^{i}t^{j-1})^{-1} \,. \nn
\eea

We then turn to the instanton partition functions. We expand both partition functions by the instanton fugacity $u$
and compare them up to several orders.
The instanton parts also contain the extra factors and they can be identified from one instanton calculation.
At $k=1$, the field theory partition function of \cite{Kim:2012gu} takes the simply form of
\be
	Z_{\rm k=1}^{SU(2)} = \frac{1}{32}\left[\frac{\prod_{a=1}^{N_f}2i\sin\frac{m_a}{2}}
	{i^2\sinh\frac{\gamma_1\pm\gamma_2}{2}\sin\frac{i\gamma_1\pm\lambda}{2}}
	+\frac{\prod_{a=1}^{N_f}2\cos\frac{m_a}{2}}
	{\sinh\frac{\gamma_1\pm\gamma_2}{2}\cos\frac{i\gamma_1\pm\lambda}{2}}
	\right] \,,
\ee
where we used a succinct notation, $\sin(a\!\pm\! b)=\sin(a\!+\!b)\sin(a\!-\!b)$.
One can then easily take the difference of two partition functions at 1-instanton 
and the result simply becomes
\bea
	 N_f=2\ &:& \quad \tilde{Z}_{\rm k=1}^{top} - Z_{\rm k=1}^{SU(2)} = \frac{qe^{-\frac{i}{2}(m_1+m_2)}}{(1-q)(1-t)} \,, \\
	 N_f=3\ &:& \quad \tilde{Z}_{\rm k=1}^{top} - Z_{\rm k=1}^{SU(2)} = \frac{qe^{-\frac{i}{2}(m_1+m_2-m_3)}}{(1-q)(1-t)} \,,\nn \\
	 N_f=4\ &:& \quad \tilde{Z}_{\rm k=1}^{top} - Z_{\rm k=1}^{SU(2)} = \frac{qe^{-\frac{i}{2}(m_1+m_2-m_3-m_4)}+te^{\frac{i}{2}(m_1+m_2-m_3-m_4)}}{(1-q)(1-t)} \,,\nn
\eea
where $\tilde{Z}^{top}_{k}$ denotes the $k$-instanton contribution of the topological string partition function computed in subsection \ref{subsec:vertex}.
We note that the differences are all independent of the Coulomb branch parameter $\lambda$.
In fact, the Plethystic exponentials of these differences are precisely equal to the $U(1)$ factors
for the {\it non-horizontal} parallel external lines discussed above\footnote{
  The Plethystic (or multi-particle) exponential of a single-particle partition function $f(x)$ is defined as $PE[f(x)]=exp\left[\sum_{n=1}^\infty\frac{1}{n}f(x^n)\right]$
  where $x$ represents all the fugacities of $f$.
  The difference at $k\!=\!1$ is interpreted as the partition function of a single-particle from
  the decoupled $U(1)$ contribution and thus the full partition function of the $U(1)$ factor
  is given by the Plethystic exponential of it.
  Note that the single-particle states of the $U(1)$ factor carry the instanton charge $+1$
  and therefore are captured by 1-instanton computation.
}.
They are given by\footnote{The five-dimensional $U(1)$ factor in the case of $N_f = 2$ was also independently obtained from the field theory analysis by Christoph A.~Keller and Jaewon Song. We thank their correspondence.}
\bea
	N_f=2\ &:& \quad Z_{U(1)}^{\parallel} = \prod_{i,j=1}^\infty(1-ue^{-\frac{i}{2}(m_1+m_2)}q^{i}t^{j-1})^{-1} \,, \\
	N_f=3\ &:& \quad Z_{U(1)}^{\parallel} = \prod_{i,j=1}^\infty(1-ue^{-\frac{i}{2}(m_1+m_2-m_3)}q^{i}t^{j-1})^{-1} \,, \nn \\
	N_f=4\ &:& \quad Z_{U(1)}^{\parallel} = \prod_{i,j=1}^\infty\left[(1-ue^{-\frac{i}{2}(m_1+m_2-m_3-m_4)}q^{i}t^{j-1})(1-ue^{\frac{i}{2}(m_1+m_2-m_3-m_4)}q^{i-1}t^{j})\right]^{-1} \,.\nn
\eea
We find that the instanton partition functions are related by $Z^{SU(2)}_{\rm inst} = \tilde{Z}^{top}_{\rm inst}/Z_{U(1)}^{\parallel}$.
We verified this relation so far up to 4-instantons\footnote{To see this relation, it is crucial to use $O(k)$ dual gauge group for the instanton moduli space integral instead of $SO(k)$ dual gauge group. }.

Thus, all in all, we have checked the relation (\ref{eq:TSA-Nekra-relation}) for
$N_f=2,3,4$ cases and also identified 
the total $U(1)$ factors, $Z_{U(1)} = Z_{U(1)}^{=}\!\cdot\! Z_{U(1)}^{\parallel}$.

\subsubsection{$N_f = 5$ from $T_3$}\label{subsec:T3}
As explained in section \ref{sec:TN}, 
the web diagram of $T_3$ theory in Figure \ref{fig:T3} can be seen as $\mathbb{P}^1\times \mathbb{P}^1$
at 5 points blow up and thus the resulting 5d theory can be identified with 
an $SU(2)$ gauge theory with $N_f=5$ fundamental hypermultiplets. 
In particular, the $E_6$ global symmetry enhancement of $T_3$ theory is realized in the $SU(2)$ gauge theory as the global symmetry enhancement from $SO(10)$ to $E_6$ at the conformal fixed point.
This $E_6$ symmetry enhancement in the gauge theory is confirmed perturbatively
using the superconformal index in \cite{Kim:2012gu,Bashkirov:2012re}. 
We will here check that the partition function of $T_3$ theory is in fact the same as that of the
$SU(2)$ gauge theory. This automatically implies that the superconformal index of $T_3$ theory has $E_6$ global symmetry and thus provides a highly non-trivial check of our suggestion of the $T_3$ partition function.

It follows from the topological vertex computation (\ref{T3-partition}) and the $U(1)$ factor prescription that the $T_3$ theory partition function is
\be
	Z_{T_3} = \tilde{Z}_{T_3} / Z_{U(1)} \,,
\ee
where $\tilde{Z}_{T_3}$ is given in (\ref{T3-partition}) and $Z_{U(1)}$ is the 
corresponding $U(1)$ factor given by
\be\label{eq:T3-U(1)factor}
	Z_{U(1)} = \prod_{i,j=1}^\infty\prod_{a=1}^3(1-Q_aq^i t^{j-1})^{-1}(1-\tilde{Q}_aq^i t^{j-1})^{-1}(1-\hat{Q}_aq^{i-1} t^{j})^{-1} \,,
\ee
with
\bea
	Q_{1,2,3} &\equiv& (e^{-im_1+im_2},e^{im_1-im_3},e^{im_2-im_3}) \,, \nn \\
	\tilde{Q}_{1,2,3} &\equiv& (u_1e^{im_4},u_2e^{-\!\frac{i}{2}(m_1\!+\!m_2\!+\!m_3\!+\!m_4)},u_1u_2e^{-\!\frac{i}{2}(m_1\!+\!m_2\!+\!m_3\!-\!m_4)}) \,, \nn \\
	\hat{Q}_{1,2,3} &\equiv& (u_1e^{-im_4},u_2e^{\!\frac{i}{2}(m_1\!+\!m_2\!+\!m_3\!+\!m_4)},u_1u_2e^{\frac{i}{2}(m_1\!+\!m_2\!+\!m_3\!-\!m_4)}) \,.
\eea
The K\"ahler parameters $Q_a, \tilde{Q}_a$ and $\hat{Q}_a$ correspond to the $U(1)$ factors
from the parallel 5-branes along the horizontal, diagonal and vertical direction, respectively.
Indeed, as noticed in section \ref{subsec:vertex}, the $T_3$ partition function takes the same form of a $U(2)\times U(1)$ quiver gauge theory partition function, up to the $U(1)$ factors: 
\be
	Z_{T_3} = Z^{U(2)\times U(1)}_{\rm Nekra}/(Z_{U(1)}^{\parallel}Z_{U(1)}^{/\!/}) \,,
\ee
where
\begin{align}
	Z_{U(1)}^{\parallel} =  \prod_{i,j=1}^\infty\prod_{a=1}^3(1-\tilde{Q}_aq^i t^{j-1})^{-1},\qquad
	Z_{U(1)}^{/\!/} = \prod_{i,j=1}^\infty\prod_{a=1}^3(1-\hat{Q}_aq^i t^{j-1})^{-1} \,.
\end{align}

The explicit comparison with the $SU(2)$ partition function is rather subtle.
In particular, the $T_3$ partition function comes with a summation over 3 Young diagrams $\nu_1,\nu_2,\nu_5$ with 2 instanton fugacities $u_1,u_2$, while the $SU(2)$ partition function comes with a summation over
2 Young diagrams corresponding to $\nu_1,\nu_2$ with an instanton fugacity $u$.
To compare two partition functions, we first identify an instanton fugacity $u_2$ of $T_3$ theory
with the instanton fugacity $u$ in the field theory such as $u=u_2e^{-\frac{i}{2}m_5}$. Then the instanton expansion by $u$ in the field theory
is realized by the expansion of $u_2$ in the $T_3$ theory.
The remaining fugacity $u_1$ is identified with the 5-th flavor fugacity, $u_1 = e^{-im_5}$.
With the identification of the fugacities, one can easily see that the $Z_{\rm inst}$
in (\ref{T3-partition}) involves non-trivial contribution from zero instanton sector at $\mathcal{O}(u^0)$. We take $|\nu_1|=|\nu_2|=0$ and read the zero instanton contribution
\be
	\sum_{\nu_5} u_1^{|\mu_5|}\prod_{s\in \mu_5}
		\frac{\prod_{\alpha=1}^22i\sin\frac{E_{5\emptyset} - \lambda_\alpha+m_4+i\gamma_1}{2}}
		{(2i)^2\sin\frac{E_{55}}{2}\sin\frac{E_{55}+2i\gamma_1}{2}} = \prod_{i,j=1}^\infty \frac{(1-u_1e^{i\lambda}q^{i-\frac{1}{2}}t^{j-\frac{1}{2}})(1-u_1e^{-i\lambda}q^{i-\frac{1}{2}}t^{j-\frac{1}{2}})}{(1-u_1e^{im_4}q^it^{j-1})(1-u_1e^{-im_4}q^{i-1}t^{j})} \,.
\ee
Here, we used the identity (\ref{eq:T2-identity}).
The numerator becomes the perturbative part of the 5-th fundamental hyper and the denominator
becomes the extra $U(1)$ factor of the K\"ahler parameters $\tilde{Q}_1,\hat{Q}_1$.
To see the agreement with the field theory results, we should define the instanton partition function of $T_3$ theory without this zero instanton contribution and also without the $U(1)$ factors.
\bea\label{eq:T3-partition4.2.3}
	Z_{T_3} \!\!&\!\!=\!\!&\!\! Z_{\rm pert}\cdot Z_{\rm inst} \,,  \\
	Z_{\rm pert} \!\!&\!\!=\!\!&\!\! \prod_{i,j=1}^\infty
			\bigg\{\frac{\prod_{b=1,2,4}(1\!-\!e^{-i\lambda+im_b}q^{i-\frac{1}{2}}t^{j-\frac{1}{2}})\!
			(1\!-\!e^{-i\lambda-im_b}q^{i-\frac{1}{2}}t^{j-\frac{1}{2}})}
			{(1-q^it^{j-1})^{\frac{1}{2}}(1-q^{i-1}t^{j})^{\frac{1}{2}}
			(1-e^{-2i\lambda}q^it^{j-1})(1-e^{-2i\lambda}q^{i-1}t^{j})} \nn \\
			&&\times \prod_{a=3,5}(1\!-\!e^{i\lambda-im_a}q^{i-\frac{1}{2}}t^{j-\frac{1}{2}})\!
			(1\!-\!e^{-i\lambda-im_a}q^{i-\frac{1}{2}}t^{j-\frac{1}{2}}) \bigg\} \,,\nn \\
		Z_{\rm inst} \!\!&\!\!=\!\!&\!\!
		\prod_{i,j=1}^\infty\bigg\{\frac{\prod_{a=2,3}(1-\tilde{Q}_aq^it^{j-1})(1-\hat{Q}_aq^{i-1}t^j)}
		{(1-e^{i\lambda-im_5}q^{i-\frac{1}{2}}t^{j-\frac{1}{2}})(1-e^{-i\lambda-im_5}q^{i-\frac{1}{2}}t^{j-\frac{1}{2}})}\bigg\}\times 
		\sum_{\nu_1,\nu_2,\mu_5}(ue^{\frac{i}{2}m_5})^{|\nu_1|+|\nu_2|}(e^{-im_5})^{|\mu_5|} \nn\\
		&&
		\times 
		\prod_{\alpha=1}^2\prod_{s\in\nu_\alpha}\left[
		\frac{\left(\prod_{a=1}^32i\sin\frac{E_{\alpha\emptyset}-m_a+i\gamma_1}{2}\right)
		(2i\sin\frac{E_{\alpha 5}-m_4+i\gamma_1}{2})}
		{\prod_{\beta=1}^2(2i)^2\sin\frac{E_{\alpha\beta}}{2}\sin\frac{E_{\alpha\beta}+2i\gamma_1}{2}} \right]
		\prod_{s\in \mu_5}
		\frac{\prod_{\alpha=1}^22i\sin\frac{E_{5\alpha}+m_4+i\gamma_1}{2}}
		{(2i)^2\sin\frac{E_{55}}{2}\sin\frac{E_{55}+2i\gamma_1}{2}}. \nn
\eea
Here, the instanton part $Z_{\rm inst}$ is normalized so that $Z_{\rm inst}=1$ at order $\mathcal{O}(u^0)$.
The perturbative part is precisely that of the $SU(2)$ gauge theory with $N_f=5$ flavors.

For the instanton part, we expand both instanton contributions by the instanton fugacity
$u$ and compare them in order.
In the field theory computation, it is obvious that, at each instanton order, the expansion terminates at finite order of $e^{-im_5}$ because the contributions involving $e^{-im_5}$ come from the fermionic zero modes and they only appear in the numerator of the instanton partition function. 
On the other hand, the instanton contribution $Z_{\rm inst}$ is the summation over all possible Young diagram configurations labeled by $\mu_5$ and thus its expansion by $e^{-im_5}$ in general does not end.
We however find that the Young diagram summation
terminates at finite order of $e^{-im_5}$ and higher order contributions become zero\footnote{At $k$-instanton sector, it appears that the sum over $|\mu_5|$ terminates at order $(e^{-im_5})^k$ and higher order terms vanish.
We checked this up to $k=3$ for several orders in $e^{-im_5}$ with a computer}.
With this observation, the $Z_{\rm inst}$ of $T_3$ theory precisely agrees with the field theory instanton partition function given in \cite{Kim:2012gu}, which is checked up to 3 instantons.

The superconformal index of the $T_3$ theory is given by
\be\label{eq:T3-superconformalindex}
	I_{T_3} = \int [d\lambda] \big|Z_{T_3}\big|^2 \,.
\ee
By expanding this index by the fugacity $x\equiv e^{-\gamma_1}$, which is related to the
conformal dimensions of BPS states, one can check that the flaver fugacities form characters of $E_6$
global symmetry implying the $E_6$ global symmetry enhancement of $T_3$ theory. See \cite{Kim:2012gu} for the details.

\section{The 5d partition function of $T_N$ theory}
\label{sec:general-TN}

\begin{figure}
\centering
\includegraphics[width=12cm]{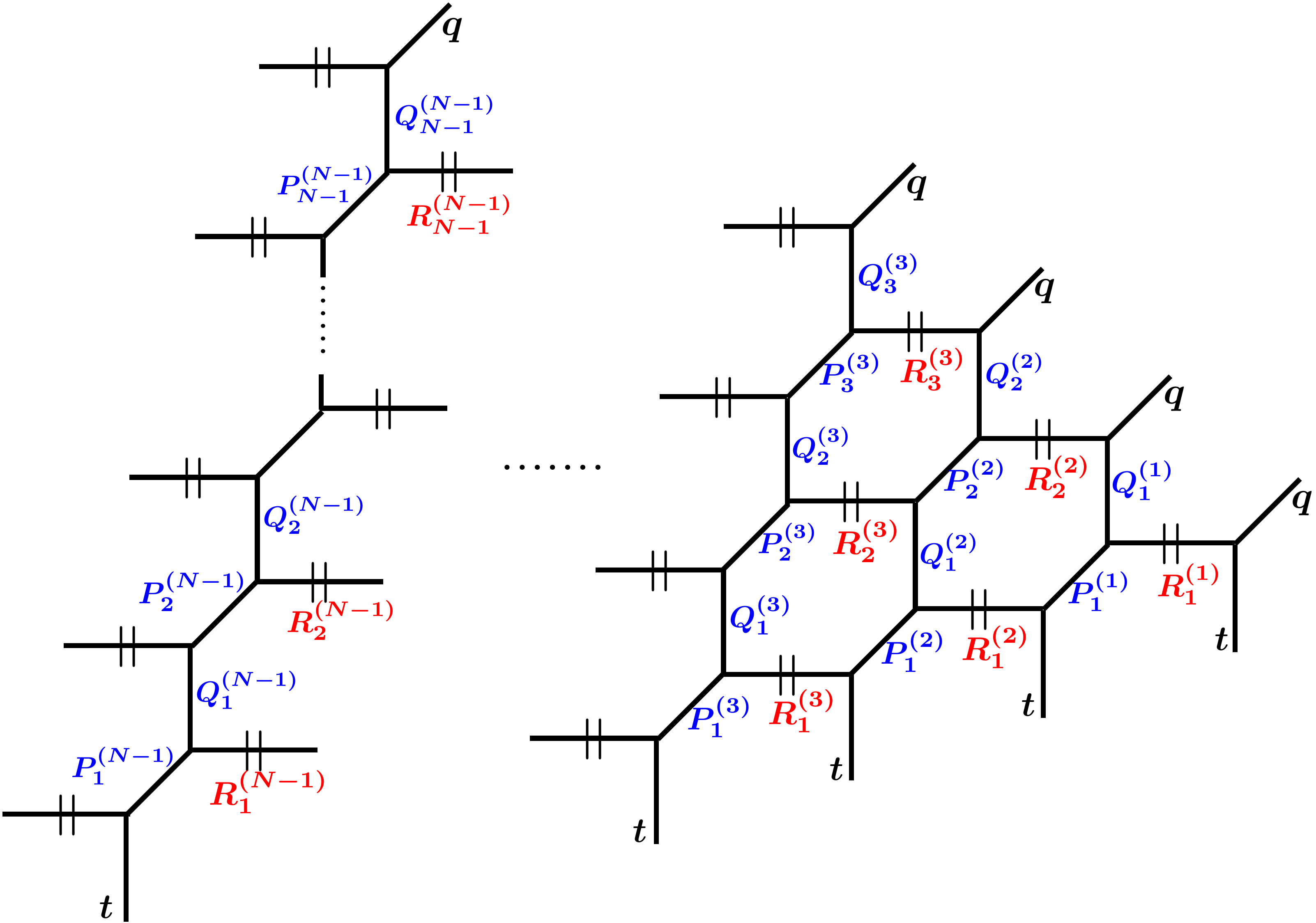}
\caption{The toric web-diagram of the blow-up of $\mathbb{C}^3/(\mathbb{Z}_N\times \mathbb{Z}_N)$. We take the horizontal directions as the preferred directions. The fugacity $R_{k}^{(n)}$ are associated with the horizontal (and therefore preferred) internal edges. The other internal edges are associated with fugacities $P_k^{(n)}$ and $Q_k^{(n)}$. There are $\frac{(N+4)(N-1)}{2}$ independent K\"ahler parameters.}
\label{fig:TN-diagram}
\end{figure}

In \ref{subsubsec:TN}, we have evaluated the topological string partition function associated with the $T_N$-theory for $N=2,3,4$, and seen their relation to linear quiver gauge theories. In this section, we generalize this to arbitrary $N$ and conjecture the Nekrasov partition function of the 5d $T_N$-theory for general $N$.

The relevant toric Calabi-Yau threefold is the blow-up of $\mathbb{C}^3/(\mathbb{Z}_N\times \mathbb{Z}_N)$, whose toric web-diagram is shown in figure \ref{fig:TN-diagram}. We associate the fugacities $P_k^{(n)}, Q_{k}^{(n)}$ and $R_k^{(n)}$ with the two-cycles as in figure \ref{fig:TN-diagram}. Due to the relations $P_k^{(n)}Q_k^{(n)} = Q_k^{(n+1)}P_{k+1}^{(n+1)}$ and $R_k^{(n)} = R_1^{(n)}(P_{1}^{(n-1)} P_{2}^{(n-1)} \cdots P_{k-1}^{(n-1)})(P_2^{(n)}\cdots P_k^{(n)})^{-1}$, there are $\frac{(N+4)(N-1)}{2}$ independent K\"ahler parameters.

The refined topological string partition function on this Calabi-Yau threefold is evaluated by using the refined topological vertex. We describe the detail of the calculation in appendix \ref{app:TN}, and here simply write the result as
\begin{eqnarray}
	\tilde{Z}_{T_N} \!\!&\!\!=\!\!&\!\! (M(t,q)M(q,t))^{\chi(X)/4}\cdot  Z_0\cdot  Z_{\rm inst}\cdot Z_{U(1)}^{=}
\end{eqnarray}
with
\begin{align}
Z_0 =&  \prod_{i,j=1}^{\infty}\Bigg\{\frac{\left[\prod_{ a\leq b}(1-e^{-i\lambda_{N-1;b} + i\tilde{m}_a}q^{i-\frac{1}{2}}t^{j-\frac{1}{2}})\prod_{b<a} (1-e^{i\lambda_{N-1;b}-i\tilde{m}_{a}}q^{i-\frac{1}{2}}t^{j-\frac{1}{2}})  \right]}{\prod_{n=1}^{N-1}\prod_{a< b}(1-e^{i\lambda_{n;a}-i\lambda_{n;b}}q^it^{j-1})(1-e^{i\lambda_{n;a}-i\lambda_{n;b}}q^{i-1}t^{j})}
\nonumber\\
& \times \prod_{n=2}^{N-1}\prod_{a\leq b}(1-e^{i\lambda_{n;a}-i\lambda_{n-1;b} + im_n} q^{i-1/2}t^{j-1/2})\prod_{ b < a}(1-e^{i\lambda_{n-1;b}-i\lambda_{n;a} - im_n}q^{i-1/2}t^{j-1/2})\Bigg\},
\nonumber\\
	\hspace{-0.5cm}Z_{\rm inst} =& \sum_{\vec{Y}_1,\vec{Y}_2,\cdots,\vec{Y}_{N-1}}
	\left[\prod_{n=1}^{N-1}u_n^{|\vec{Y}_n|}z_{\rm vec}(n)\right]\times 
	\left[\prod_{a=1}^Nz_{\rm fund}(N\!-\!1,\tilde{m}_a)\right] \times
	\left[\prod_{n=2}^{N-1}z_{\rm bifund}(n-1,n,m_n)\right],
\nonumber \\
Z_{U(1)}^{=} =& \prod_{1\leq a< b\leq N} \prod_{k,\ell=1}^\infty (1-e^{i\tilde{m}_a-i\tilde{m}_b}q^k t^{\ell-1})^{-1}.
\label{eq:Nekrasov-quiver}
\end{align}
The definitions of $z_\text{vec},\, z_\text{fund}$ and $z_\text{bifund}$ are given in appendix \ref{app:Nekrasov}, in which the ranks of the gauge group are set to be $N_n=n$. 
Here $\lambda_{n;k}$ for $n=2,\cdots,N-1$ and $k=1,\cdots,n$ are defined by
\begin{eqnarray}
P^{(n-1)}_kQ^{(n-1)}_k = \exp(-i\lambda_{n;k+1}+i\lambda_{n;k})
\end{eqnarray}
and $\sum_{k=1}^{n}\lambda_{n;k} = 0$. The parameters $m_n$ for $n=2,\cdots,N-1$ are defined by
\begin{eqnarray}
P^{(n-1)}_k = \exp(i\lambda_{n;k}-i\lambda_{n-1;k} + im_n),
\end{eqnarray}
where we set $\lambda_{1;k}=0$. We also define $\tilde{m}_k$ for $k=1,\cdots, N$ so that
\begin{eqnarray}
P^{(N-1)}_kQ^{(N-1)}_k = \exp(-i\tilde{m}_{k+1}+i\tilde{m}_{k}),\qquad P^{(N-1)}_k = \exp(i\tilde{m}_k - i\lambda_{N-1;k}).
\end{eqnarray}
The above relations reparameterize $P_k^{(n)}$ and $Q_k^{(n)}$ in terms of $\lambda_{n;k},m_n$ and $\tilde{m}_k$. The remaining parameters $u_n$ are defined by
\begin{eqnarray}
u_n = R_1^{(n)}Q_n^{(n)\frac{1}{2}}P_1^{(n)\frac{1}{2}}(P_2^{(n)}P_3^{(n)}\cdots P_{n}^{(n)})^{-\frac{1}{2}}(P_1^{(n-1)}P_2^{(n-1)}\cdots P_{n-1}^{(n-1)})^{\frac{1}{2}}.
\end{eqnarray}

Note that $\tilde{Z}_{T_N}/Z_{U(1)}^{=}$ is regarded as the Nekrasov partition function of a gauge theory described by the quiver diagram
	\begin{center}
	\includegraphics[scale=0.4]{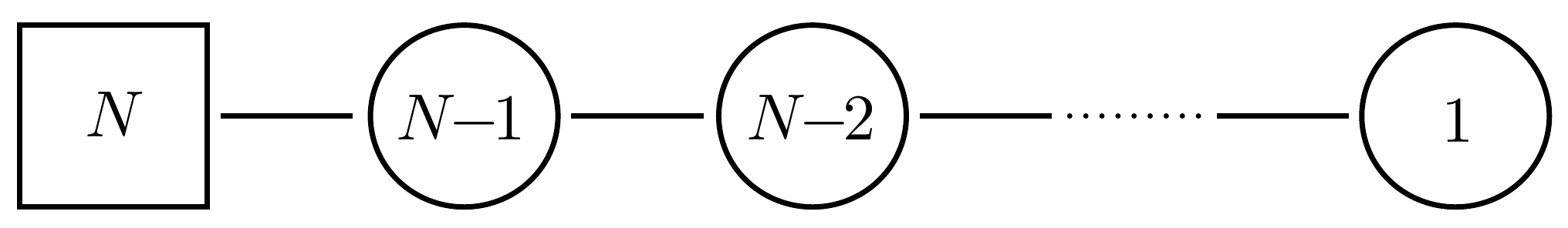}\
	\end{center}
where the circle nodes denote gauge groups and the square node denotes a global symmetry group.
We have bi-fundamental hypermultiplets between two nodes and $N$ fundamental hypermultiplets
for the leftmost $U(N\!-\!1)$ gauge group.
 Here $\lambda_{k,\alpha}$ are identified with the Coulomb branch parameters for
the $U(N\!-\!1)\times \cdots \times U(1)$ gauge group, $\tilde{m}_{a=1,\cdots,N}$ are the mass parameters
of the fundamental hypermultiplets for $U(N\!-\!1)$, and $m_{k=2,\cdots,N\!-\!1}$ are
the masses of the $N-2$ bi-fundamental hypermultiplets.
 The $Z_0$ is the perturbative part where the numerator comes from
the $N$ fundamental and $N-2$ bi-fundamental matter contributions while the denominator comes from
the $N-1$ vector multiplet contributions. The instanton part $Z_{\rm inst}$ is the summation over
all possible instanton contributions labeled by Young diagrams. The explicit expressions for$z_{\rm vect}, z_{\rm fund}$ and $z_{\rm bifund}$ are given in Appendix.


\begin{figure}
\centering
\includegraphics[width=10cm]{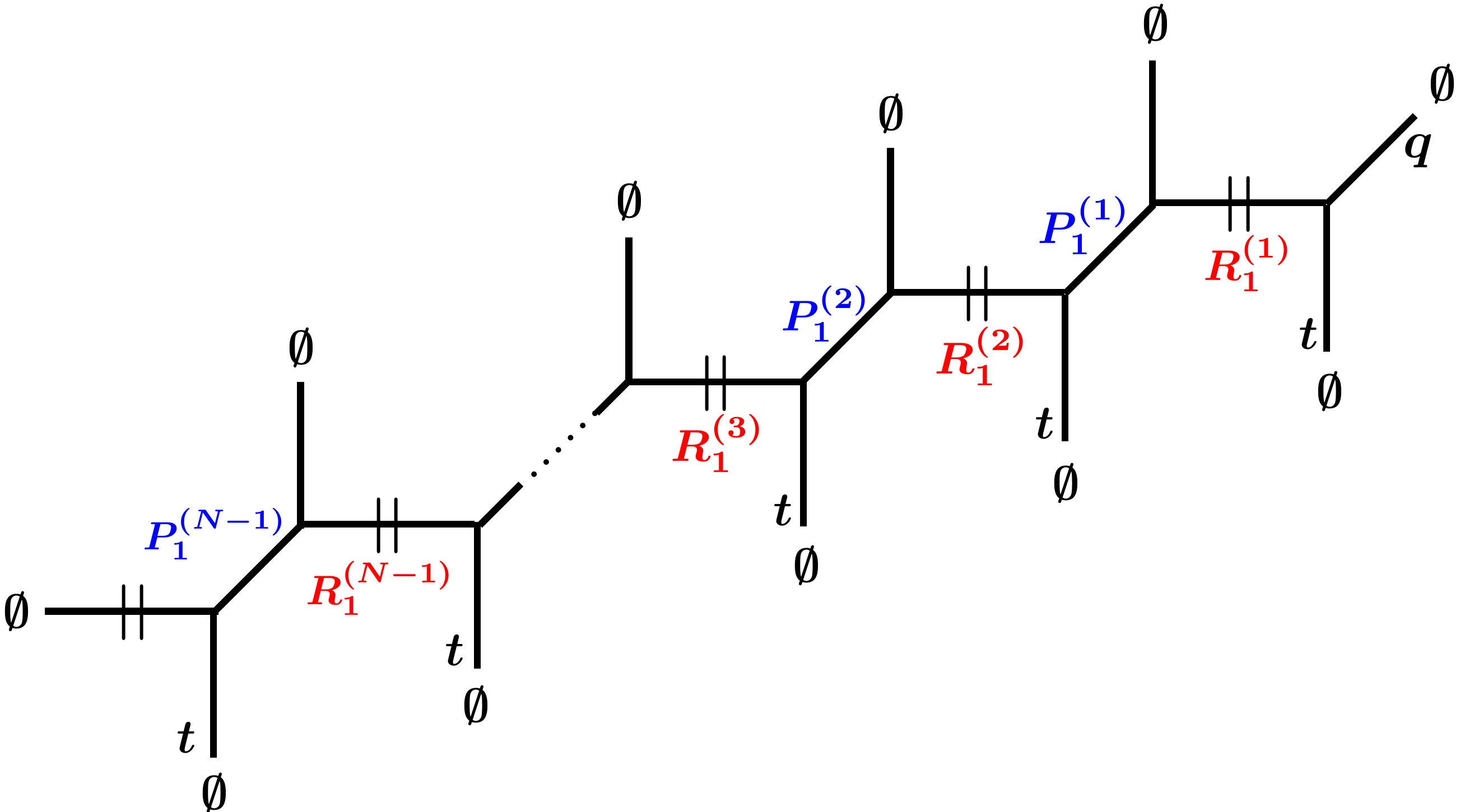}
\caption{The tree diagram which includes the contribution from M2-branes on $\tilde{\beta}_{ab}$.}
\label{fig:U1-bottom}
\end{figure}

While $\mathbb{C}^3/(\mathbb{Z}_N\times \mathbb{Z}_N)$ is expected to engineer the $T_N$-theory, we have obtained the Nekrasov partition function of the linear quiver gauge theory. The reason for this is that $\tilde{Z}_{T_N}/Z_{U(1)}^{=}$ still includes ``U(1)-factors'' coming from decoupled M2-branes. As discussed in \ref{subsec:U1-factors}, a decoupled M2-brane is wrapping two-cycle associated with a pair of parallel external lines of the toric web-diagram. In our web-diagram in figure \ref{fig:TN-diagram}, there are $N$ parallel external lines in each of the left, bottom and upper-right directions. Let $\beta_{ab}$ be the two-cycle associated with the $a$-th and $b$-th external lines extending in the {\it left} of figure \ref{fig:TN-diagram}. We similarly define $\tilde{\beta}_{ab}$ and $\hat{\beta}_{ab}$ for the {\it bottom} and {\it upper-right} directions, respectively. Then we find that the contribution from M2-branes on $\beta_{ab}$ is precisely given by $Z_{U(1)}^{=}$. On the other hand, contributions from M2-branes on $\tilde{\beta}_{ab}$ or $\hat{\beta}_{ab}$ are not included in $Z_{U(1)}^{=}$. Since the central charges of such M2-branes depend on $R_k^{(n)}$, they are included in $Z_\text{inst}$. In order to obtain the partition function of the $T_N$-theory, we have to eliminate all such contributions.

Let us first consider the $U(1)$-factor coming from M2-branes on $\tilde{\beta}_{ab}$. From the general argument in subsections \ref{subsec:comparison} and \ref{subsec:refined-vertex}, we know that it is of the form
\begin{eqnarray}
\prod_{i,j=1}^\infty(1-Q_{\tilde{\beta}_{ab}} q^{i+n_1/2}t^{j+n_2/2})^m
\label{eq:factor-bottom}
\end{eqnarray}
 for some $n_1,n_2,m\in \mathbb{Z}$.\footnote{Since $\tilde{\beta}_{ab}$ is genus zero, it is sufficient to consider $j_L=0$.} Here $Q_{\tilde{\beta}_{ab}}$ is the fugacity for M2-branes on $\tilde{\beta}_{ab}$, and given by
\begin{eqnarray}
Q_{\tilde{\beta}_{ab}} = \prod_{n=a}^{b-1}(R_1^{(n)}P_1^{(n)}).
\end{eqnarray}
To determine the values of $n_1,n_2$ and $m$, we have to look at the topological vertex calculation \eqref{eq:TN-vertex}, which is essentially an infinite sum of monomials of fugacities $P_k^{(n)},Q_k^{(n)}$ and $R_k^{(n)}$. Since we are only interested in the factor \eqref{eq:factor-bottom}, we can set any fugacity to be zero except for $Q_{\tilde{\beta}_{ab}}$. In particular, we can set $Q_k^{(n)}=0$ for all $k$ and $n$. Then the relevant amplitude comes from the tree diagram shown in figure \ref{fig:U1-bottom}. 

\begin{figure}
\centering
\includegraphics[width=10cm]{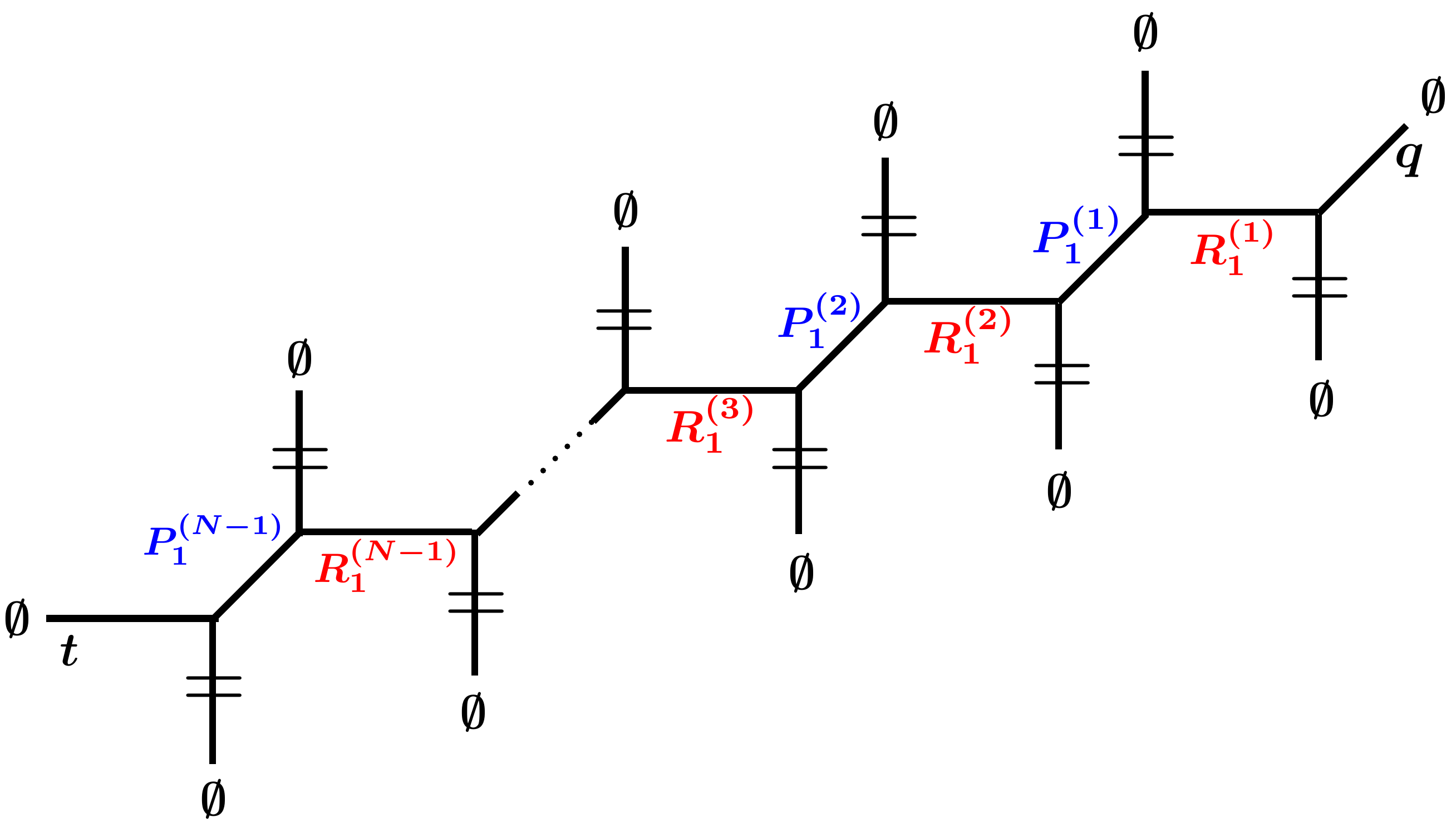}
\caption{A different choice of the preferred direction.}
\label{fig:U1-bottom-2}
\end{figure}

Since some of the internal edges are preferred directions in figure \ref{fig:U1-bottom}, it is not straightforward to evaluate the amplitude of this diagram. However, it was conjectured in \cite{Iqbal:2007ii} that the refined topological string amplitude is independent of the choice of the preferred direction.\footnote{In terms of a different refinement of topological vertex studied in \cite{Awata:2005fa, Awata:2008ed}, it is expected that this invariance is related to the conjecture given in \cite{Awata:2009yc}.} In particular, the diagram in figure \ref{fig:U1-bottom} is expected to give the same amplitude as that in figure \ref{fig:U1-bottom-2}. We here assume that this conjecture is true, and evaluate the amplitude of the diagram in figure \ref{fig:U1-bottom-2}.
This diagram is essentially equivalent to that in figure \ref{fig:tree} if we set $Y_{n;k} =  Y_{n-1;k}= \emptyset$ and exchange $q$ and $t$. Then, by the same argument as in appendix \ref{app:TN}, we can easily evaluate the amplitude as
\begin{align}
& \prod_{1\leq a< b\leq N}\prod_{i,j=1}^\infty\bigg(1-\Big(\prod_{n=a}^{b-2}R_1^{(n)}P_1^{(n)}\Big)R_1^{(b-1)}t^{i-1/2}q^{j-1/2}\bigg)\bigg(1-\Big(\prod_{n=a}^{b-2}P_1^{(n)}R_{1}^{(n+1)}\Big) P_1^{(b-1)}t^{i-1/2}q^{j-1/2}\bigg)
\nonumber\\
&\times \prod_{1\leq a< b\leq N-1}\prod_{i,j=1}^\infty \bigg(1-\Big(\prod_{n=a}^{b-1}P_1^{(n)}R_{1}^{(n+1)}\Big)t^{i-1}q^{j}\bigg)^{-1}  \prod_{1\leq a< b\leq N}\prod_{i,j=1}^\infty\bigg(1-\Big(\prod_{n=a}^{b-1}R_1^{(n)}P_1^{(n)}\Big)t^{i}q^{j-1}\bigg)^{-1}.
\end{align}
Here only the final product
\begin{eqnarray}
Z_{U(1)}^{||} \equiv \prod_{1\leq a< b\leq N}\prod_{i,j=1}^\infty\bigg(1-\Big(\prod_{n=a}^{b-1}R_1^{(n)}P_1^{(n)}\Big)t^{i}q^{j-1}\bigg)^{-1}
\label{eq:bottom-1}
\end{eqnarray}
is of the form of \eqref{eq:factor-bottom}. The other products are contributions from M2-branes which have some electric charge in 5d gauge theory. We therefore identify \eqref{eq:bottom-1} with the $U(1)$-factor from M2-branes on $\tilde{\beta}_{ab}$.

In the same way, we can also identify the $U(1)$-factor from M2-branes on $\hat{\beta}_{ab}$ with
\begin{eqnarray}
Z_{U(1)}^{/\!/} \equiv \prod_{1\leq a< b\leq N}\prod_{i,j=1}^\infty\bigg(1-\Big(\prod_{n=a}^{b-1} R_{n}^{(n)}Q_{n}^{(n)}\Big)q^{i}t^{j-1}\bigg)^{-1}.
\end{eqnarray}
Therefore the total $U(1)$-factor is written as
\begin{align}
 Z_{U(1)} =& Z_{U(1)}^{=} Z_{U(1)}^{||} Z_{U(1)}^{/\!/} \,.
\end{align}
Now, we conjecture that the 5d Nekrasov partition function of the $T_N$-theory is given by
\begin{eqnarray}
\tilde{Z}_{T_N}/Z_{U(1)}.
\end{eqnarray}
Note here that $Z_{U(1)}^{=}, Z_{U(1)}^{||}$ and $Z_{U(1)}^{/\!/}$ satisfy the rule described in \ref{subsec:U1-factors}.

\section{Low energy partition functions on Higgs vacua
}
\label{sec:Higgs}
As discussed in section \ref{sec:TN}, 
the Higgs branch of the $T_N$-theory is also understood in terms of the 5-brane web-diagrams. An infinite $(p,q)$ 5-brane in the web diagram can be
considered as a semi-infinite 5-brane ending on a orthogonal $(p,q)$ 7-brane, which we denoted by $\otimes$ in the web diagram, and thus the web diagram has $N$ 7-branes at each end of the trivalent legs. The Higgs branch 
opens up when some of the parallel external 5-branes are coincident, which requires some of the mass parameters of the $T_N$-theory to be tuned. In particular, the positions of the 5-branes suspended between 7-branes, together with a part of the gauge field on the 5-branes, parameterize the Higgs branch moduli space. Now let us consider a far infrared region in the Higgs branch, by taking the vev's of the hypermultiples to be very large. We end up with 5-brane diagrams in which some of the 5-branes terminate at the same 7-brane, as in figure \ref{fig:HiggsT4}. The resulting diagram describes another isolated theory $\mathcal{T}_{IR}$\footnote{We denote by $\mathcal{T}_{IR}$ the infrared theory living on a Higgs vacuum.}.

We note that the web diagrams are in general non-toric, as drawn in Figure \ref{fig:Higg} and \ref{fig:HiggsT4}. Therefore we cannot expect that 
the computations of the partition functions using the topological vertex method  give rise to correct results.
However, since the topological string amplitude is independent of the complex structure of the Calabi-Yau threefold, it is expected to be evaluated at the intersection of the Coulomb and Higgs branch. The only subtlety is that there is an extra contribution from hypermultiplets associated with the 5-branes between 7-branes.

\begin{figure}[h]
	\centering
	\includegraphics[scale=0.4]{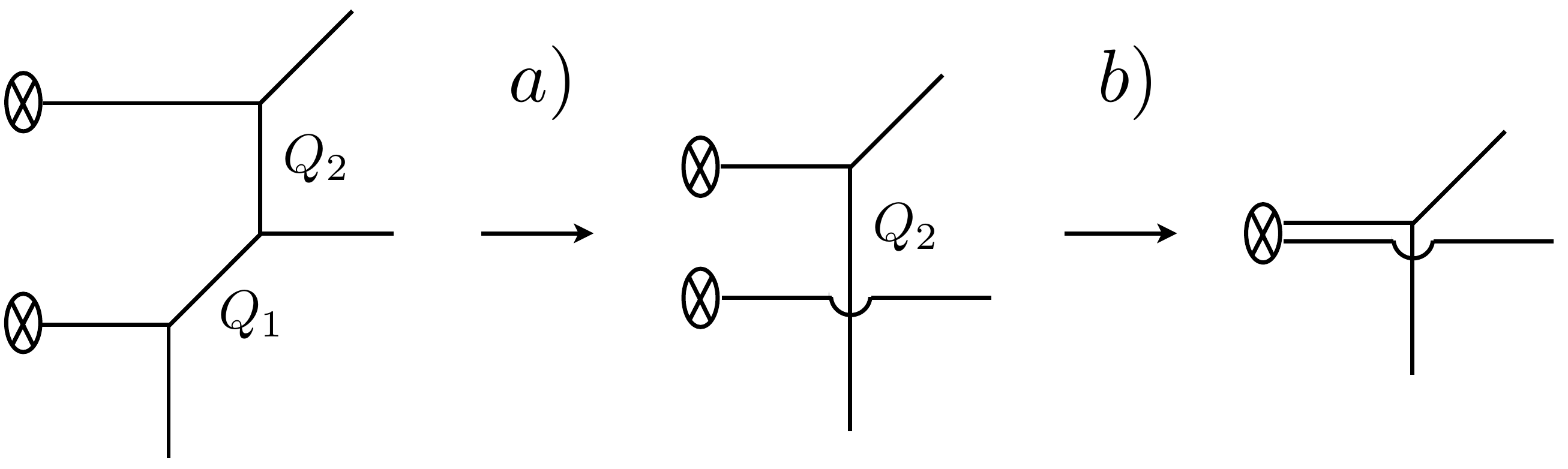}
	\caption{Two step description to move along the Higgs branch in web diagram. $Q_1$ and $Q_2$ represent the K\"ahler parameters for the 2 two-cycles.}
	\label{fig:SimpleHiggs}
\end{figure}

In this section, we shall show a prescription to evaluate the partition function of $\mathcal{T}_{IR}$ 
by taking certain limit of the $T_N$ theory partition function.
The general Higgs branch in web diagram is generated by repeating the two steps as drawn in the Figure \ref{fig:SimpleHiggs}.

The step $(a)$ is related to the geometric transition from closed string geometry on the resolved conifold to open string geometry for Lagrangian branes at the external leg \cite{Dimofte:2010tz,Taki:2010bj, Aganagic:2011sg,Bonelli:2011fq,Bonelli:2011wx}.
The Figure 19 in \cite{Dimofte:2010tz} illustrates the geometric transition of the resolved conifold  with a K\"ahler parameter $Q_1=q^{\frac{1}{2}}t^{-\frac{3}{2}}$\footnote{The Omega deformation parameters in their paper \cite{Dimofte:2010tz} are related as $(p_1,p_2)=(q,t)$.} to the one Lagrangian brane in the open string geometry.
More generally, one can choose the K\"ahler parameter such as
\be
	Q_1=q^{r-\frac{1}{2}}t^{\frac{1}{2}-s} \,,
\ee
where $r,s\ge 1$ and then it corresponds to a general Lagrangian brane on the bottom leg after the transition. This setting can be interpreted in the field theory as the insertion of a surface operator
which has support on the surface
\be
	w_1^{r-1}w_2^{s-1} = 0 \quad \subset \mathbb{R}^4 \,,
\ee
where $w_1$ and $w_2$ are the complex coordinates for $\mathbb{R}^{1,2}$ and $\mathbb{R}^{3,4}$, respectively.
We note that the case in step (a) of Figure \ref{fig:SimpleHiggs} is the open string geometry with no Lagrangian brane, and it corresponds to taking $(r,s)=(1,1)$, namely $Q_1=q^{\frac{1}{2}}t^{-\frac{1}{2}}$.
It would be also interesting to study the surface defects with general choice of $(r,s)$.

The step $(b)$ is achieved by a certain limit of the K\"ahler parameter $Q_2$ as the final diagram has no 2-cycle.
It turns out that we need to choose $Q_2 =q^{\frac{1}{2}}t^{-\frac{1}{2}}$. 

We can understand the above prescription in terms of superconformal index.
In four-dimensions, the index of a class $\mathcal{S}$ theory endowed with a surface operator can be obtained by residue
calculation of the superconformal index of a larger UV theory in which the IR class $\mathcal{S}$ theory is embedded \cite{Gaiotto:2012uq,Gaiotto:2012xa}.
The same method turns out to be applicable to our 5d examples, though we only consider the trivial
surface operator in this paper.
To obtain the the index of $\mathcal{T}_{IR}$, we first start with embedding $\mathcal{T}_{IR}$ to a UV theory.
The index of the UV theory has simple poles in flavor fugacities which arise from bosonic zero modes of the hypermultiplets. 
The residue at a simple pole is related to the index of the infrared theory which lives at the end
of the RG flow triggered by non-zero vev of the scalar zero mode.
The examples in this section correspond to the low energy indices of residues at the pole
\be
	Q_1Q_2 = q/t \,,
\ee
in the UV superconformal indices.
The result usually contains the index of extra free hypermultiplets. We should strip it off to get the correct IR index as the free hypermultiplets are decoupled.
After dividing the extra index from the residue calculation, we can compute the superconformal
index of the theory $\mathcal{T}_{IR}$.
We will relate this IR limit of a superconformal index with that of a Nekrasov partition function below with explicit examples.


\subsection{Free theory from $T_3$ theory}
\label{subsec:HiggsT3}

\begin{figure}[h]
	\centering
	\includegraphics[scale=0.4]{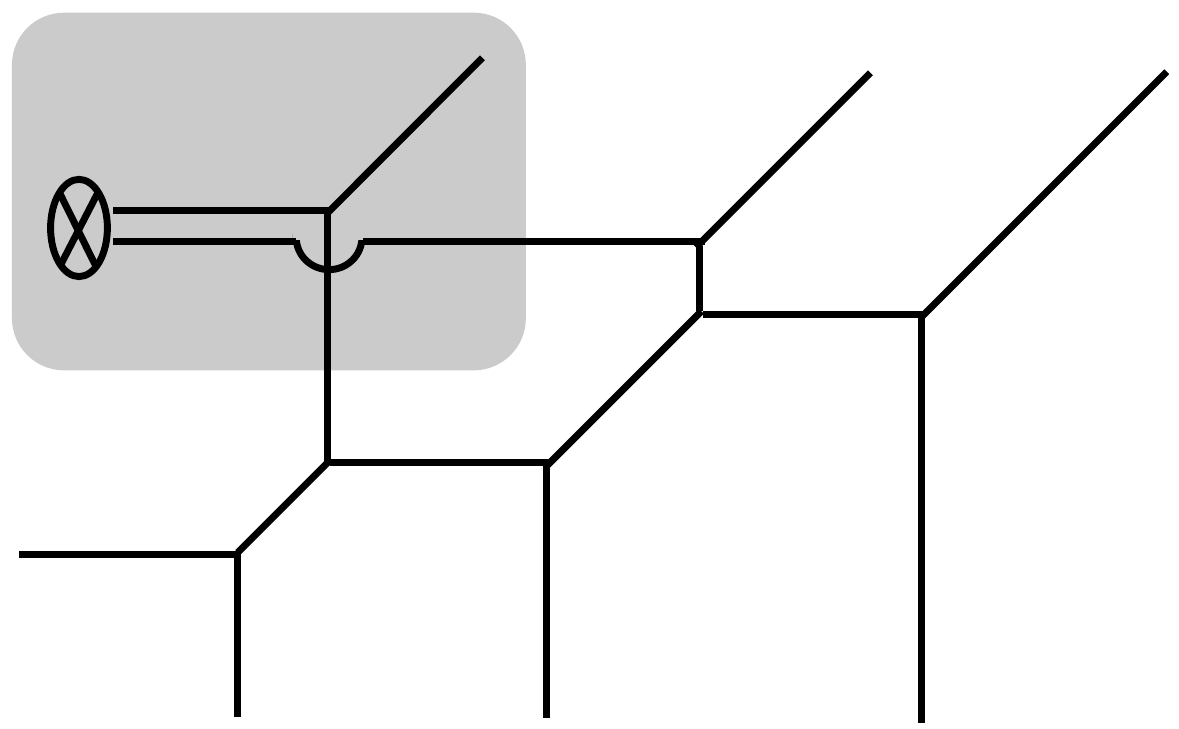}
	\caption{A Higgs vacuum of $T_3$ theory. }
	\label{fig:T3-Higgs}
\end{figure}

The first example is the Higgs vacuum of $T_3$ theory which breaks the $E_6$ global symmetry to $SU(3)\!\times\! SU(3)\!\times\! U(1)$ and gives rise to
the free theory $\mathcal{T}_{IR}$ with 9 hypermultiplets in infrared.
The corresponding web diagram is depicted in Figure \ref{fig:T3-Higgs}.

In the above diagram, we have deformed the shaded part following procedure in Figure \ref{fig:SimpleHiggs}, which breaks $SU(3)$ global symmetry of the three (1,0) 5-branes to  $U(1)$  at the left leg.
Under this deformation, we set the K\"ahler parameters $Q_1,Q_3$ in Figure \ref{fig:T3} such as\footnote{In fact, the tuning 
\be
	Q_1 = Q_3 = t^{\frac{1}{2}}q^{-\frac{1}{2}} 
\ee
also gives the same partition function $Z_{\mathcal{T}_{IR}}$ of the infrared theory $\mathcal{T}_{IR}$ appearing in the Higgsed $T_3$ theory. This is due to the symmetry under the exchange between $m_1$ and $m_3$ in the partition function $Z_{T_3}$ up to two perturbative factors. The different perturbative terms will turn out to be irrelevant extra factors after the tuning, and hence we will get the same partition function $Z_{\mathcal{T}_{IR}}$.}
\be
	Q_1 = Q_3 = q^{\frac{1}{2}}t^{-\frac{1}{2}} \,,
\ee
or equivalently
\be\label{T3-Higgs-parameters}
	m_1 = \lambda+i\gamma_1, \quad m_3 = \lambda -i\gamma_1 \,.
\ee

To obtain the partition function of $\mathcal{T}_{IR}$, we plug these parameters into the $T_3$ partition function (\ref{eq:T3-partition4.2.3}).
Note that, as we are interested in the Higgs branch of $T_3$ theory, we should use the partition function (\ref{eq:T3-partition4.2.3}) which is different from the original partition function (\ref{T3-partition}) by the $U(1)$ factor.
Inserting the parameters, we obtain
\bea \label{T3-Higgs-partition}
	\hspace{-1cm}\tilde{Z}_{\mathcal{T}_{IR}}  \!\!&\!\!=\!\!&\!\! Z_0\cdot Z_{\rm inst} \,, \nn \\
	\hspace{-1cm}Z_0 \!\!&\!\!=\!\!&\!\! \prod_{i,j=1}^\infty \bigg[
	\frac{
	(1-q^it^{j-1})^{\frac{3}{2}}\prod_{a=2,4}
	(1-e^{-i\lambda+im_a}q^{i-\frac{1}{2}}t^{j-\frac{1}{2}})
	(1-e^{-i\lambda-im_a}q^{i-\frac{1}{2}}t^{j-\frac{1}{2}})}
	{(1-q^{i-1}t^{j})^{\frac{1}{2}}} \nn \\
	&&\times (1\!-\!u_1e^{im_4}q^{i}t^{j-1})(1-u_1e^{-im_4}q^{i-1}t^{j})
		(1\!-\!u_2e^{-i\lambda-\frac{i}{2}(m_2+m_4)}q^{i}t^{j-1})(1\!-\!u_2e^{i\lambda-\frac{i}{2}(m_2+m_4)}q^{i-1}t^{j}) \nn \\
	&&\times (1-u_1u_2e^{-i\lambda-\frac{i}{2}(m_2-m_4)}q^{i}t^{j-1})
		(1-u_1u_2e^{i\lambda+\frac{i}{2}(m_2-m_4)}q^{i-1}t^{j}) \bigg] \,, \nn \\
	\hspace{-1cm}Z_{\rm inst} \!\!&\!\!=\!\!&\!\!\sum_{\nu_2,\nu_5}u_2^{|\nu_2|}u_1^{|\nu_5|}
		\prod_{s\in\nu_2}\!
		\frac{(2i\sin\frac{E_{2\emptyset}-m_2+i\gamma_1}{2})
		(2i\sin\frac{E_{2 5}-m_4+i\gamma_1}{2})}
		{(2i)^2\sin\frac{E_{22}}{2}\sin\frac{E_{22}+2i\gamma_1}{2}} \prod_{s\in \nu_5}\!
		\frac{(2i\sin\frac{E_{5\emptyset}-\lambda+m_4+i\gamma_1}{2})(2i\sin\frac{E_{52}+m_4+i\gamma_1}{2})}
		{(2i)^2\sin\frac{E_{55}}{2}\sin\frac{E_{55}+2i\gamma_1}{2}} \nn \\
		\!\!&\!\!=\!\!&\!\!\prod_{i,j=1}^\infty\!\frac{(1\!-\!u_1e^{i\lambda}q^{i-\frac{1}{2}}t^{j-\frac{1}{2}})
		(1\!-\!u_1e^{-i\lambda}q^{i-\frac{1}{2}}t^{j-\frac{1}{2}})
		(1\!-\!u_2e^{\frac{i}{2}(m_2-m_4)}q^{i-\frac{1}{2}}t^{j-\frac{1}{2}})
		(1\!-\!u_2e^{-\frac{i}{2}(m_2-m_4)}q^{i-\frac{1}{2}}t^{j-\frac{1}{2}})}
		{(1\!-\!u_1e^{im_4}q^{i}t^{j-1})(1-u_1e^{-im_4}q^{i-1}t^{j})
		(1\!-\!u_2e^{-i\lambda-\frac{i}{2}(m_2+m_4)}q^{i}t^{j-1})(1\!-\!u_2e^{i\lambda-\frac{i}{2}(m_2+m_4)}q^{i-1}t^{j})} \nn \\
		&&\times\prod_{i,j=1}^\infty\frac{(1-u_1u_2e^{\frac{i}{2}(m_2+m_4)}q^{i-\frac{1}{2}}t^{j-\frac{1}{2}})(1-u_1u_2e^{-\frac{i}{2}(m_2+m_4)}q^{i-\frac{1}{2}}t^{j-\frac{1}{2}})}
		{(1-u_1u_2e^{-i\lambda-\frac{i}{2}(m_2-m_4)}q^{i}t^{j-1})
		(1-u_1u_2e^{i\lambda+\frac{i}{2}(m_2-m_4)}q^{i-1}t^{j})} \,.
\eea
In the instanton part $Z_{\rm inst}$, the Young diagram summation over $\nu_1$ becomes trivial
because of the sine factors from fundamental hypermultiplets in the numerator, namely the sine factor $\prod_{s\in \nu_1}\sin\frac{E_{1\emptyset}(s)-m_1+i\gamma_1}{2}$ always has zero contribution unless $|\nu_1|=0$. From the definition of $E_{\alpha\beta}$ in (\ref{definition-E}), one can easily see that there is alway a position $s\in \nu_\alpha$ where $l_{\nu_\alpha}(s) = a_{\emptyset}(s)+1=0$ and therefore $E_{1\emptyset}(s)-m_1+i\gamma_1=0$ at the $s$.
The 2nd equality for $Z_{\rm inst}$ is verified up to fourth order in the flavor fugacities $u_1,u_2$ with a computer.


Then the partition function for the infrared theory $\mathcal{T}_{IR}$ becomes
\bea
	\tilde{Z}_{\mathcal{T}_{IR}}&=& \frac{q^{-\frac{1}{24}}\eta(q)}{t^{-\frac{1}{24}}\eta(t)}
	\prod_{i,j=1}^\infty(1-q^{i}t^{j-1})^{\frac{1}{2}}(1-q^{i-1}t^{j})^{\frac{1}{2}}
	\cdot\prod_{f=1}^{10}(1-Q_fq^{i-\frac{1}{2}}t^{j-\frac{1}{2}}) \,,\label{infraT3}
\eea
where the 10 K\"ahler parameters are defined as
\be
	\hspace{-0.5cm}Q_f \equiv \left(e^{-i(\mu_{1}+\tilde{\mu}_{a=1,2}+\mu)},\, e^{i(\mu_{a=1,2}+\tilde{\mu}_{3}+\mu)},\ e^{-i(\mu_{2}+\tilde{\mu}_{a=1,2}+\mu)}, \ 
	e^{-i(\mu_{3}+\tilde{\mu}_{1}+\mu)}, \ e^{i(\mu_{3}+\tilde{\mu}_{a=2,3}+\mu)}, \
	e^{-3i\mu}\right)\,,
\ee
and $\eta(q)\!\equiv\!q^{\frac{1}{24}}\prod_{i=1}^\infty(1-q^i)$ is Dedekind eta function.
We here used the the new mass parameters $\mu_{a=1,2,3}, \,\tilde\mu_{a=1,2,3}$ and $\mu$. 
$\mu_{a=1,2,3}$ are associated to the first $SU(3)$ realized by the $(0,1)$ 7-branes attached to the vertical external legs. $\tilde\mu_{a=1,2,3}$ are associated to the second $SU(3)$ realized by the $(1,1)$ 7-branes attached to the diagonal external legs. Lastly, $\mu$ is associated to the $U(1)$ symmetry realized by the $(1,0)$ 7-branes attached to the horizontal external legs. More explicitly, we defined
\be
Q_2=e^{i(\mu_1+\tilde{\mu}_3+\mu)}, Q_4=e^{-i(\mu_2+\tilde{\mu}_2+\mu)}, Q_5= e^{i(\mu_3 + \tilde{\mu}_2 + \mu)}, Q_b = e^{i(-\tilde{\mu}_2 + \tilde{\mu}_3)}, Q_f=e^{i(\mu_1+\tilde{\mu}_3-2\mu)}.
\label{chargeHiggsedT3}
\ee
Note that the charge assignment \eqref{chargeHiggsedT3} after the Higgsing is systematically determined by the requirement that $Q_1$ and $Q_3$ do not have a charge under the remaining global symmetry $SU(3) \times SU(3) \times U(1)$. 

Eq.~\eqref{infraT3} is the partition function of 9 free hypermultiplets with masses $\mu_{1,2,3}+\tilde{\mu}_{1,2,3}+\mu$ and the extra hypermultiplet contribution
\be\label{eq:T3-extra}
	Z_{\rm extra} =\frac{q^{-\frac{1}{24}}\eta(q)}{t^{-\frac{1}{24}}\eta(t)}
	\prod_{i,j=1}^\infty(1-q^{i}t^{j-1})^{\frac{1}{2}}(1-q^{i-1}t^{j})^{\frac{1}{2}}(1-e^{-3i\mu}q^{i-\frac{1}{2}}t^{j-\frac{1}{2}})\,.
\ee
Apart from this extra factor, the global symmetry $SU(3)\!\times\! SU(3)\!\times\! U(1)$ is manifest. 
Therefore the partition function of
$\mathcal{T}_{IR}$ theory defined as
\be 
	Z_{\mathcal{T}_{IR}} = \tilde{Z}_{\mathcal{T}_{IR}}/Z_{\rm extra} 
\ee
produces that of 9 fundamental free hypermultiplets charged under the $SU(3)\!\times\! SU(3)\!\times\! U(1)$ global symmetry.
Ignoring the eta function factors, the $Z_{\rm extra}$ corresponds to the extra free hypermultiplets of the IR theory.
The first term is precisely the inverse of the Nekrasov partition function of a free vector multiplet,
 i.e. $(Z_{{\rm pert}}^{\rm vm})^{-1}$ in (\ref{eq:1-loop-vector}), and the second term is the free hypermultiplet partition function with the $U(1)$ flavor chemical potential $3\mu$.
In the 4d residue computation \cite{Gaiotto:2012uq,Gaiotto:2012xa}, the index of $\mathcal{T}_{IR}$ is expected to have exactly the same
free hypermultiplet factors.\footnote{
	The free HL index for the case obtained in \cite{Gaiotto:2012uq} is $I^{free} = \frac{1}{(1-\tau^2)(1-\tau a)(1-\tau/a)}$ where $a$ denotes the $U(1)$ flavor fugacity.
	The first factor is the same as the inverse of the free vector multiplet index, $\mathcal{I}_V^{-1}$, and the latter two factors are the index of a free hypermultiplet with $U(1)$ charge $\pm1$. We thank Davide Gaiotto for bringing this reference to our attention.
}

The extra hypermultiplet contribution can be also understood from the web diagram. 
Let us consider the extra factors before any cancellation by the contribution of the Cartan part of the vector multiplet. Then, we originally have the following extra factors
\be
\prod_{i,j=1}^{\infty}(1-q^{i}t^{j-1})^2(1-e^{-3i\mu}q^{i-\frac{1}{2}}t^{j-\frac{1}{2}}).
\label{HiggsedU(1)}
\ee
Let us first focus on the contribution of the last factor of \eqref{HiggsedU(1)}. The fugacity $e^{-3i\mu}$ is associated with the K\"ahler parameter $Q_fQ_2^{-1}$. This is originally the K\"ahler parameter of the two-cycle between the top internal horizontal line (to which we assign $\nu_1$ in Figure \ref{fig:T3}) and the bottommost external horizontal line going in the left direction of Figure \ref{fig:T3}. However, after the Higgsing corresponding to the tuning \eqref{T3-Higgs-parameters}, the top internal horizontal line becomes an external horizontal line. Therefore, the K\"ahler parameter $Q_fQ_2^{-1}$ is now assigned to a line between the parallel external legs in the Higgsed $T_3$ diagram of Figure \ref{fig:T3-Higgs}. Then the contribution coming from strings between the parallel external legs going in the left direction only depends on the charge of the $U(1)$ symmetry. This may be also considered as what we call the $U(1)$ factor in the Higgsed diagram. Namely, the extra factor comes from the contribution associated with the new parallel external legs in the Higgsed diagram. The other two contributions in \eqref{HiggsedU(1)} can be thought in the same way. Originally, they are associated with the contributions of M2-branes wrapping the two-cycles with the K\"ahler parameters $Q_1, Q_3$. Hence after the Higgsing, the contributions can be understood as the ones coming from the parallel external legs which are on top of each other.

Let us briefly explain the 5d superconformal index of $\mathcal{T}_{IR}$ from the index of $T_3$ theory.
As explained above, the index of the infrared theory $\mathcal{T}_{IR}$ can be computed by residue calculation of the UV index at poles of flavor fugacities.
The superconformal index of $T_3$ is given in (\ref{eq:T3-superconformalindex}).
The limit (\ref{T3-Higgs-parameters}) for $\mathcal{T}_{IR}$ corresponds to the pole at
\be
	e^{i(m_1-m_3)}=Q_1Q_3 = q/t
\ee
of the $T_3$ index. This pole arises when two simple poles at
\be
	e^{\pm i\lambda} = e^{im_1}(t/q)^{\frac{1}{2}} \,, \quad e^{\pm i\lambda} = e^{im_3}(q/t)^{\frac{1}{2}}
\ee
collide together in the contour integral of the Coulomb branch parameter $\lambda$.  
The residue computation gives
\be
	I_{\mathcal{T}_{IR}} = \big|\tilde{Z}_{\mathcal{T}_{IR}}\big|^2 =
	\big|Z_{\mathcal{T}_{IR}}\big|^2\cdot \big|Z_{\rm extra}\big|^2 \,.
\ee
Note that the 5d superconformal index of the $\mathcal{T}_{IR}$ theory includes the extra hypermultiplet index expected from the 4d $\mathcal{T}_{IR}$ theory: a singlet hypermultiplet
and a $U(1)$ charged hypermultiplet.

\begin{figure}[h]
	\centering
	\includegraphics[scale=0.4]{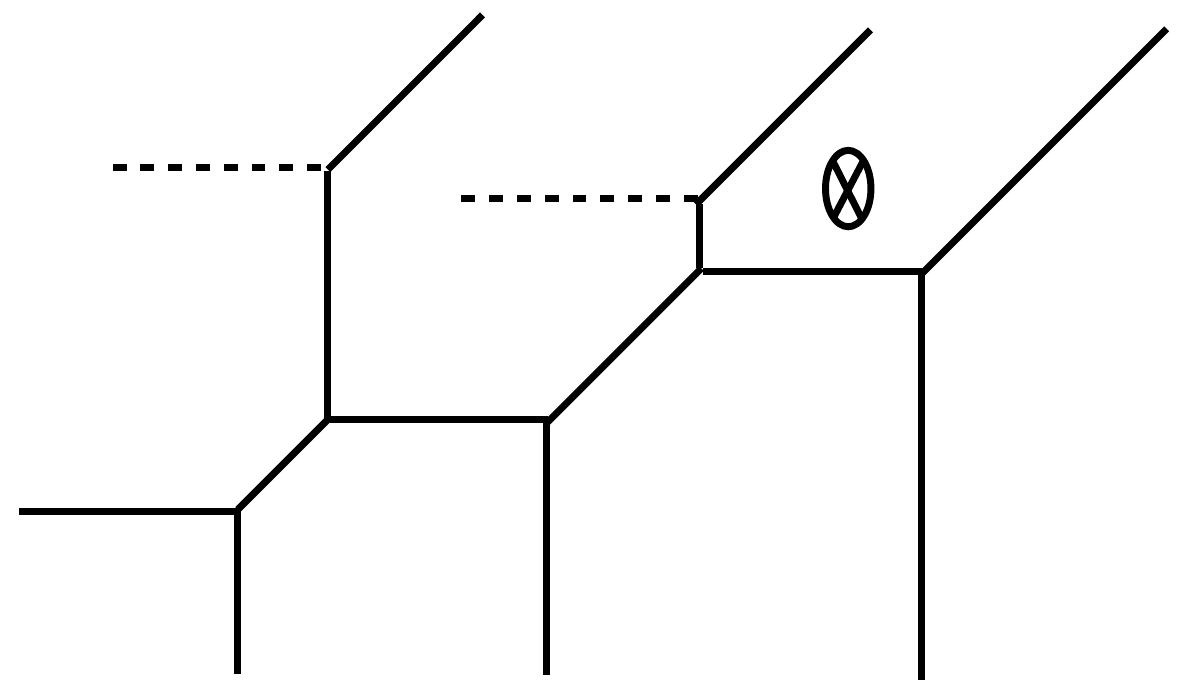}
	\caption{Web diagram for a Higgs vacuum with a D7-brane at the right-center }
	\label{fig:T3-Higgs2}
\end{figure}

We also note that the partition function $\tilde{Z}_{\mathcal{T}_{IR}}$ is almost identical to the 5d partition function of
the $U(1)\!\times\! U(1)$ quiver gauge theory with a fundamental hypermultiplet for the first $U(1)$ and for the second $U(1)$ and a bi-fundmamental hypermultiplet of masses $m_2,-\lambda+m_4,m_4$, respectively.
Two partition functions differ only by the terms in the 2nd and 3rd line of the perturbative part $Z_0$.
In order to understand this non-trivial relation, we apply the generalized s-rule formulated in \cite{Bergman:1998ej,Benini:2009gi} to the Higgs branch web diagram and deform  Figure \ref{fig:T3-Higgs} by dragging the D7-brane horizontally to the center like Figure \ref{fig:T3-Higgs2}.
The dotted line denotes a branch cut of the monodromy \cite{Bergman:1998ej,Benini:2009gi}. As the D7-brane moves across the 5-branes
from left to the center, the
5-branes ending on D7-brane disappear because of the brane creation/annihilation mechanism \cite{Hanany:1996ie} and leaves no D5-brane on it at the final diagram.
Then the final diagram, Figure \ref{fig:T3-Higgs2}, implies that the corresponding gauge theory is the aforementioned  $U(1)\!\times \!U(1)$ quiver gauge theory.
The fundamental hypermultiplet for the second $U(1)$ (for the right-most D5-brane) comes from the string modes connecting the D5-brane and the D7-brane.

\subsection{$E_7$ theory from $T_4$ theory}
\label{subsec:HiggsT4}
We now consider the Higgs branch of $T_4$ theory.
An interesting theory which lives at the end of a Higgs branch is the theory with an $E_7$ global symmetry. The web diagram is drawn in Figure \ref{fig:HiggsT4}.
One can see that in order to get the web diagram we need to apply the Higgsing prescription explained previously twice for upper- and lower-left legs.
In $T_3$ theory, the IR limit for $\mathcal{T}_{IR}$ is achieved by setting the K\"ahler parameters as 
(\ref{T3-Higgs-parameters}). Similarly, for $T_4$ theory, we set the parameters as\footnote{We can also use 
\be
m_1 =\lambda_1-i\gamma_1,\quad m_2 = \lambda_1 +i\gamma_1 ,
\ee
instead of the first two tunings in \eqref{T4-Higgs-parameters}. This is again due to the fact that the $T_4$ partition function $Z_{T_4}$ is symmetric under the exchange between $m_1$ and $m_2$ up to two perturbative factors which eventually become irrelevant extra factors. Similarly, the tunings 
 \be
m_3 = \lambda_3 - i\gamma_1 ,\quad m_4 = \lambda_3 +i\gamma_1
\ee
instead of the last two tunings in \eqref{T4-Higgs-parameters} also give the same partition function $Z_{\mathcal{T}_{IR}}$ of the infrared theory $\mathcal{T}_{IR}$ arising in the Higgsed $T_4$ theory.}
\be\label{T4-Higgs-parameters}
	m_1 =\lambda_1+i\gamma_1,\quad m_2 = \lambda_1 -i\gamma_1 ,\quad m_3 = \lambda_3 +i\gamma_1 ,\quad m_4 = \lambda_3 -i\gamma_1
\ee
to go to the Higgs branch.
The infrared theory $\mathcal{T}_{IR}$ has the global symmetry $SU(4)^2\times SU(2)$ visible in the web diagram and it is believed to enhance to $E_7$ as explained in section \ref{sec:TN}.
Its partition function should respect this symmetry enhancement.
In particular, a 5d $SU(2)$ gauge theory with $N_f=6$ fundamental flavors has the same properties.
The dimensions of Coulomb and Higgs branch of this theory, dim$_{\mathbb{C}}(\mathcal{M}_{\rm Coulomb})\!=\!1$ and dim$_{\mathbb{H}}(\mathcal{M}_{\rm Higgs})\!=\!17$, agrees with those of 
$\mathcal{T}_{IR}$ and its global symmetry $SO(12)$ is also enhanced to $E_7$ at the conformal fixed point \cite{Seiberg:1996bd}.
We expect that two theories are identical and thus the partition function of $\mathcal{T}_{IR}$ reproduces that of the $SU(2)$ theory with $N_f=6$ fundamental hypermultiplets.

Following the Higgsing prescription and by applying the parameters (\ref{T4-Higgs-parameters}) to the $T_4$ partition function, $Z_{T_4}\equiv\tilde{Z}_{T_4}/Z_{U(1)}$, we find
\bea\label{eq:T4-Higgs-partition}
	\hspace{-1cm}\tilde{Z}_{\mathcal{T}_{IR}}\!\!&\!\!=\!\!&\!\! Z_{\rm pert}\cdot Z_{\rm inst}\cdot Z_{\rm extra} \,,\nn \\
	\hspace{-1cm}Z_{\rm pert} \!\!&\!\!=\!\!&\!\! \prod_{i,j=1}^\infty\!\!\bigg\{\!
	\frac{
	\prod_{\alpha'=1}^2\prod_{a=1,3,5,6}(1\!-\!e^{i(\lambda_{\alpha'}'\!-\!\mu_a)}q^{i\!-\!\frac{1}{2}}t^{j\!-\!\frac{1}{2}})}
	{(1-q^{i}t^{j-1})^{\frac{1}{2}}(1-q^{i-1}t^{j})^{\frac{1}{2}}
	(1-e^{-2i\lambda'}q^{i}t^{j-1})(1-e^{-2i\lambda'}q^{i-1}t^{j})}\nn \\
	&&\times (1\!-\!e^{-i(\lambda'\!+\!\mu_4)}q^{i\!-\!\frac{1}{2}}t^{j\!-\!\frac{1}{2}})
	(1\!-\!e^{-i(\lambda'\!-\!\mu_4)}q^{i\!-\!\frac{1}{2}}t^{j\!-\!\frac{1}{2}})\times
	\prod_{\alpha'=1}^2
	(1\!-\!e^{-i(\lambda_{\alpha'}'\!-\!\mu_2)}q^{i\!-\!\frac{1}{2}}t^{j\!-\!\frac{1}{2}})\bigg\}  \,, \nn \\
	\hspace{-1cm}Z_{\rm inst}^{\rm Higgs} \!\!&\!\!=\!\!&\!\!
	\prod_{i,j=1}^\infty \bigg\{
	\frac{(1\!-\!e^{-i(\mu_3+\mu_5)}q^{i\!-\!1}t^{j}) (1\!-\!e^{i(\mu_3-\mu_5)}q^{i}t^{j\!-\!1})(1\!-\!e^{-i(\mu_4+\mu_6)}q^{i\!-\!1}t^{j}) (1\!-\!e^{i(\mu_4-\mu_6)}q^{i}t^{j\!-\!1}) }
	{\prod_{\alpha'=1}^2\prod_{a=5,6}(1\!-\!e^{i(\lambda_{\alpha'}'\!-\!\mu_a)}q^{i\!-\!\frac{1}{2}}t^{j\!-\!\frac{1}{2}})}\nn \\
	\!\!&\!\!&\!\! \times 
	(1\!-\!ue^{\frac{i(-\!\mu_1\!-\!\mu_2\!+\!\mu_3\!+\!\mu_4\!-\!\mu_5\!-\!\mu_6)}{2}}q^{i}t^{j\!-\!1})
	(1\!-\!ue^{\frac{i(-\!\mu_1\!-\!\mu_2\!-\!\mu_3\!+\!\mu_4\!+\!\mu_5\!-\!\mu_6)}{2}}q^{i}t^{j\!-\!1})
	(1\!-\!ue^{\frac{i(-\!\mu_1\!-\!\mu_2\!+\!\mu_3\!-\!\mu_4\!-\!\mu_5\!+\!\mu_6)}{2}}q^{i}t^{j\!-\!1}) \nn \\
	\!\!&\!\!&\!\! \times 
	(1\!-\!ue^{\frac{i(-\!\mu_1\!-\!\mu_2\!-\!\mu_3\!-\!\mu_4\!+\!\mu_5\!+\!\mu_6)}{2}}q^{i}t^{j\!-\!1})
	(1\!-\!ue^{\frac{i(\mu_1\!+\!\mu_2\!-\!\mu_3\!-\!\mu_4\!-\!\mu_5\!-\!\mu_6)}{2}}q^{i\!-\!1}t^{j})
	(1\!-\!ue^{\frac{i(\mu_1\!+\!\mu_2\!+\!\mu_3\!-\!\mu_4\!+\!\mu_5\!-\!\mu_6)}{2}}q^{i\!-\!1}t^{j})\nn \\
	\!\!&\!\!&\!\! \times (1\!-\!ue^{\frac{i(\mu_1\!+\!\mu_2\!-\!\mu_3\!+\!\mu_4\!-\!\mu_5\!+\!\mu_6)}{2}}q^{i\!-\!1}t^{j})
	(1\!-\!ue^{\frac{i(\mu_1\!+\!\mu_2\!+\!\mu_3\!+\!\mu_4\!+\!\mu_5\!+\!\mu_6)}{2}}q^{i\!-\!1}t^{j})\bigg\} \nn \\
	\!\!&\!\!&\!\! \times \!\!\!\sum_{\nu_2,\tilde{\nu}_1,\tilde{\nu}_2,\nu_8}\!\!\!\!u_1^{|\nu_2|}u_2^{|\nu'_1|+|\nu'_2|}u_3^{|\nu_8|}\! \prod_{s\in\nu_2}\!\!\frac{\prod_{\alpha'=1}^22i\sin\frac{E_{\nu_2 \alpha'}+\mu_3+i\gamma_1}{2}}{(2i)^2\sin\frac{E_{\nu_2\nu_2}}{2}\sin\frac{E_{\nu_2\nu_2}+2i\gamma_1}{2}}\prod_{s\in \nu_8}\!\!\frac{\prod_{\alpha'=1}^22i\sin\frac{E_{\nu_8\alpha'} + \mu_4+i\gamma_1}{2}}{(2i)^2\sin\frac{E_{\nu_8\nu_8}}{2}\sin\frac{E_{\nu_8\nu_8}+2i\gamma_1}{2}}\nn \\
	\!\!&\!\!&\!\! \times \!\!\!\prod_{\alpha'=1}^2\prod_{s\in\nu'_{\alpha'}}\!\!\frac{\left(\prod_{a=1}^22i\sin\frac{E_{\alpha' \emptyset} -\mu_a+i\gamma_1}{2}\right)\!\left(2i\sin\frac{E_{\alpha' \nu_2} -\mu_3+i\gamma_1}{2}\right)\!\left(2i\sin\frac{E_{\alpha' \nu_8}- \mu_4+i\gamma_1}{2}\right)}{\prod_{\beta'=1}^2(2i)^2\sin\frac{E_{\alpha'\beta'}}{2}\sin\frac{E_{\alpha'\beta'}+2i\gamma_1}{2}} \,,\nn \\
	Z_{\rm extra} \!\!&\!\!=\!\!&\!\!\frac{t^{-\frac{1}{24}}\eta(t)}{q^{-\frac{1}{24}}\eta(q)}\prod_{i,j=1}^\infty(1\!-\!q^{i-1}t^j)^2(1\!-\!e^{-i(\mu_1-\mu_2)}q^it^{j-1})(1\!-\!e^{-i(\mu_1-\mu_2)}q^{i-1}t^{j})\,,
\eea
with
\be
	\mu_1 = \lambda_1-\tilde{m}_1 \,, \ \ \mu_2 = \lambda_3-\tilde{m}_1, \ \ \mu_3 = \lambda_2-\tilde{m}_1 , \  \ \mu_4 = \tilde{m}_2 , \ \ e^{-i\mu_5} = u_1, \ \ e^{-i\mu_6} = u_3
	, \ \ u = u_2e^{-\frac{i}{2}(\mu_5+\mu_6)},
\ee
and $\lambda_2$ is absent in $E_{\alpha\beta}$'s.
We here introduced new mass parameters $\mu_{a=1,\cdots,6}$ and rearranged the partition function
into three part, $Z_{\rm pert},Z_{\rm inst},Z_{\rm extra}$ for the explicit comparison with 
the gauge theory partition function. The partition function of $\mathcal{T}_{IR}$ is defined 
without the extra factor $Z_{\rm extra}$ by
\be
  Z_{\mathcal{T}_{IR}} = \tilde{Z}_{\mathcal{T}_{IR}}/Z_{\rm extra} \,.
\ee
The extra factor $Z_{\rm extra}$ will be discussed below.

We now compare the partition function of $\mathcal{T}_{IR}$ with the results in \cite{Nekrasov:2004vw,Kim:2012gu}, which were obtained from the field theory using localization computation.
This partition function agrees precisely with the partition function of $SU(2)$ gauge theory with $6$ fundamental flavors. 
One can easily see that the perturbative part $Z_{\rm pert}$ is equivalent to the perturbative part of the field theory result. The denominator comes from the $SU(2)$ vector
multiplet and the numerator comes from the $6$ fundamental hypermultiplets with masses $\mu_{a=1,\cdots,6}$.

The $Z_{\rm inst}$ corresponds to the instanton partition function of the gauge theory. We identify the instanton fugacity $u_2$ of $\tilde{Z}_{\mathcal{T}_{IR}}$ with the instanton fugacity $u$ of the $SU(2)$ theory as $u = u_2e^{\frac{i}{2}(\mu_5+\mu_6)}$.
The infinite product terms in the first 4 lines give rise to the correct normalization so that $Z_{\rm inst}$ counts the only instanton sectors, namely $Z_{\rm inst}=1$ at order $\mathcal{O}(u^0)$.
However the explicit comparison of two partition functions is somewhat sophisticated.
Firstly, 
we note that the $Z_{\rm inst}$ is given by the sum over 4 Young diagrams with
3 fugacities $u_1,u_2,u_3$, while the instanton partition function of the field theory is given
by a sum over 2 Young diagrams with an instantons fugacity $u$ of the $SU(2)$ gauge group. 
This subtlety is similar to what we have encountered for the $T_3$ theory case in section (\ref{subsec:T3}).
Moreover the explicit computations for higher instanton numbers $k>1$ in the field theory are not completely 
done due to technical reasons.

We first compare the 1-instanton partition functions, which corresponds to contributions at order $\mathcal{O}(u^1)$.
We expand both partition functions by the flavor fugacities $u_1=e^{-i\mu_5},u_3=e^{-i\mu_6}$.
It is again obvious that the expansion of the field theory instanton partition function terminates at finite order of $e^{-i\mu_5}$ and $e^{-i\mu_6}$, while
that of $Z_{\rm inst}$ in general goes to infinite order.
However, similarly to the $T_3$ case, we again see that the Young diagram summation
terminates at order $u_1u_3$ and higher order contributions become zero, which is checked up to fourth order in fugacity $u_1,u_3$.
Assuming this observation holds for all order in $u_1,u_3$, we find that the $Z_{\rm inst}$ agrees with the field theory partition function at 1-instanton level.  

For higher instantons at instanton fugacity $u^{k\ge 2}$, we use the superconformal index and check whether
the BPS states captured by the index form representations of the $E_7$ global symmetry.
The superconformal index is given by
\be
	I_{\mathcal{T}_{IR}} = 1+\chi_{\bf 133}^{E_7}x^2+\chi_2(y)\left[1+\chi^{E_7}_{\bf 133}\right]x^3+
	\left[1+\chi_{\bf 7371}^{E_7}+\chi_3(y)(1+\chi^{E_7}_{\bf 133})\right]x^4 + \cdots\,,
\ee
with the branching rules
\bea
	E_7 \!\!&\!\!\supset\!\!&\!\! SO(12)\times U(1) \nn \\
	 {\bf 133}\!\!&\!\!=\!\!&\!\!{\bf 66}_0\!+\!{\bf 32}_1+{\bf 32}_{-1}+{\bf 1}_2+{\bf 1}_0 +{\bf 1}_{-2} \nn \\
	 {\bf 7371}\!\!&\!\!=\!\!&\!\! {\bf 1728}_1\!+\!{\bf 1728}_{-1}\!+\!{\bf 1638}_0 \!+\! {\bf 495}_0 \!+\! {\bf 462}_{2}\!+\!{\bf 462}_0\!+\!{\bf 462}_{-2} \!+\!{\bf 66}_2\!+\!{\bf 66}_0\!+\!{\bf 66}_{-2}\nn \\
	&&\!+{\bf 32}_{3}\!+\!2\times{\bf 32}_{1}\!+\!2\times{\bf 32}_{-1}\!+\!{\bf 32}_{-3}\!+\!{\bf 1}_{4}\!+\!{\bf 1}_{2}\!+\!2\times{\bf 1}_{0}+\!{\bf 1}_{-2}\!+\!{\bf 1}_{-4}\,.
\eea
This is checked up to three instantons with a computer.
The flavor fugacities take the form of the characters of $E_7$ representations.
This provides the strong evidence that the theory $\mathcal{T}_{IR}$ has the enhanced $E_7$
global symmetry and also the partition function we obtained is correct.

The extra factor $Z_{\rm extra}$ in (\ref{eq:T4-Higgs-partition}) is interpreted as
the partition function of the free hypermultiplets in IR theory. 
It follows from the 4d superconformal index computation in \cite{Gaiotto:2012uq,Gaiotto:2012xa} that we have free hypermultiplets in the IR: two singlets and two charged hypermultiplets  under the unbroken $SU(2)$
global symmetry\footnote{The 4d index calculation following \cite{Gaiotto:2012uq} yields the free HL index $\mathcal{I}^{free} = \frac{1}{(1-\tau^2)^2}\prod_{i\neq j}^2\frac{1}{(1-a_i/a_j)(1-\tau^2a_i/a_j)}$ where $a_i$'s are the $SU(2)$ flavor fugacities. The first factor, $\mathcal{I}_V^{-2}$, is from 2 free hypermultiplets and the other factors are from 2 hypermultiplets charged under the $U(1)\subset SU(2)$ with fugacities $\tau a_1/a_2$ and $\tau^{-1}a_1/a_2$.}. The extra factor captures the contribution of the IR free hypermultiplets.
The factor independent of the K\"ahler parameters corresponds to the two singlet free hypermultiplets
and the other two factors correspond to the two charged hypermultiplets with masses $\mu_1\!-\!\mu_2\!\pm\! i\gamma_1$. 

As in the case of the extra hypermultiplet factors which arise in the infrared theory appearing in the Higgs branch of the $T_3$ theory, the extra factors $Z_{\rm extra}$ in \eqref{eq:T4-Higgs-partition} also have an interpretation from the web diagram in Figure \ref{fig:HiggsT4}. Namely, they are related to the contributions of strings between the new parallel external legs going in the left direction in Figure \ref{fig:HiggsT4}. There are two types of such new parallel external legs in the Higgsed $T_4$ diagram in Figure \ref{fig:HiggsT4}. One type is the parallel external legs in Figure \ref{fig:HiggsT4} which originally come from one internal horizontal line and one external horizontal line in the $T_4$ diagram. From the web diagram in Figure \ref{fig:HiggsT4}, we infer the extra factors from the first kind of the parallel external legs as 
\be
\prod^{\infty}_{i,j=1}(1-q^it^{j-1})^4(1-e^{-i(\mu_1-\mu_2)}q^it^{j-1})^2(1-e^{-i(\mu_1-\mu_2)}q^{i-1}t^{j})^2.
\label{U(1)Higgsed1}
\ee
The other type is the parallel external legs in Figure \ref{fig:HiggsT4} which originally come from the two internal horizontal lines in the $T_4$ diagram. The extra factors from the second kind of the parallel external legs can be inferred as 
\be
\prod^{\infty}_{i,j=1}\frac{1}{(1-e^{-i(\mu_1-\mu_2)}q^it^{j-1})(1-e^{-i(\mu_1-\mu_2)}q^{i-1}t^{j})}. \label{U(1)Higgsed2}
\ee
Putting the contributions of \eqref{U(1)Higgsed1} and \eqref{U(1)Higgsed2} together gives 
\be
\prod^{\infty}_{i,j=1}(1-q^it^{j-1})^4(1-e^{-i(\mu_1-\mu_2)}q^it^{j-1})(1-e^{-i(\mu_1-\mu_2)}q^{i-1}t^{j}). \label{U(1)Higgsed}
\ee
One factor of the first four factors in \eqref{U(1)Higgsed} is canceled by the contribution from a part of the Cartan parts of the vector multiplets in the partition function $Z_{T_4}$ of the $T_4$ theory. By adding the remaining Cartan contribution except for the contribution from the Cartan part of the $SU(2)$ vector multiplet , the extra factors precisely agree with $Z_{\rm extra}$ in (\ref{eq:T4-Higgs-partition}).

\begin{figure}[h]
	\centering
	\includegraphics[scale=0.4]{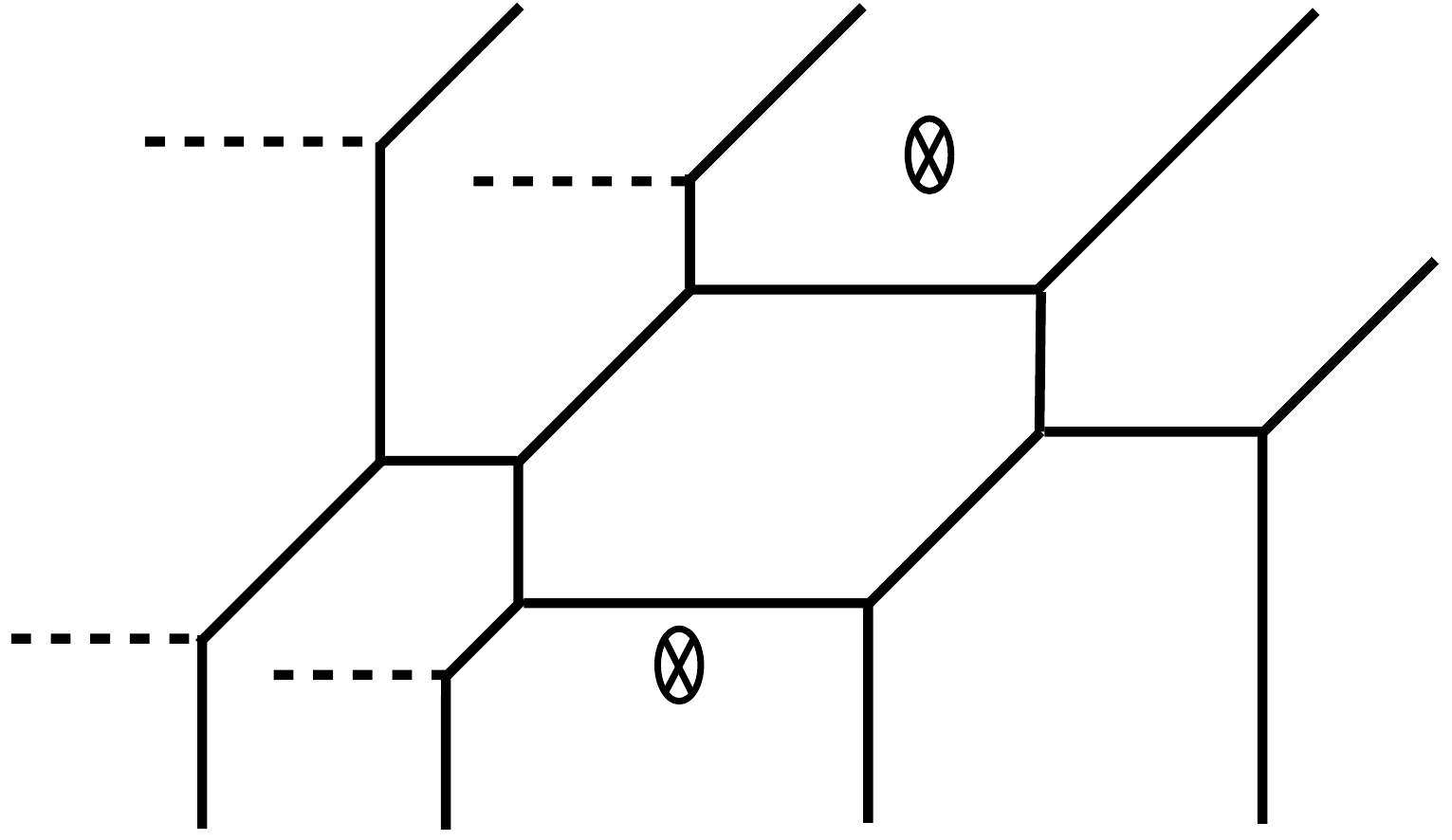}
	\caption{Web diagram of a Higgs vacuum of $T_4$ with 2 D7-branes at the center }
	\label{fig:T4-Higgs2}
\end{figure}

The partition function $Z_{\mathcal{T}_{IR}}$ can also be identified with the partition function of the quiver gauge theory.
We deform the web diagram in Figure \ref{fig:HiggsT4} by moving the D7-branes toward the center and finally get the web diagram in Figre \ref{fig:T4-Higgs2}.
This diagram implies that the corresponding gauge theory is the quiver gauge theory 
of $U(1)\times U(2)\times U(1)$ gauge groups with a bi-fundamental hypermultiplet for each quiver
and two fundamental hypermultiplets of the $U(2)$ gauge group.
We can actually identify the partition function $Z_{\mathcal{T}_{IR}}$ to the partition function
of the quiver theory by a simple relation
\be
	Z_{\mathcal{T}_{IR}} = Z^{U(2)\times U(1)^2} / (Z_{U(1)}^{\parallel}Z_{U(1)}^{/\!/}) \,,
\ee
where $Z_{U(1)}^{\parallel}$ and $Z_{U(1)}^{/\!/}$ are the $U(1)$ factors of $T_4$ theory
at the limit (\ref{T4-Higgs-parameters}) of the Higgs branch.

\subsection{Direct computation from web diagrams}\label{subsec:non-toric}
Let us try to compute the partition functions of $\mathcal{T}_{IR}$ theories using more direct approach.
We will try to use the web diagrams of the Higgs vacua and extract their
partition functions using the refined topological vertex method.
However, 
it is known that the topological vertex formalism can be used to compute the topologcial
string partition function only for toric Calabi-Yau threefolds.
The web diagrams for Higgs vacua are non-toric as depicted in Figure \ref{fig:HiggsT4}
and Figure \ref{fig:T3-Higgs}.
Thus if we apply the topological vertex method directly to the non-toric diagrams, we will in general get wrong results.
In this subsection, we show a prescription to apply the topological vertex formalism to the non-toric diagrams for the Higgs vacua of $T_3$ and $T_4$ theories.

\begin{figure}[h]
	\centering
	\includegraphics[scale=0.4]{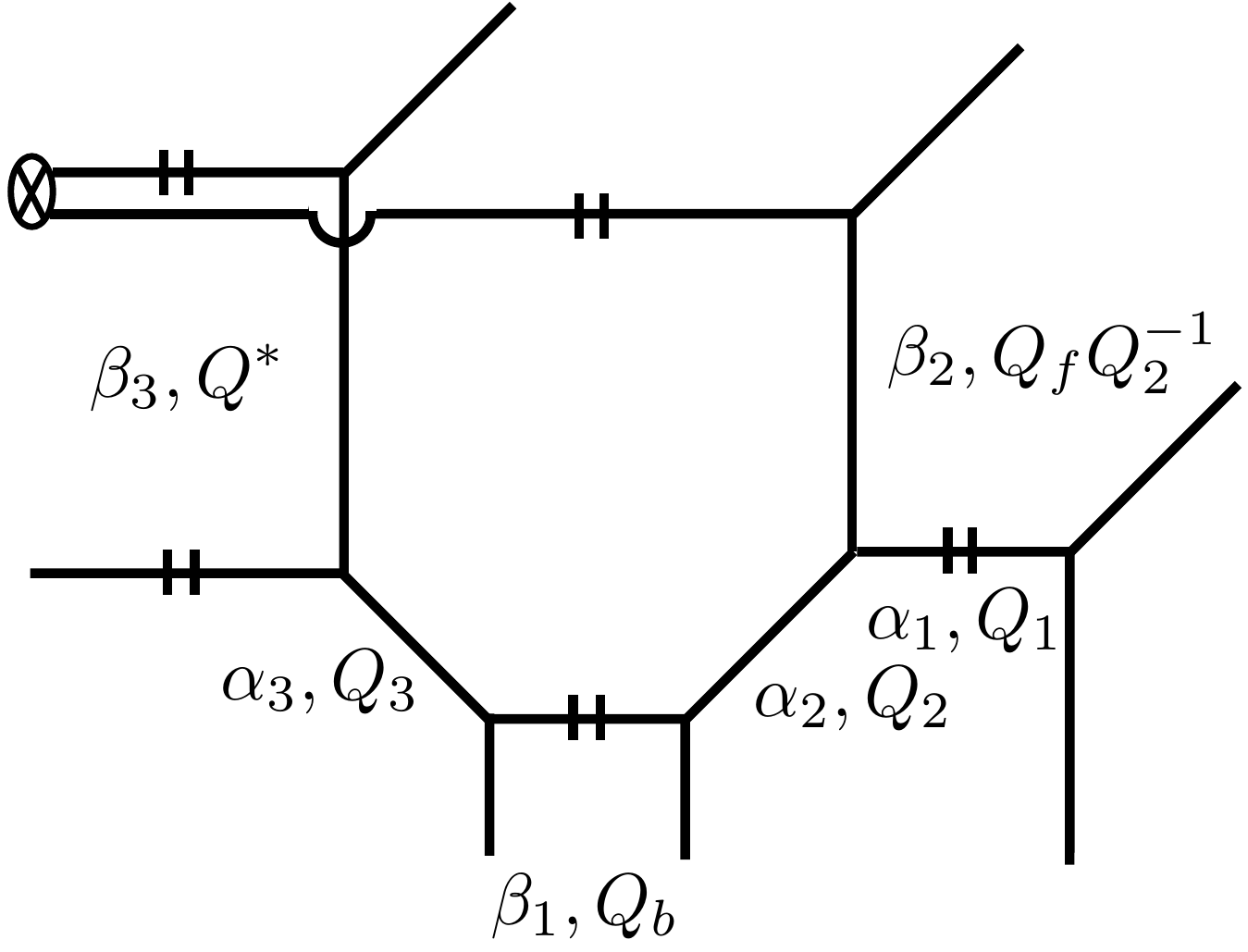}
	\caption{A Higgs vacuum of $T_3$ theory. }
	\label{fig:T3-Higgs3}
\end{figure}

Let us start with the web diagram of a Higgs vacuum of $T_3$ theory drawn in Figure \ref{fig:T3-Higgs3}.

Here we associate an independent K\"ahler parameter to each internal edge.
It makes sense for other internal edges corresponding to independent 2-cycles, but not for the edge
with $Q^*$. 
Since two upper D5-branes are attached to the same D7-brane, the K\"ahler parameter $Q^*$
for the left-vertical edge should be related to the K\"ahler parameter $Q_fQ_3^{-1}$.
We find that the following relation gives rise to the expected result.
\be
	Q^* = Q_fQ_3^{-1}\left(\frac{q}{t}\right)^{\frac{1}{2}} \,.
\ee
The topological string partition function with this relation is given by
\bea\label{eq:T3-Higgs-partition3}
	\hspace{-.5cm}Z \!\!&\!\!=\!\!&\!\! \sum_{\vec\alpha,\vec\beta} (-Q_1)^{|\alpha_1|}(-Q_2)^{|\alpha_2|}(-Q_3)^{|\alpha_3|}(-Q_b)^{|\beta_1|}
				(-Q^*)^{|\beta_3|}(-Q_fQ_2^{-1})^{|\beta_2|} 
				f_{\beta_1^t}(t,q)\tilde{f_{\beta_3^t}}(q,t) \nn \\
		&&\times C_{\emptyset\beta_2\emptyset}(q,t)C_{\alpha_2\beta_2^t\alpha_1^t}(t,q)C_{\emptyset\emptyset\alpha_1}(q,t)C_{\alpha_2^t\emptyset\alpha_3}(q,t)
		C_{\emptyset\beta_3^t\emptyset}(q,t)C_{\beta_3\alpha_3^t\emptyset}(q,t)C_{\emptyset\alpha_3\beta_1^t}(t,q) \nn \\
	\!\!&\!\!=\!\!&\!\!\prod_{i,j=1}^\infty\frac{(1-e^{-i\lambda-im_2}t^{i-\frac{1}{2}}q^{j-\frac{1}{2}})(1-e^{-i\lambda-im_4}t^{i-\frac{1}{2}}q^{j-\frac{1}{2}})(1-e^{-i\lambda+im_4}t^{i-\frac{1}{2}}q^{j-\frac{1}{2}})}
	{(1-e^{-i(\lambda-m_2)}q^{i+\frac{1}{2}}t^{j-\frac{3}{2}})} \nn \\
	&&\times \!\sum_{\alpha_1,\beta_1}u_1^{|\alpha_1|}u_2^{|\beta_1|}
		\!\prod_{s\in\alpha_1}\!\!
		\frac{\sin\!\frac{E_{\alpha_1\emptyset}-\lambda+m_4+i\gamma_1}{2}
		\sin\!\frac{E_{\alpha_1\beta_1}+m_4+i\gamma_1}{2}}
		{\sin\frac{E_{\alpha_1\alpha_1}}{2}\sin\frac{E_{\alpha_1\alpha_1}+2i\gamma_1}{2}} \!\prod_{s\in \beta_1}\!\!
		\frac{\sin\!\frac{E_{\beta_1\emptyset}-m_2+i\gamma_1}{2}\sin\!\frac{E_{\beta_1\alpha_1}-m_4+i\gamma_1}{2}}
		{\sin\frac{E_{\beta_1\beta_1}}{2}\sin\frac{E_{\beta_1\beta_1}+2i\gamma_1}{2}}\,, \nn \\
\eea
where we identified the K\"ahler parameters as
\be
	Q_1 = -u_1e^{i\lambda} \,, \quad Q_2 = e^{-i\lambda-im_4} \,, \quad Q_3 = e^{-i\lambda-im_2}\,, \quad Q_f = e^{-2i\lambda} \,, \quad Q_bQ_2^{\frac{1}{2}}Q_3^{\frac{1}{2}} = u_2
\ee
to compare with (\ref{T3-Higgs-partition}). Here we set $\chi(X) = 0$ since the web diagram has no compact 4-cycle, i.e. no Coulomb branch.
One can  see that the last line in (\ref{eq:T3-Higgs-partition3}) agrees with  the instanton part $Z_{\rm inst}$ in (\ref{T3-Higgs-partition}).
The ratio of two partition functions is then given by
\bea
	\frac{\tilde{Z}_{\mathcal{T}_{IR}}}{Z} \!\!&\!\!=\!\!&\!\! 
	\prod_{i,j=1}^\infty\!\bigg[(1\!-\!e^{-i\lambda+im_2}q^{i-\frac{1}{2}}t^{j-\frac{1}{2}})
	(1\!-\!e^{-i\lambda+im_2}q^{i+\frac{1}{2}}t^{j-\frac{3}{2}})
	(1\!-\!u_1e^{im_4}q^{i}t^{j-1})(1-u_1e^{-im_4}q^{i-1}t^{j}) \nn \\
	&&\times
		(1\!-\!u_2e^{-i\lambda-\frac{i}{2}(m_2+m_4)}q^{i}t^{j-1}) 
		(1\!-\!u_2e^{i\lambda-\frac{i}{2}(m_2+m_4)}q^{i-1}t^{j})\nn \\
	&&\times 
		(1\!-\!u_1u_2e^{-i\lambda-\frac{i}{2}(m_2-m_4)}q^{i}t^{j-1})
		(1\!-\!u_1u_2e^{i\lambda+\frac{i}{2}(m_2-m_4)}q^{i-1}t^{j}) \bigg] \,.
\eea
This ratio is exactly the $U(1)$ factor of $T_3$ theory in (\ref{eq:T3-U(1)factor}) at 
the Higgs vacuum with the parameter setting (\ref{T3-Higgs-parameters}), up to the factor for $e^{im_1-im_3}\!=\!q/t$ which is independent of the K\"ahler parameters.

\begin{figure}[h]
	\centering
	\includegraphics[scale=0.4]{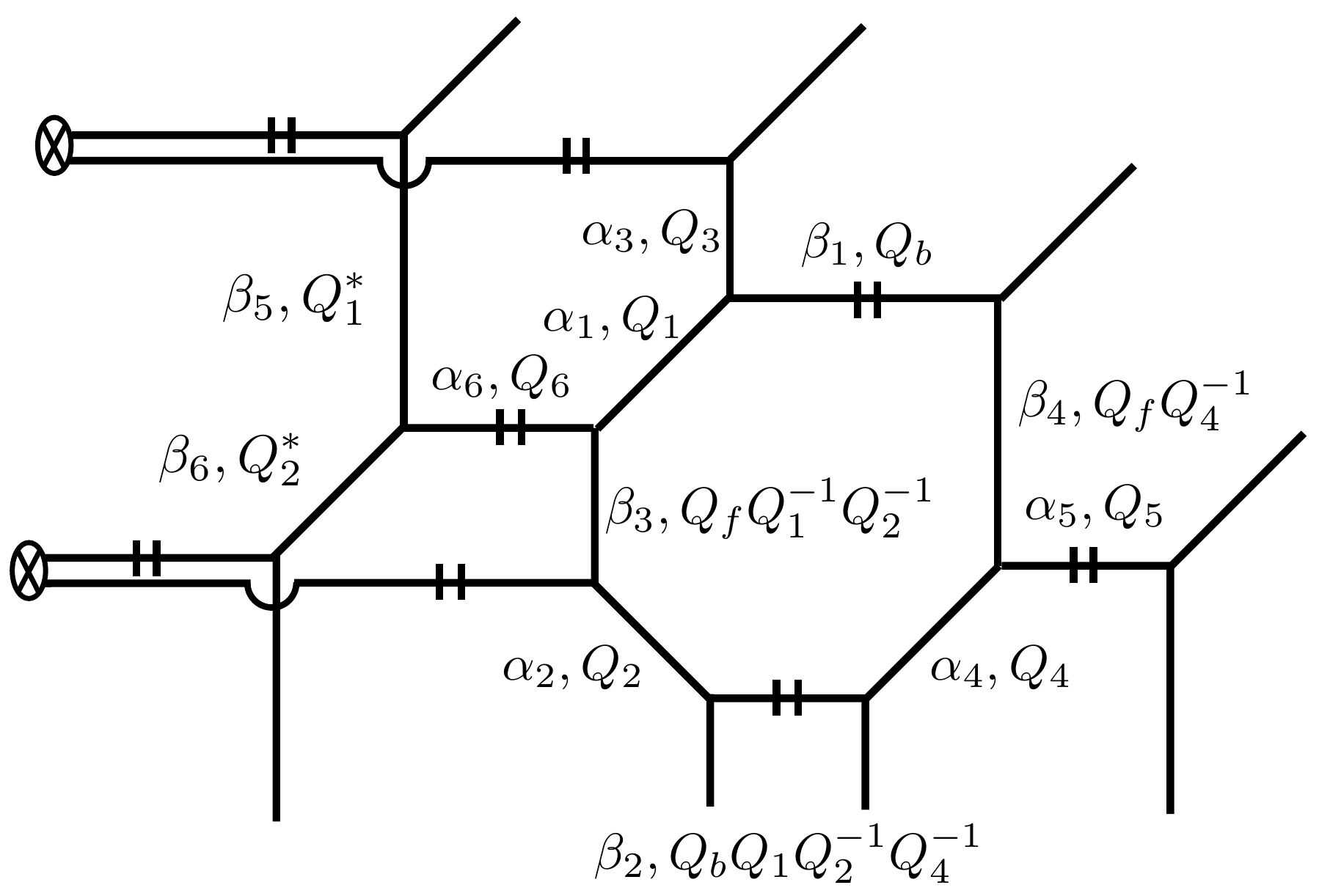}
	\caption{Web diagram of $E_7$ theory. }
	\label{fig:T4-Higgs3}
\end{figure}

Following a similar logic, we now consider the partition function of $E_7$ theory.
The web diagram with K\"ahler parameters is shown in Figure \ref{fig:T4-Higgs3}.
The K\"ahler parameters $Q_1^*$ and $Q_2^*$ are not independent parameters and thus  determined by other K\"ahler parameters.
We identify them as follows:
\be
	Q_1^* = Q_1Q_3\left(\frac{q}{t}\right)^{\frac{1}{2}}
	\,, \quad Q_2^* = Q_fQ_1^{-1}Q_2^{-1}\left(\frac{q}{t}\right)^{\frac{1}{2}} \,.
\ee
Using the topological vertex formalism, we obtain the partition function 
\bea
	\hspace{-1cm}Z\!\!&\!\!=\!\!&\!\!\left(M(t,q)M(q,t)\right)^{\frac{1}{2}}  \nn \\
	\hspace{-1cm}&&\times \sum_{\vec\alpha,\vec\beta} (-Q_1)^{|\alpha_1|}(-Q_2)^{|\alpha_2|}(-Q_3)^{|\alpha_3|}(-Q_4)^{|\alpha_4|}(-Q_5)^{|\alpha_5|}
	(-Q_6)^{|\alpha_6|}(-Q_b)^{|\beta_1|}(-Q_bQ_1Q_2^{-1}Q_4^{-1})^{|\beta_2|}
				\nn \\
	\hspace{-1cm}&& \times 
		(-Q_fQ_1^{-1}Q_2^{-1})^{|\beta_3|}(-Q_fQ_4^{-1})^{|\beta_4|}(-Q_1^*)^{|\beta_5|}(-Q_2^*)^{|\beta_6|}f_{\beta_2^t}(t,q)\tilde{f_{\beta_3^t}}(q,t)
		C_{\emptyset\beta_4\beta_1^t}(q,t)C_{\alpha_4\beta_4^t\alpha_5^t}(t,q) \nn \\
	\hspace{-2cm}	&& \times C_{\emptyset\emptyset\alpha_5}(q,t)C_{\alpha_4^t\emptyset\beta_2}(q,t) C_{\alpha_1^t\alpha_3^t\beta_1}(t,q)C_{\emptyset\alpha_3\emptyset}(q,t)C_{\alpha_1\beta_3^t\alpha_6^t}(q,t)C_{\beta_3\alpha_2^t\emptyset}(q,t)C_{\emptyset\alpha_2\beta_2^t}(t,q)\nn\\
	\hspace{-1cm}\!\!&\!\!=\!\!&\!\! \prod_{i,j=1}^\infty\!\!\frac{\!\!\left\{\!\prod_{a=2,3,4}(1\!-\!e^{-i\lambda+i\mu_a}q^{i-\frac{1}{2}}t^{j-\frac{1}{2}})(1\!-\!e^{-i\lambda-i\mu_a}q^{i-\frac{1}{2}}t^{j-\frac{1}{2}})\!\right\}\!
	(1\!-\!e^{-i\lambda-i\mu_1}q^{i-\frac{1}{2}}t^{j-\frac{1}{2}})
	(1\!-\!e^{i\lambda-i\mu_1}q^{i-\frac{1}{2}}t^{j-\frac{1}{2}})}
	{(1\!-\!q^it^{j-1})^{\!\frac{1}{2}}(1\!-\!q^{i-1}t^j)^{\!\frac{1}{2}}(1\!-\!e^{-2i\lambda}q^it^{j-1})\!(1\!-\!e^{-2i\lambda}q^{i-1}t^j)\!
	(1\!-\!e^{-i(\mu_1-\mu_2)}q^it^{j-1})\!(1\!-\!e^{-i(\mu_1-\mu_2)}q^{i+1}t^{j-2})} \nn \\
	\hspace{-1cm}\!\!&\!\!&\!\! \times \!\!\!\sum_{\nu_2,\tilde{\nu}_1,\tilde{\nu}_2,\nu_8}\!\!\!\!u_1^{|\nu_2|}u_2^{|\nu'_1|+|\nu'_2|}u_3^{|\nu_8|}\! \prod_{s\in\nu_2}\!\!\frac{\prod_{\alpha'=1}^22i\sin\frac{E_{\nu_2 \alpha'}+\mu_3+i\gamma_1}{2}}{(2i)^2\sin\frac{E_{\nu_2\nu_2}}{2}\sin\frac{E_{\nu_2\nu_2}+2i\gamma_1}{2}}\prod_{s\in \nu_8}\!\!\frac{\prod_{\alpha'=1}^22i\sin\frac{E_{\nu_8\alpha'} + \mu_4+i\gamma_1}{2}}{(2i)^2\sin\frac{E_{\nu_8\nu_8}}{2}\sin\frac{E_{\nu_8\nu_8}+2i\gamma_1}{2}}\nn \\
	\hspace{-1cm}\!\!&\!\!&\!\! \times \prod_{\alpha'=1}^2\prod_{s\in \nu'_{\alpha'}}\!\!\frac{\left(\prod_{a=1}^22i\sin\frac{E_{\alpha' \emptyset} -\mu_a+i\gamma_1}{2}\right)\!\left(2i\sin\frac{E_{\alpha' \nu_2} -\mu_3+i\gamma_1}{2}\right)\!\left(2i\sin\frac{E_{\alpha' \nu_8}- \mu_4+i\gamma_1}{2}\right)}{\prod_{\beta'=1}^2(2i)^2\sin\frac{E_{\alpha'\beta'}}{2}\sin\frac{E_{\alpha'\beta'}+2i\gamma_1}{2}}\,,
\eea
where 
\bea
	&&Q_f=e^{-2i\lambda} \,, \quad Q_1 = e^{-i\lambda+i\mu_3}\,, 
	\quad Q_2 = e^{-i\lambda-i\mu_2} \,, \quad Q_3 = e^{i\lambda-i\mu_1} \,, \quad Q_4 =e^{-i\lambda-i\mu_4} \nn \\
	&& Q_bQ_1^{\frac{1}{2}}Q_2^{-\frac{1}{2}}Q_3^{\frac{1}{2}}Q_4^{-\frac{1}{2}} = u_2
	\,, \quad Q_5 = -u_3e^{i\lambda}\,, \quad Q_6 = -u_1e^{i\lambda}\,,
\eea
and $\lambda_{1}' = -\lambda_2'=\lambda\,, \ \lambda_2=\lambda_8=0$.

One can compare this partition function with  $E_7$ partition function in (\ref{eq:T4-Higgs-partition}). The ratio of two partition functions is given by
\bea
	\hspace{-.5cm}\frac{\tilde{Z}_{\mathcal{T}_{IR}}}{Z} \!\!&\!\!=\!\!&\!\! \prod_{i,j=1}^\infty(1-q^{i-1}t^j)^2(1-e^{-i(\mu_1-\mu_2)}q^it^{j-1})^2(1-e^{-i(\mu_1-\mu_2)}q^{i-1}t^j)(1-e^{-i(\mu_1-\mu_2)}q^{i+1}t^{j-2}) \nn \\
	\!\!&\!\!&\!\!\times(1\!-\!e^{-i(\mu_3+\mu_5)}q^{i\!-\!1}t^{j}) (1\!-\!e^{i(\mu_3-\mu_5)}q^{i}t^{j\!-\!1})(1\!-\!e^{-i(\mu_4+\mu_6)}q^{i\!-\!1}t^{j}) (1\!-\!e^{i(\mu_4-\mu_6)}q^{i}t^{j\!-\!1}) \nn \\
	\!\!&\!\!&\!\!\times(1\!-\!ue^{\frac{i(-\!\mu_1\!-\!\mu_2\!+\!\mu_3\!+\!\mu_4\!-\!\mu_5\!-\!\mu_6)}{2}}q^{i}t^{j\!-\!1})
	(1\!-\!ue^{\frac{i(-\!\mu_1\!-\!\mu_2\!-\!\mu_3\!+\!\mu_4\!+\!\mu_5\!-\!\mu_6)}{2}}q^{i}t^{j\!-\!1})
	(1\!-\!ue^{\frac{i(-\!\mu_1\!-\!\mu_2\!+\!\mu_3\!-\!\mu_4\!-\!\mu_5\!+\!\mu_6)}{2}}q^{i}t^{j\!-\!1}) \nn \\
	\!\!&\!\!&\!\! \times 
	(1\!-\!ue^{\frac{i(-\!\mu_1\!-\!\mu_2\!-\!\mu_3\!-\!\mu_4\!+\!\mu_5\!+\!\mu_6)}{2}}q^{i}t^{j\!-\!1})
	(1\!-\!ue^{\frac{i(\mu_1\!+\!\mu_2\!-\!\mu_3\!-\!\mu_4\!-\!\mu_5\!-\!\mu_6)}{2}}q^{i\!-\!1}t^{j})
	(1\!-\!ue^{\frac{i(\mu_1\!+\!\mu_2\!+\!\mu_3\!-\!\mu_4\!+\!\mu_5\!-\!\mu_6)}{2}}q^{i\!-\!1}t^{j})\nn \\
	\!\!&\!\!&\!\! \times (1\!-\!ue^{\frac{i(\mu_1\!+\!\mu_2\!-\!\mu_3\!+\!\mu_4\!-\!\mu_5\!+\!\mu_6)}{2}}q^{i\!-\!1}t^{j})
	(1\!-\!ue^{\frac{i(\mu_1\!+\!\mu_2\!+\!\mu_3\!+\!\mu_4\!+\!\mu_5\!+\!\mu_6)}{2}}q^{i\!-\!1}t^{j}) \,,
\eea
after omitting the divergent factors like (\ref{eq:divergent-factor}).
The ratio is the same as the $U(1)$ factor of the $T_4$ theory at the Higgs vacuum of
(\ref{T4-Higgs-parameters}).

\section{Conclusion}
In this paper, we have shown how to compute the Nekrasov partition functions of five-dimensional $T_N$ theories using the refined topological vertex formalism on the toric Calabi-Yau threefolds of the resolved $\mathbb{C}^3/(\mathbb{Z}_N\!\times\! \mathbb{Z}_N)$ proposed in \cite{Benini:2009gi}. For that, we have identified the contribution from decoupled M2-branes to the topological string amplitude, which also enables us to evaluate the Nekrasov partition function of an $SU(N)$ gauge group rather than $U(N)$.
We have also shown that the
partition functions of the low energy theories at a Higgs vacuum of $T_N$ theories can be obtained by taking certain limits of the $T_N$ partition functions.

In principle, our method can be used to compute the exact partition functions of the theories from any web diagrams and also the theories realized as the low energy descriptions of their Higgs branches. Furthermore, our expressions of the partition functions are technically tractable since they do not involve any contour integrals such as the $Sp(1)$ Nekrasov partition function obtained from the field theory technique in \cite{Marino:2004cn, Nekrasov:2004vw, Kim:2012gu}. It would be also interesting to perform further checks of our proposal from a field theory analysis, for example for theories which have Lagrangian descriptions such as $SU(N)$ gauge theories.

The Higgs branch results we obtained in this paper could be a prototype for the refined topological vertex on non-toric web diagrams, which may be also related to non-toric Calabi-Yau threefolds.
The results provided non-trivial checks that the IR theory at a particular Higgs vacuum of $T_4$ theory is dual to the $Sp(1)$ gauge theory with 6 flavors and has the $E_7$ global symmetry.
It is possible to extend the Higgsing technique to other $T_N$ theories in order to get the
partition functions of the $E_8$ theory or higher rank $E_{6,7,8}$ theories which are constructed by non-toric web diagrams. 
Relatedly, an attractive future direction would be to develop a systematic way to apply the topological
vertex formalism directly to non-toric diagrams. 
A primitive version of this work was done in section \ref{subsec:non-toric}.

When compactified on $S^1$, the class of 5d theories we are considering reduces to the 6d $(2,0)$ theory compactified on a sphere with three regular punctures. For example, the $T_N$ theory is associated with a sphere with three maximal punctures \cite{Gaiotto:2009we}. The example of subsection \ref{subsec:HiggsT3} is associated with a shpere with one minimal and two maximal punctures. All such theories are classified in \cite{Chacaltana:2010ks}. It would be interesting to evaluate the 5d Nekrasov partition function of such theories, by using our Higgsing method.

We also have found that the Nekrasov partition function for $T_N$ theory is exactly the same as
that of a quiver gauge theory, up to decoupled free factors. This result may imply that two theories are somehow related in five dimensions.
It would be interesting to work out this relation clearly from the field theory view point.
The decoupled $U(1)$ factor may play an important role. 

The four dimensional reduction of the partition functions we computed can be considered along two different directions, i.e. ``time'' $S^1$ direction or large $n$ limit of orbifold $\mathbb{R}^4/Z_n$ (or $S^4/\mathbb{Z}_n$ for superconformal index), but it turns out that both are subtle.
Note that, in 5d, the instanton charge plays as an extra flavor charge and its fugacity should not be
identified with the gauge coupling of 4d theory because $T_N$ theories have no marginal coupling.
Under the reduction along the $S^1$ circle\footnote{To take this limit, we first restore the $S^1$ radius $\beta$ in the partition function such as $u=e^{i\beta m}$ where $u$ is the flavor fugacity and take $\beta\rightarrow0$ limit.}, the instanton expansion by the fugacity $u$ diverges since $u\rightarrow1$.
On the other hand, the reduction by large orbifold limit introduced in \cite{Mekareeya:2013ija} appears to cause a partial flavor symmetry breaking in 4d\footnote{In \cite{Mekareeya:2013ija}, the instanton contribution to the partition function is suppressed and thus instanton states are dropped out in 4d partition function.}. 
To resolve the second subtlety, we may need to carefully take into account the effect of the orbifold singularity.

Finally, our studies may provide new insights into the 5d extension of AGT conjecture \cite{Alday:2009aq,Wyllard:2009hg,Awata:2009ur,Awata:2010yy,Schiappa:2009cc,Mironov:2011dk,Nieri:2013yra,Tan:2013tq,Tan:2013xba}. 
The 5d $T_N$ partition function allows us to study the three-point correlator of $q$-deformed $\mathcal{W}_N$ algebra though its 4d counterpart is not known yet. 
For the generic correlators, we can consider gluing the external legs in the web diagram
and compute the corresponding partition functions.


\subsection*{Acknowledgements}

\noindent
{We would like to thank Jaume Gomis, Amer Iqbal, Katsushi Ito, Christoph Keller, Seok Kim, Jaewon Song and Yuji Tachikawa for useful and enlightening discussions, and especially Davide Gaiotto for insightful discussions and comments on the manuscript.
The work of HH is supported by the REA grant agreement PCIG10-GA-2011-304023 from the People Programme of FP7 (Marie Curie Action), the grant FPA2012-32828 from the MINECO, the ERC Advanced Grant SPLE under contract ERC-2012-ADG-20120216-320421 and the grant SEV-2012-0249 of the ``Centro de Excelencia Severo Ochoa'' Programme. The work of HK was supported by the Perimeter Institute for Theoretical Physics. Research at Perimeter Institute is supported by the Government of Canada through Industry Canada and by the Province of Ontario through the Ministry of Research and Innovation.
The work of TN is supported by the U.S. Department of Energy under grant DE-FG02-96ER40959.}

\newpage

\centerline{\Large \bf Appendix}

\appendix

\section{Instanton partition function}
\label{app:Nekrasov}

In this appendix, we briefly summarize the formulae for the vector and various hypermultiplet contributions to the instanton partition function of $U(N_l)\!\times\! U(N_n)$ quiver gauge theory.
The Nekrasov instanton partition function \cite{Nekrasov:2002qd,Nekrasov:2003rj,Bruzzo:2002xf} takes the form of
\be
  Z_{\rm inst} = \sum_{\vec{Y}_l, \vec{Y}_n}u_l^{|\vec{Y}_l|}u_n^{|\vec{Y}_n|}Z(\vec{Y}_l,\vec{Y}_n) \,,
\ee
where $u_l,u_n$ denote the instanton fugacities for the gauge groups.
The summation runs over all possible Young diagram configurations for the colored Young diagrams $\vec{Y}_l$ and $\vec{Y}_n$. $|\vec{Y}_l|$ denotes the total number of boxes in $\vec{Y}_l$.
The function $Z(\vec{Y}_l,\vec{Y}_n)$ is given by the product of all the matter contributions.
The contributions of the hypermultiplets in the fundamental and the anti-fundamental representation of $U(N_l)$ gauge group are given by
\bea
  z_{\rm fund}(l;m) \!\!&\!\!=\!\!&\!\! 
  \prod_{\alpha=1}^{N_l}\prod_{s\in Y_{l,\alpha}}\left[2i\sin\frac{E(l,\emptyset,\alpha,\emptyset)-m+i\gamma_1}{2}\right] \,,\nn \\
  z_{\rm anti-fund}(l;m) \!\!&\!\!=\!\!&\!\! 
  \prod_{\alpha=1}^{N_l}\prod_{s\in Y_{l,\alpha}}\left[2i\sin\frac{-E(l,\emptyset,\alpha,\emptyset)-m-i\gamma_1}{2}\right] \,,
\eea
where $m$ is the mass parameter (or $U(1)$ flavor chemical potential) for
hypermultiplet and $s$ denotes the $(i,j)$ position of the corresponding Young diagram.
The bi-fundamental hypermultiplet of $U(N_l) \!\times\! U(N_n)$ gauge group contributes to the partition function by
\be
	\hspace{-.5cm}z_{\rm bifund}(l,n;m)\!=\!
	\prod_{\alpha_l=1}^{N_l}\prod_{\beta_n=1}^{N_n}\prod_{s\in Y_{l,\alpha_l}}
	\!\!\!\left[2i\sin\!\frac{E(l,n,\alpha_l,\beta_n,s)\!-\!m\!+\!i\gamma_1}{2}\right]
	\!\!\prod_{\tilde{s}\in Y_{n,\beta_n}}\!\!\!\left[2i\sin\!\frac{E(n,l,\beta_n,\alpha_l,\tilde{s})\!+\!m\!+\!i\gamma_1}{2}\right] \,.
\label{eq:component}
\ee
The function $E$ is defined by
\be
	E(l,n,\alpha_l,\beta_n,s) = \lambda_{l,\alpha_l}-\lambda_{n,\beta_n}
	+i(\gamma_1+\gamma_2)\ell_{Y_{l,\alpha_l}}(s) -i(\gamma_1-\gamma_2)(a_{Y_{n,\beta_n}}(s)+1) \,,
\ee
where $\lambda_{l,\alpha}$ is the Coulomb branch parameter for $U(N_l)$ gauge group.

The vector multiplet contribution is simply given by 
\begin{align}
z_{\rm vec}(l) = \frac{1}{z_{\rm bifund}(l,l;+i\gamma_1)} \,. \nn
\end{align}

\section{The topological string partition function on $\mathbb{C}^3/({\mathbb{Z}_N\times \mathbb{Z}_N})$}
\label{app:TN}

\begin{figure}
\centering
\includegraphics[width=6.5cm]{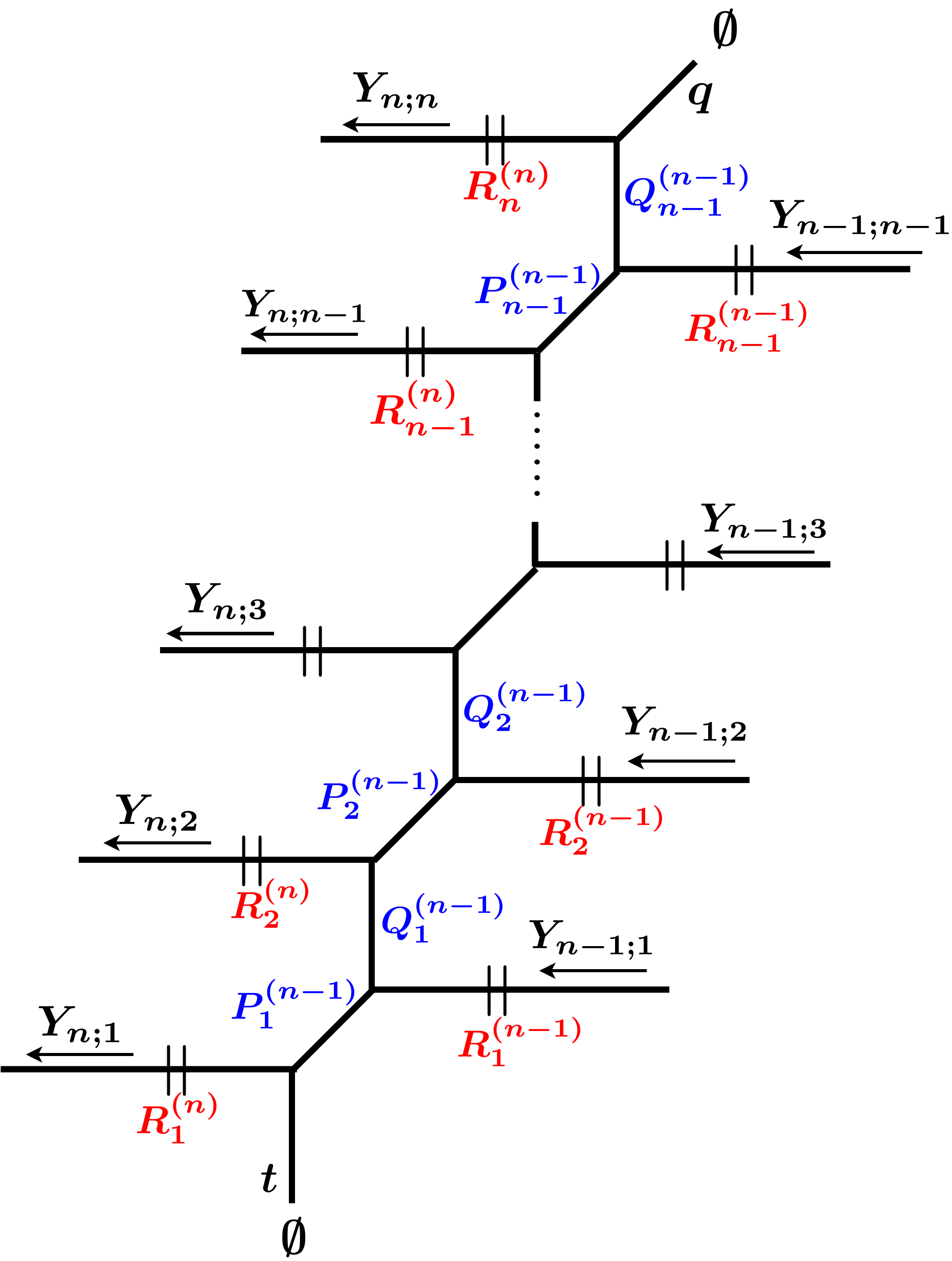}
\caption{The component diagram which gives the factor $A_n(\vec{Y}_n,\vec{Y}_{n-1})$.}
\label{fig:tree}
\end{figure}
We here give a explicit proof of the expression \eqref{eq:Nekrasov-quiver} for the refined topological string partition function on the resolved $\mathbb{C}^3/(\mathbb{Z}_N\times \mathbb{Z}_N)$. The relevant toric web-diagram is shown in figure \ref{fig:TN-diagram}.
From the toric data, it follows that  $P_{k}^{(n)}Q_{k}^{(n)} = Q_{k}^{(n+1)}P_{k+1}^{(n+1)}$ and $R_a^{(n)} = R_{1}^{(n)}(P_{a-1}^{(n-1)}P_{a-2}^{(n-1)}\cdots P_{1}^{(n-1)})(P_a^{(n)}P_{a-1}^{(n)}\cdots P_{2}^{(n)})^{-1},$. Note that all the framing factors are trivial here. 

The refined topological string partition function is given by 
\begin{eqnarray}
Z_{\rm ref} = (M(t,q)M(q,t))^{\frac{\chi(X)}{4}}\times Z \,,
\end{eqnarray}
where $\chi(X)$ is the Euler characteristics of the Calabi-Yau threefold, and $Z$ is evaluated via the refined topological vertex. An essential technique to compute $Z$ is the refinement of \cite{Iqbal:2004ne}.\footnote{For a similar computation for $U(N)$ linear quivers, see \cite{Bao:2011rc}.} The toric web-diagram in figure \ref{fig:TN-diagram} implies that
\begin{eqnarray}
Z = \sum_{\vec{Y}_1,\cdots,\vec{Y}_{N-1}} A_{N}(\emptyset,\vec{Y}_{N-1})\prod_{n=1}^{N-1}\bigg( B_{n+1,n}(\vec{Y}_{n})A_{n}(\vec{Y}_{n},\vec{Y}_{n-1})\bigg) \,,
\label{eq:TN-vertex}
\end{eqnarray}
where
\begin{eqnarray}
B_{n+1,n}(\vec{Y}_{n}) = \prod_{a=1}^{n}(-R_i^{(n)})^{|Y_{n,a}|}\,.
\end{eqnarray}
and $\vec{Y}_n = (Y_{n;1},\cdots, Y_{n;n})$ is the set of $n$ Young diagrams.
The factor $A_{n}(\vec{Y}_n,\vec{Y}_{n-1})$ is the contribution from the diagram in figure \ref{fig:tree}, and evaluated as
\begin{eqnarray}
A_{n}(\vec{Y}_n, \vec{Y}_{n-1})= \sum_{\vec{\mu},\vec{\rho}}C_{\mu_1\emptyset Y_{n;1}}(q,t)\prod_{i=1}^{n-1}\bigg[C_{\mu_i^t\rho_i Y_{n-1;i}^{t}}(t,q)\, C_{\mu_{i+1}\rho_{i}^t Y_{n;i+1}}(q,t)\; \big(-P_{i}^{(n-1)}\big)^{|\mu_i|}\big(-Q_{i}^{(n-1)}\big)^{|\rho_i|}\bigg].
\nonumber\\
\end{eqnarray}
Here $\vec{\mu}=(\mu_1,\cdots,\mu_{n-1})$ and $\vec{\rho}=(\rho_1,\cdots,\rho_{n-1})$ are two sets of Young diagrams.  Note that we implicitly set $\vec{Y}_{0} = \emptyset$.
To be more explicit, we have
\begin{eqnarray}
&&A_{n}(\vec{Y}_n, \vec{Y}_{n-1}) \;=\; t^{\frac{||Y_{n;1}||^2}{2}}\tilde{Z}_{Y_{n;1}}(q,t)\prod_{i=1}^{n-1} t^{\frac{||Y_{n;i+1}||^2}{2}}\widetilde{Z}_{Y_{n;i+1}}(q,t)q^{\frac{||Y_{n-1;i}^{t}||^2}{2}}\widetilde{Z}_{Y_{n-1;i}^{t}}(t,q)
\nonumber\\
&&\qquad \times \sum_{\eta_i,\xi_i;\xi_0=\xi_{n-1}=\emptyset;}\prod_{i=1}^{n-1} \left(\frac{q}{t}\right)^{\frac{|\eta_i|-|\xi_i|}{2}} \sum_{\mu_i} \big(-P_{i}^{(n-1)}\big)^{|\mu_i|}s_{\mu_i/\eta_i}(t^{-\rho}q^{-Y_{n-1;i}^{^t}})s_{\mu_{i}^t/\xi_{i-1}}(q^{-\rho}t^{-Y_{n;i}})
\nonumber\\
&&\qquad\qquad\qquad\qquad\qquad\qquad\times \sum_{\rho_i}\big(-Q_{i}^{(n-1)}\big)^{|\rho_i|} s_{\rho_i/\eta_i}(t^{-Y_{n-1;i}}q^{-\rho}) s_{\rho_i^t/\xi_i}(q^{-Y_{n;i+1}^{t}}t^{-\rho}).
\end{eqnarray}
 We can evaluate this by generalizing the method of \cite{Iqbal:2004ne} to the refined topological vertex. We first use the identities
\begin{eqnarray}
\sum_{\mu} s_{\mu^t/\eta}({\bf x})s_{\mu/\xi}({\bf y}) &=& \prod_{i,j=1}^\infty (1+x_iy_j)\sum_{\mu}s_{\xi^t/\mu}({\bf x})s_{\eta^t/\mu^t}({\bf y}),\\
\sum_{\mu} s_{\mu/\eta}({\bf x})s_{\mu/\xi}({\bf y}) &=& \prod_{i,j=1}^\infty(1-x_iy_j)^{-1}\sum_{\mu}s_{\xi/\mu}({\bf x})s_{\eta/\mu}({\bf y}), 
\end{eqnarray}
several times
to obtain
\begin{eqnarray}
A_n(\vec{Y}_n,\vec{Y}_{n-1})
&=& \left(\prod_{a=1}^{n} t^{\frac{ ||Y_{n;a}||^2}{2}}\widetilde{Z}_{Y_{n;a}}(q,t)\right) \left(\prod_{a=1}^{n-1} q^{\frac{ ||Y_{n-1;a}^{t}||^2}{2}}\widetilde{Z}_{Y_{n-1;a}^{t}}(t,q)\right)
\nonumber\\
&& \times \prod_{1\leq a\leq b\leq n-1}\prod_{i,j=1}^\infty\left(1-P_{a}^{(n-1)}Q_{a}^{(n-1)}\cdots P_{b}^{(n-1)} q^{i-1/2-Y_{n-1;b,j}^{t}}t^{j-1/2-Y_{n;a,i}}\right)
\nonumber\\
&&\times \prod_{1\leq a\leq b\leq n-1}\prod_{i,j=1}^\infty\left(1- Q_{a}^{(n-1)}P_{a+1}^{(n-1)}\cdots Q_{b}^{(n-1)}q^{i-1/2-Y_{n;b+1,j}^{t}}t^{j-1/2-Y_{n-1;a,i}}\right)
\nonumber\\
&&\times \prod_{1\leq a\leq b\leq n-1}\prod_{i,j=1}^\infty \left(1-P_{a}^{(n-1)}Q_{a}^{(n-1)}\cdots Q_{b}^{(n-1)}q^{i-Y_{n;b+1,j}^{t}}t^{j-1-Y_{n;a,i}} \right)^{-1}
\nonumber\\
&&\times \prod_{1\leq a\leq b\leq n-2}\prod_{i,j=1}^\infty\left(1-Q_{a}^{(n-1)}P_{a+1}^{(n-1)}\cdots P_{b+1}^{(n-1)}q^{i-1-Y_{n-1;b+1,j}^{t}}t^{j-Y_{n-1;a,i}}\right)^{-1}.
\nonumber\\
\end{eqnarray}
By using the identity
\begin{eqnarray}
\prod_{i,j=1}^\infty(1-Qq^{i-1-Y_{1,j}}t^{j-Y_{2,i}}) &=& \prod_{i,j=1}^\infty(1-Qq^{i-1}t^{j})\prod_{s\in Y_2}(1-Qq^{-a_{Y_{1}^t(s)}-1}t^{-\ell_{Y_2}(s)})
\nonumber\\
&&\qquad \times \prod_{s\in Y_1^t}(1-Qq^{a_{Y_2}(s)}t^{\ell_{Y_1^t}(s)+1}),
\end{eqnarray}
we can further rewrite this as
\begin{align}
A_n(\vec{Y}_n,\vec{Y}_{n-1}) = (g_n\, h_n)\cdot A_n^{(\rm pert)} \cdot A_n^{(\rm matter)}(\vec{Y}_n,\vec{Y}_{n-1}) \cdot A_n^{(\rm gauge)}(\vec{Y}_n,Y_{n-1}),
\end{align}
where
\begin{eqnarray}
g_n &=& 
 \prod_{a=1}^{n}q^{-\frac{||Y_{n;a}^{t}||^2}{4}}t^{\frac{||Y_{n;a}||^2}{4}} \prod_{a=1}^{n-1}q^{\frac{||Y_{n-1;a}^{t}||^2}{4}}t^{-\frac{||Y_{n-1;a}||^2}{4}}
\nonumber\\
&&\times\prod_{1\leq a\leq b\leq n}\bigg[q^{-\frac{ |Y_{n;b}|}{4} - \frac{ |Y_{n;a-1}|}{4}}t^{\frac{|Y_{n;a-1}|}{4}+ \frac{ |Y_{n;b}|}{4}}\bigg] \prod_{1\leq a\leq b\leq n-1}\bigg[q^{\frac{|Y_{n-1;b}|}{4}+ \frac{ |Y_{n-1;a-1}|}{4}}t^{-\frac{ |Y_{n-1;a-1}|}{4} - \frac{ |Y_{n-1;b}|}{4}}\bigg],
\nonumber\\
\\
h_n &=& \prod_{a=1}^{n-1}(-1)^{|Y_{n-1;a}|} 
 \prod_{1\leq a\leq b\leq n-1}\Bigg[\bigg(P_{a}^{(n-1)}\bigg)^{\frac{|Y_{n-1;b}|-|Y_{n;b+1}|}{2}} \bigg(Q_{b}^{(n-1)}\bigg)^{\frac{|Y_{n-1;a}|-|Y_{n;a}|}{2}}\Bigg]\,,
\nonumber\\
\end{eqnarray}
and
\begin{align}
A_n^{(\rm pert)} =&  \prod_{i,j=1}^\infty\bigg[\prod_{1\leq a\leq b\leq n-1}(1-P_{a}^{(n-1)}Q_{a}^{(n-1)}\cdots P_{b}^{(n-1)} q^{i-1/2}t^{j-1/2})
\nonumber\\
&\times \prod_{1\leq a\leq b\leq n-1}(1-Q_{a}^{(n-1)}P_{a+1}^{(n-1)}\cdots Q_{b}^{(n-1)} q^{i-1/2}t^{j-1/2})
\nonumber\\
&  \times \prod_{1\leq a\leq b\leq n-1}(1-P_{a}^{(n-1)}Q_{a}^{(n-1)}\cdots Q_{b}^{(n-1)} q^{i}t^{j-1})^{-1}
\nonumber\\
&\times \prod_{1\leq a\leq b\leq n-2}(1-Q_{a}^{(n-1)}P_{a+1}^{(n-1)}\cdots P_{b+1}^{(n-1)} q^{i-1}t^{j})^{-1}\bigg],
\nonumber\\
\\
A_n^{(\rm matter)}
 =& \prod_{a=1}^{n}\prod_{b=1}^{n-1}\bigg\{\prod_{s\in Y_{n;a}}\left[2i\sin \bigg(\frac{E(n,n-1,a,b,s)  + m_n - \frac{\epsilon_1 + \epsilon_2}{2}}{2}\bigg)\right]
\nonumber\\
&\times \prod_{s\in Y_{n-1;b}}\left[2i\sin \bigg(\frac{ E(n-1,n,b,a,s) - m_n - \frac{\epsilon_1+\epsilon_2}{2}}{2}\bigg)\right],
\\[2mm]
A_n^{(\rm gauge)}
 =& \prod_{a=1}^{n}\prod_{s\in Y_{n;a}}\left[2i \sin\left(\frac{ E(n,n,a,a,s) }{2}\right)\right]^{-1}
\nonumber\\
& \times \prod_{a=1}^{n-1}\prod_{s\in Y_{n-1;a}}\left[2i \sin\left(\frac{E(n-1,n-1,a,a,s) - (\epsilon_1+\epsilon_2)}{2}\right)\right]^{-1}
\nonumber\\
&\times \prod_{1\leq a<b\leq n}\bigg\{\prod_{s\in Y_{n;a}}\left[2i \sin\left(\frac{E(n,n,a,b,s) - (\epsilon_1+\epsilon_2)}{2}\right)\right]^{-1}
\nonumber\\
&\qquad\qquad\qquad\times \prod_{s\in Y_{n;b}}\left[2i \sin\left(\frac{ E(n,n,b,a,s) }{2}\right)\right]^{-1}\bigg\}
\nonumber\\
&\times \prod_{1\leq a<b\leq n-1}\bigg\{\prod_{s\in Y_{n-1;a}}\left[2i \sin\left(\frac{E(n-1,n-1,a,b,s)}{2}\right)\right]^{-1}
\nonumber\\
&\qquad\qquad\qquad\times \prod_{s\in Y_{n-1;b}}\left[2i \sin\left(\frac{E(n-1,n-1,b,a,s) - (\epsilon_1+\epsilon_2)}{2}\right)\right]^{-1}\bigg\}.
\nonumber\\
\end{align}
In the last two factors, we defined $\lambda_{n;k}$ and $m_n$ so that $\sum_{k=1}^{n}\lambda_{n;k} =0$ and
\begin{align}
P^{(n-1)}_{k}Q^{(n-1)}_{k} = \exp(-i\lambda_{n;k+1} +i\lambda_{n;k} ),\qquad  P^{(n-1)}_{k} = \exp (i\lambda_{n;k}- i\lambda_{n-1;k} + im_{n}).
\end{align}
and used
\begin{eqnarray}
E(n,m,a,b,s) = \lambda_{n;a} - \lambda_{m;b} - \epsilon_1 \ell_{Y_{n;a}}(s) + \epsilon_2(a_{Y_{m;b}}(s) + 1).
\end{eqnarray}
Recall that $q=e^{-i\epsilon_2}$ and $t=e^{i\epsilon_1}$.
Here $n$ and $k$ of $\lambda_{n;k}$ and $m_n$ run over $n=2,\cdots,N$ and $k=1,\cdots,n$.
Note that this parameterization properly satisfies $P_{k}^{(n-1)}Q_{k}^{(n-1)} = Q_{k}^{(n)}P_{k+1}^{(n)}$. We now have
\begin{eqnarray}
P^{(n-1)}_{a}Q^{(n-1)}_{a}\cdots P^{(n-1)}_{b} &=& \exp( i\lambda_{n;a}-i\lambda_{n-1;b}  +im_{n}),
\\
Q^{(n-1)}_{a}P^{(n-1)}_{a+1}\cdots Q^{(n-1)}_{b} &=& \exp(i\lambda_{n-1;a}-i\lambda_{n;b+1} -im_n),
\\
P^{(n-1)}_{a}Q^{(n-1)}_{a}\cdots Q^{(n-1)}_{b} &=& \exp(i\lambda_{n;a}-i\lambda_{n;b+1}),
\\
Q^{(n-1)}_{a}P^{(n-1)}_{a+1}\cdots P^{(n-1)}_{b+1} &=& \exp(i\lambda_{n-1;a}-i\lambda_{n-1;b+1}).
\end{eqnarray}

By putting all together and use the fact that $\prod_{n=1}^N g_n = 1$, we find 
\begin{align}
A_{N}&(\emptyset,\vec{Y}_{N-1})\prod_{n=1}^{N-1}\bigg(B_{n+1,n}(\vec{Y}_n)\,A_n(\vec{Y}_{n},\vec{Y}_{n-1}) \bigg)
\nonumber\\
&= \prod_{m=1}^{N}(A_{m}^{\rm pert}h_m)\times \prod_{n=1}^{N-1}\Big(B_{n+1,n}(\vec{Y}_n)\,z_{\rm vec}(n)\Big)
\nonumber\\
&\qquad\times \prod_{a=1}^{N}z_{\rm fund}(N-1;\lambda_{N;a}+m_{N})\times \prod_{n=2}^{N-1}z_{\rm bifund}(n-1,n;m_n),
\end{align}
where $z_{\rm vec},\, z_{\rm fund}$ and $z_{\rm bifund}$ are defined in \eqref{eq:component}. Now recall that, from the toric data, we have the relation
\begin{align}
R_a^{(n)} = R_{1}^{(n)}(P_{a-1}^{(n-1)}P_{a-2}^{(n-1)}\cdots P_{1}^{(n-1)})(P_a^{(n)}P_{a-1}^{(n)}\cdots P_{2}^{(n)})^{-1},
\end{align}
which implies
\begin{align}
\prod_{m=1}^Nh_m \prod_{n=1}^{N-1}B_{n+1,n}(\vec{Y}_n)
= \prod_{n=1}^{N-1}(u_n)^{\sum_{a=1}^{n}|Y_{n,a}|}
\end{align}
with
\begin{align}
u_n \equiv R_{1}^{(n)}Q_{n}^{(n)\frac{1}{2}}P_{1}^{(n)\frac{1}{2}}( P_{2}^{(n)}\cdots P_{n}^{(n)})^{-\frac{1}{2}}(P_{1}^{(n-1)}\cdots P_{n-1}^{(n-1)})^{\frac{1}{2}}.
\end{align}
Therefore we finally obtain
\begin{align}
Z =  \left(\prod_{n=1}^{N}A_{n}^{\rm (pert)}\right)\sum_{\vec{Y}_1,\cdots\vec{Y}_{N-1}}\Bigg\{ \bigg[\prod_{n=1}^{N-1} &(u_n)^{|\vec{Y}_n|}z_{\rm vec}(n)\bigg] \bigg[\prod_{a=1}^{N}z_{\rm fund}(N-1;\tilde{m}_a)\bigg]
\nonumber\\
&\times \bigg[\prod_{n=2}^{N-1}z_{\rm bifund}(n-1,n;m_n)\bigg]\Bigg\}.
\label{eq:TN-final}
\end{align}
where $|\vec{Y}_{n}|\equiv \sum_{a=1}^{n}|Y_{n,a}|$ and $\tilde{m}_a = \lambda_{N,a} + m_N$. Here the sum over Young diagrams in \eqref{eq:TN-final} correctly gives the instanton partition function of the corresponding gauge theory. On the other hand, the first product $\prod_{n=1}^{N}A_{n}^{(\rm pert)}$ is written as
\begin{align}
\prod_{n=1}^NA_n^{\rm (pert)}
= Z_0\, Z_{U(1)}^{=},
\end{align}
with
\begin{align}
Z_0 &= \prod_{i,j=1}^{\infty}\Bigg\{\frac{\left[\prod_{ a\leq b}(1-e^{-i\lambda_{N-1;b} + i\tilde{m}_a}q^{i-\frac{1}{2}}t^{j-\frac{1}{2}})\prod_{b<a} (1-e^{i\lambda_{N-1;b}-i\tilde{m}_{a}}q^{i-\frac{1}{2}}t^{j-\frac{1}{2}})  \right]}{\prod_{n=1}^{N-1}\prod_{a< b}(1-e^{i\lambda_{n;a}-i\lambda_{n;b}}q^it^{j-1})(1-e^{i\lambda_{n;a}-i\lambda_{n;b}}q^{i-1}t^{j})}
\nonumber\\
&\qquad \times \prod_{n=1}^{N-1}\prod_{a\leq b}(1-e^{i\lambda_{n;a}-i\lambda_{n-1;b} + im_n} q^{i-1/2}t^{j-1/2})\prod_{ b < a}(1-e^{i\lambda_{n-1;b}-i\lambda_{n;a} - im_n}q^{i-1/2}t^{j-1/2})\Bigg\},
\nonumber\\
Z_{U(1)}^{=} &=  \prod_{1\leq a<b\leq N}\prod_{k,\ell=1}^\infty(1-e^{i\tilde{m}_a-i\tilde{m}_b}q^k t^{\ell-1})^{-1}.
\end{align}
Note that $(M(t,q)M(q,t))^{\frac{\chi(X)}{4}}Z_0$ gives the perturbative part of the Nekrasov partition function, while the extra factor
\begin{align}
Z_{U(1)}^{=} =  \prod_{1\leq a<b\leq N}\prod_{k,\ell=1}^\infty(1-e^{i\tilde{m}_a-i\tilde{m}_b}q^k t^{\ell-1})^{-1}.
\end{align}
is a contribution from decoupled M2-branes associated with pairs of horizontal external lines in figure \ref{fig:TN-diagram}. Thus we have proved the result \eqref{eq:Nekrasov-quiver}.

\section{$SU(N)$ gauge theory with $N_f = 2N$ flavors}
\label{app:SUN}

Our proposal of the $U(1)$ factor $Z_{U(1)}$ in section \ref{subsec:U1-factors} can be straightforwardly generalized to a higher rank $SU(N)$ gauge group. As an example, let us compute the partition function of an $SU(N)$ gauge theory with $N_f = 2N$ hypermultiplets in the fundamental representation. 

\begin{figure}[tb]
	\begin{center}
	\includegraphics[width=100mm]{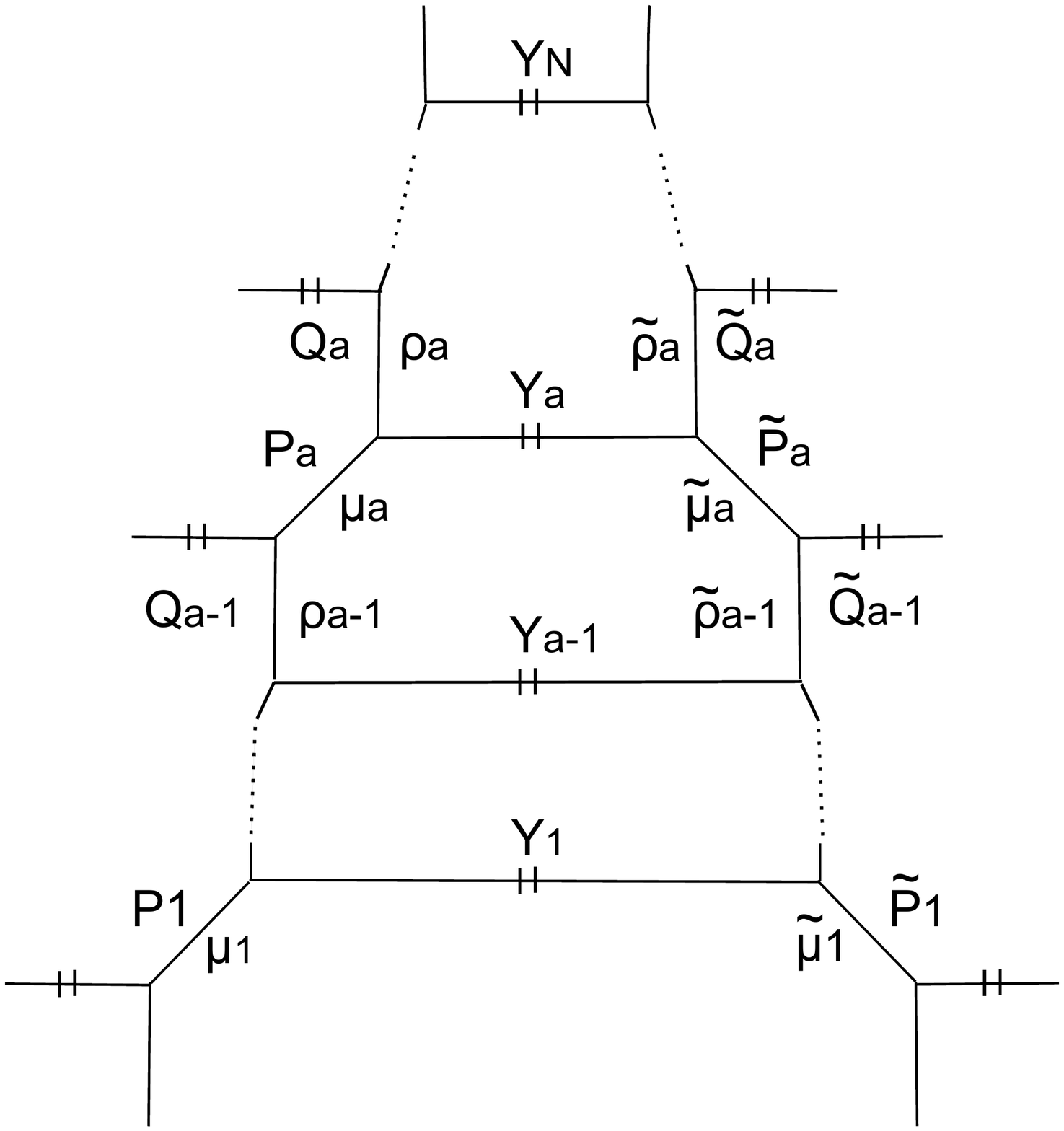}
	\caption{A web diagram for an $SU(N)$ gauge theory with $N_f = 2N$ hypermultiplets in the fundamental representation. Not all the K\"ahler parameters $P_a, \tilde{P}_a, a = 1, \cdots, N$ and $Q_a, \tilde{Q}_a, a = 1, \cdots, N-1$ are independent but subject to $Q_aP_{a+1} = \tilde{Q}_a\tilde{P}_{a+1}$. }
\label{fig:SUN}
	\end{center}
\end{figure}

The web diagram which geometrically engineers an $SU(N)$ gauge theory with $N_f = 2N$ flavors is depicted in Figure \ref{fig:SUN}. 
The partition function of the web diagram Figure \ref{fig:SUN} can be again computed by the refined topological vertex
\bea
\tilde{Z}_{SU(N)} = (M(t, q)M(q, t))^{\frac{N-1}{2}}Z(t, q, R, P_a, Q_a, \tilde{P}_a, \tilde{Q}_a),
\eea
where 
\bea
Z(t, q, R, P_a, Q_a, \tilde{P}_a, \tilde{Q}_a) \!\!&\!\!=\!\!&\!\! \sum_{\vec{\mu}, \vec{\rho}, \vec{Y}}\prod_{a=1}^N\Big[ C_{\mu_a\rho_{a-1}^t\emptyset}(q, t)(-P_a)^{|\mu_a|} C_{\mu_a^t\rho_a Y_a^t}(t, q) (-Q_a)^{|\rho_a|} f_{Y_a}(q, t)\\
&&(-R P_a^{-1}\tilde{P}_a^{-1}\cdots P_N^{-1}\tilde{P}_N^{-1})^{|Y_a|}
C_{\tilde{\rho}_{a-1}\tilde{\mu}_a\emptyset}(t, q)(-\tilde{P}_a)^{|\mu_a|}C_{\tilde{\rho}_a\tilde{\mu}_a^t Y_a}(q, t)(- Q_a)^{|\tilde{\rho}_a|}\Big] \,. \nn \label{SUN}
\eea
Here we define $\rho_0 = \rho_{N} = \emptyset$ and $\vec{\mu} = (\mu_1, \cdots, \mu_N), \vec{\rho} = (\rho_1, \cdots, \rho_{N-1}), \vec{Y} = (Y_1, \cdots, Y_N)$. 

The explicit computation of the amplitude of \eqref{SUN} can be carried out in the same manner as in appendix \ref{app:TN}. The result has been obtained in \cite{Taki:2007dh}, and in terms of our convention, Eq.~\eqref{SUN} becomes
\be
\tilde{Z}_{SU(N)} = Z_{U(1)}^{=}\cdot Z_0 \cdot Z_{\text{inst}}\,,
\ee
where the $U(1)$ factor for the horizontal parallel external lines is 
\be
Z_{U(1)}^{=} = \prod_{1 \leq a \leq b \leq N-1}\prod_{i,j=1}^{\infty}\left(1 - e^{im_a - im_{b+1}}t^{i-1}q^j\right)^{-1}\left(1 - e^{i\tilde{m}_a - i\tilde{m}_{b+1}}t^{i}q^{j-1}\right)^{-1}\,. \label{SUN-U1}
\ee
The perturbative part $Z_0$ is 
\bea
Z_0 &=& \prod_{i,j=1}^{\infty}\Big[\frac{\prod_{1\leq a \leq b \leq N}\left(1-e^{-i\lambda_b + im_a}t^{i-\frac{1}{2}}q^{j-\frac{1}{2}}\right)\prod_{1\leq a \leq b \leq N-1}\left(1-e^{i\lambda_a - im_{b+1}}t^{i-\frac{1}{2}}q^{j-\frac{1}{2}}\right)}{(1-t^iq^{j-1})^{\frac{N-1}{2}}(1-t^{i-1}q^{j})^{\frac{N-1}{2}}\prod_{1\leq a\leq b \leq N-1}\left(1-e^{i\lambda_a - i\lambda_{b+1}}t^iq^{j-1}\right)}\nn\\
&& \times \frac{\prod_{1\leq a \leq b \leq N}\left(1-e^{-i\lambda_b + i\tilde{m}_a}t^{i-\frac{1}{2}}q^{j-\frac{1}{2}}\right)\prod_{1\leq a \leq b \leq N-1}\left(1-e^{i\lambda_a - i\tilde{m}_{b+1}}t^{i-\frac{1}{2}}q^{j-\frac{1}{2}}\right)}{\prod_{1\leq a\leq b \leq N-1}\left(1-e^{i\lambda_a - i\lambda_{b+1}}t^{i-1}q^{j}\right)} \Big].
\eea
Finally, the instanton part $Z_{\text{inst}}$ becomes 
\bea
Z_{\text{inst}} = \sum_{\vec{Y}}u^{|Y_1| + \cdots + |Y_N|}\prod_{a=1}^N\prod_{Y_a}\frac{\prod_{b=1}^{N}(2i)^2\sin\left(\frac{E_{a\emptyset} - m_b + i\gamma_1}{2}\right)\sin\left(\frac{E_{a\emptyset} - \tilde{m}_b + i\gamma_1}{2}\right)}{\prod_{b=1}^N(2i)^2\sin\left(\frac{E_{ab}}{2}\right)\sin\left(\frac{E_{ab}+2i\gamma_1}{2}\right)}. \label{SUN-inst}
\eea
Note that we have chosen the gauge theory parameters as 
\bea
P_a &=& e^{-\lambda_a + im_a}, Q_a = e^{i\lambda_a - im_{a+1}}, u=RP_1^{\frac{1}{2}}P_2^{-\frac{1}{2}}\cdots P_N^{-\frac{1}{2}}\tilde{P}_1^{\frac{1}{2}}\tilde{P}_2^{-\frac{1}{2}}\cdots \tilde{P}_N^{-\frac{1}{2}},\\
\tilde{P}_a &=& e^{-\lambda_a + i\tilde{m}_a}, \tilde{Q}_a = e^{i\lambda_a - i\tilde{m}_{a+1}},
\eea
where $\sum_{a=1}^N\lambda_a = 0$. We can see that 
\be
\tilde{Z}_{SU(N)}/Z_{U(1)}^{=} = Z_0 \cdot Z_{\text{inst}}
\ee
gives the $U(N)$ partition function with $Nf = 2N$ flavors up to the perturbative $U(1)$ part.

As argued in section \ref{subsec:U1-factors}, The $U(1)$ factor of \eqref{SUN-U1} is not the only $U(1)$ factor. We also have the other two $U(1)$ factors from the parallel vertical lines at the top and the bottom of Figure \ref{SUN}. The structure of the web diagram regarding the vertical lines is essentially the same as that of the $SU(2)$ gauge theory with $N_f = 4$ flavors in section \ref{sec:Nf4}. Hence, the other $U(1)$ factors can be obtained in the same manner as in section \ref{sec:Nf4}, and the result is 
\be
Z_{U(1)}^{||} = \prod_{i,j=1}^{\infty}\left(1 - u e^{\frac{i}{2}(\sum_a m_a + \sum_a\tilde{m}_a)}q^{i-1}t^j\right)^{-1}\left(1 - u e^{-\frac{i}{2}(\sum_a m_a + \sum_a\tilde{m}_a)}q^{i}t^{j-1}\right)^{-1}.
\ee
The total $U(1)$ factor then becomes 
\be
Z_{U(1)} = Z_{U(1)}^{=} Z_{U(1)}^{||}.
\ee
Therefore, we propose that the partition function of the $SU(N)$ gauge theory with $Nf = 2N$ flavors is 
\be
Z_{SU(N)} = \tilde{Z}_{SU(N)}/Z_{U(1)}.
\ee

\providecommand{\href}[2]{#2}\begingroup\raggedright\endgroup

\end{document}